\documentclass[]{aastex631}

\usepackage{color}
\usepackage{amsmath}
\usepackage{ulem}

\begin{document}

\title{On the Detection and Characterization of Quasiperiodic Oscillations in Astronomical Time Series: Gamma-ray Burst X-ray Light Curves as a Test Case}

\author[0009-0004-9835-353X]{Fei-Fan Song}
\affiliation{Yunnan Observatories, Chinese Academy of Sciences, 650011 Kunming, Yunnan Province, People's Republic of China; jirongmao@mail.ynao.ac.cn}
\affiliation{University of Chinese Academy of Sciences, 100049 Beijing, People’s Republic of China}
\author[0000-0002-7077-7195]{Jirong Mao}
\affiliation{Yunnan Observatories, Chinese Academy of Sciences, 650011 Kunming, Yunnan Province, People's Republic of China; jirongmao@mail.ynao.ac.cn}
\affiliation{Center for Astronomical Mega-Science, Chinese Academy of Sciences, 20A Datun Road, Chaoyang District, 100012 Beijing, People's Republic of China}
\affiliation{Key Laboratory for the Structure and Evolution of Celestial Objects, Chinese Academy of Sciences, 650011 Kunming, People's Republic of China}
\begin{abstract}
The study of temporal properties of variable sources
can elucidate their physical processes. In this context, we present a critical study comparing three approaches to periodic or
quasiperiodic behavior: Gaussian process, power spectrum, and 
wavelet analysis, using celerite, Lomb-Scargle periodograms, and 
weighted wavelet-Z transforms, respectively. We use 15 Swift-X-ray Telescope light curves of short gamma-ray bursts (sGRBs) as examples. A comprehensive analysis of two sGRB X-ray light curves is performed. The results reveal the importance
of artifacts, largely in the form of false quasiperiodic oscillation signals, possibly introduced by preprocessing (such as detrending) or other aspects of the analysis.
The exploration described in this paper can be helpful for future studies of variability in GRBs, active galactic nuclei, and other astronomical sources.
\end{abstract}

\keywords{Gamma-ray bursts (629)}


\section{Introduction} \label{sec:intro}

Exploring the properties of the astronomical time series is interesting. For example, gamma-ray bursts (GRBs) are among the most violet explosions in the Universe. 
Based on the duration of the temporal pulse, GRBs can be categorized into long GRBs, lasting more than 2 s, and short GRBs, lasting less than 2 s. It is widely accepted that long GRBs originate from the collapse of massive stars, while short GRBs originate from the merger of two compact objects \citep{2015PhR...561....1K}.

Some possible periodic signals in GRB light curves have been discovered in recent studies. The studies can help us to understand the physical properties of the progenitor system and the central engine. For example, \cite{2022arXiv220502186X} found the quasiperiodic oscillation (QPO) at 22 Hz in the precursor of GRB 211211A. The duration of GRB 211211A is about 34.3 s in the 10-1000 keV band measured by Fermi. However, 
GRB 211211A has been suggested to have originated from the merger of two compact stars \citep{2022arXiv220502186X, 2022Natur.612..232Y,2022Natur.612..223R}. 
\cite{2022arXiv220502186X} discussed two interpretations of QPOs in the precursor of GRB 211211A. First, in the GRB 211211A event, it is most likely that a magnetar participated in the merger. The QPO at 22 Hz could arise from the torsional oscillations or crust oscillations of a magnetar. 
Second, the QPO could arise from the premerger interaction of magnetospheres, although this explanation is challenged due to the short lifetime of magnetars and the long spiraling time before the merger of two neutron stars. \cite{2021Natur.600..621C} discovered two significant high-frequency ($f_1$ = 2132 Hz and $f_2$ = 4250 Hz) oscillation signals during the peak stage of a giant gamma-ray flare from a magnetar. They suggested that such high-frequency QPOs could arise from neutron star oscillations or magnetic reconnection events in the magnetosphere. 
However, the underlying physical origin of QPOs in GRBs is still under debate.

Early GRB afterglow in the X-ray band might be regarded as an extension of GRB prompt emission. GRB X-ray flares can be generated by the collision of two shells with very close Lorentz factors emitted from the central engine \citep{2018A&A...615A..80P}. It is indicated that some features shown in the prompt emission might also be shown in the early X-ray afterglow as relics of the prompt emission \citep{Oganesyan2018A&A...616A.138O}. Therefore, it is speculated that the QPOs in the gamma-ray light curve might be presented in the early X-ray light curve. 
Searching QPOs in the GRB X-ray light curves could be useful in exploring the physics of the GRB progenitor. \cite{2021ApJ...921L...1Z} discovered a few QPOs in the X-ray afterglow of GRB 101225A and explained them as being dipole radiation from a newborn magnetar with precession. 
Millisecond magnetars, i.e., neutron stars with extremely strong magnetic fields of about $10^{15}$ G, could be formed during GRB explosion \citep{1992Natur.357..472U,1992ApJ...392L...9D}. Rotating neutron stars can generate magnetic dipole radiation, and the luminosity of magnetic dipole radiation is related to the inclination angle between the rotational axis and the magnetic symmetry axis \citep{2019ApJ...886....5S,2020ApJ...892L..34S,2021ApJ...921L...1Z}. The inclination angle undergoes $wobbles$ due to the precession effect, leading to oscillations in the luminosity of the dipole radiation \citep{2015MNRAS.451..695Z, 2020ApJ...892L..34S}. However, it is essential to note that these oscillations are not strictly periodic signals, as the precession frequency decreases with the spindown of the magnetar. 
On the other hand, if a magnetar is formed after the burst, the material falling back to the surface of the magnetar may generate quasiperiodic signals \citep{nsacc}. 

The periodic signals in astrophysical time series are usually studied by frequency domain methods, such as the Lomb-Scargle periodogram(LSP) method. 
In recent years, Gaussian process (GP) analysis has become popular in the astrophysical research field (a brief introduction is provided in Appendix \ref{appendix: GP}). 
The time complexity of directly computing the GP likelihood is $O(N^3)$, where $N$ is the number of data points. This means a large data set at an extremely computational cost. In this work, we utilize a fast and scalable GP method called celerite \citep{celerite}. Scalable GP methods generally fall into two classes. One is approximation, and the other is restrictions on the form of the kernel. Celerite belongs to the latter one and is specifically applicable to one-dimensional data. Unlike to the approximation method, celerite offers exact solutions within the restricted kernel space. The method restricts the kernel to the sum of arbitrary exponentials.  It can reduce the time complexity of computing the GP likelihood function from $O(N^3)$ to $O(NJ^2)$, where $J$ is the number of exponential terms in the kernel. Therefore, the computational cost scales linearly with the length of the data, given a fixed kernel. In addition, the kernel space of celerite includes the kernels of all continuous autoregressive moving average (CARMA) models. 
Using celerite is computationally less expensive for CARMA models than using a Kalman filter solver, even though they have the same time complexity of $O(NJ^2)$. Additionally, another widely used GP method restricts the kernel matrix to a Toeplitz structure, which has a time complexity of $O(N^2)$ and requires uniformly sampled data and homoscedasticity \citep{2022arXiv220908940A}. However, celerite does not require the data to be uniformly sampled.

In order to examine the results obtained by the celerite model, we apply two methods to the GRB X-ray light curves. One traditional analysis for the time series is the LSP method \citep{1976Ap&SS..39..447L, 1982ApJ...263..835S} . Another widely applied method is wavelet transform. In this paper, we treat the GP celerite, LSP, and wavelet transform methods uniformly in the examination of the GRB X-ray light curves.

We stress that our aim is not to claim a confirmed QPO signal in the GRB X-ray light curve. In this paper, we attempt to comprehensively investigate the possible QPO signal identified by celerite. We illustrate the diversities obtained by the different celerite models in detail. We also use the LSP method and wavelet transform methods to analyze the GRB X-ray light curves.
The structure of this paper is as follows. The sample selection is introduced in Section \ref{sec:data_selection}. In Section \ref{sec:GP model}, we present the kernel and the mean function of the celerite model that we used, and the details of searching QPOs are given. 
In Section \ref{disscu}, we discuss the GP results compared to the LSP results and the analysis of the X-ray light curve of GRB 101225A. Our summary is given in Section \ref{summary}.

\section{Sample Selection} \label{sec:data_selection}
We focus on the data analysis of the short gamma-ray burst (sGRB) X-ray light curves from Swift observations. 
Swift is a multiwavelength observatory with three scientific instruments: the Burst Alert Telescope (BAT), X-ray Telescope (XRT) and Ultraviolet/Optical Telescope (UVOT) \citep{2004ApJ...611.1005G}. The X-ray data can be provided by the XRT observation in the energy range of 0.3-10 keV \citep{Evans2007A&A...469..379E,Evans2009MNRAS}. Here, we consider the sGRB sample selected by \cite{2021ApJS..252...16L}.  \cite{2021ApJS..252...16L} selected 124 sGRBs in the third Swift-BAT catalog covering the period from 2004 December to 2019 July. 
Further examination of the QPO in the BAT light curves will be performed after we complete this work. 
In order to combine the analysis of the X-ray and gamma-ray data of sGRBs in future studies, our sGRB sample includes only 67 sGRBs with a signal-to-noise ratio larger than 3 in the BAT light curves \citep{2021ApJS..252...16L}. 
Furthermore, in this catalog, many sGRBs do not have sufficient data points in the X-ray light curves to enable us for the temporal analysis. We 
finally select two sGRBs (GRB 050724 and GRB 060614) with sufficient X-ray light-curve data points for our analysis.
The X-ray light curves are collected from the Swift-XRT light-curve repository\footnote{see https://www.swift.ac.uk/xrt$\_$curves/.}.

\section{Methodology}\label{sec:GP model}
\subsection{GP Celerite Model} 
\subsubsection{Kernel Selection}
\begin{figure}[h]
   \centering
   {}
   \includegraphics[width=0.55\textwidth]{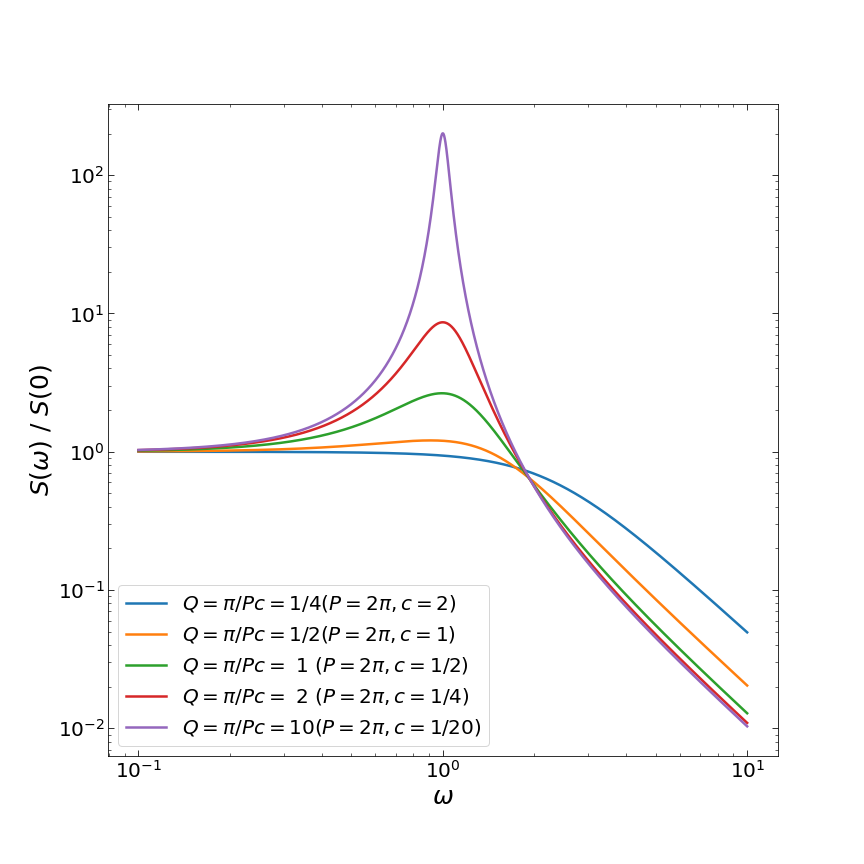}
   {}
   \includegraphics[width=0.55\textwidth]{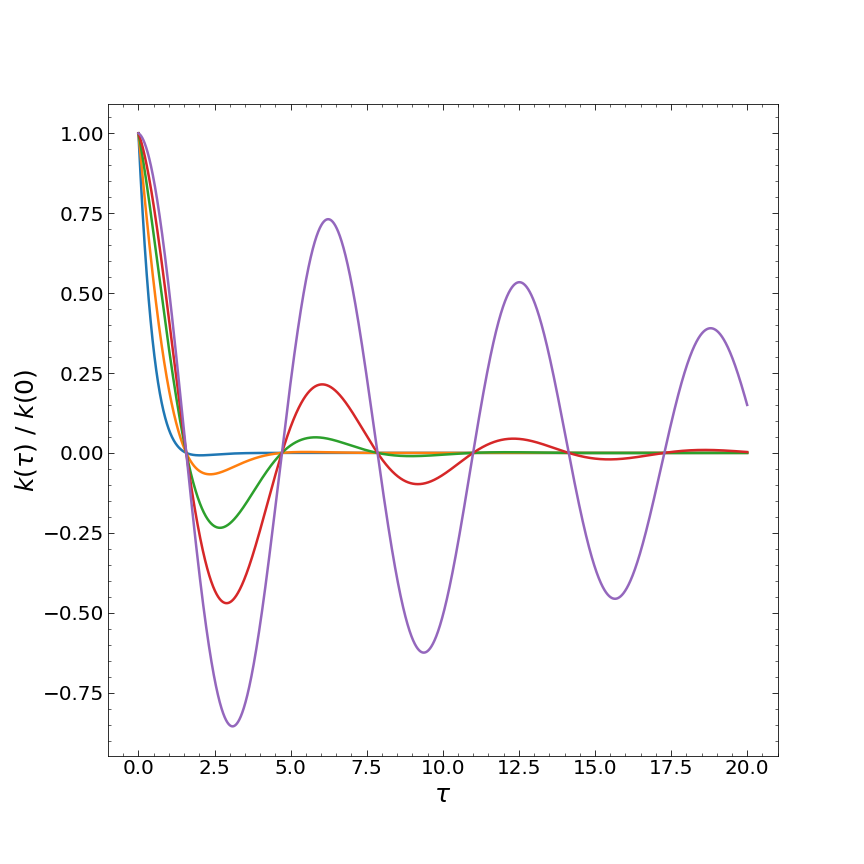}
    \caption{Top panel: PSDs of a quasiperiodic oscillator are plotted for several values of the quality factor $Q$. Bottom panel: autocorrelation functions corresponding to the cases in the top panel.}
    \label{fig: GPkernel}
\end{figure}

Celerite is a powerful tool applied in the time-domain astronomy, as it enables one to perform the analysis of the complex-structured data \citep{celerite}. 
The kernel of celerite can be constructed by products or sums of arbitrary exponentials, allowing us to construct effective celerite models to fit real data. Analyzing periodic signals in time series is possible by inferring the posterior power spectral density (PSD) from a certain celerite model. 
In this work, we aim to search for QPO signals in the sGRB X-ray light curves. For this purpose, we choose a quasiperiodic kernel, which is the combination of an exponential and a cosine function. This kernel can be interpreted as a quasiperiodic oscillator presented by 
\begin{equation}\label{kerneleq}
        k_{\rm QPO}(\tau_{nm})=ae^{-c\tau_{nm}}\cos(\frac{2\pi }{P} \tau_{nm}),
\end{equation}
where $a$ is the amplitude, $P$ is the period of the quasiperiodic oscillator, $c$ is reciprocal of the damping timescale, and $\tau_{nm}$ = $|t_n-t_m|$ is time lag between data points $t_n$ and $t_m$. The quality factor of the quasiperiodic oscillator is  $Q$ = $\pi/Pc$. According to Wiener-Khintchine theorem, if the PSD exists, the kernel and the spectral density are Fourier duals of each other \citep{GP}. Thus, we can obtain the PSD of the quasiperiodic kernel as
\begin{equation}
        S(\omega)=\frac{1}{\sqrt{2\pi}}\frac{a}{c}[\frac{1}{1+(\frac{\omega-\frac{2\pi}{P}}{c})^2}+\frac{1}{1+(\frac{\omega+\frac{2\pi}{P}}{c})^2}],
\end{equation}
where $\omega$ is the angular frequency. We set $a$ = 1, $P$ = $2\pi$, and $c$  = 1/4, 1/2, 1, 2, 10 to generate the kernel and the PSD, which are shown in Figure \ref{fig: GPkernel}. From the figure, one can see that for $Q$ $\leq$ 1/2 (overdamped), the quasiperiodic oscillator exhibits no periodicity. In this case, the model can be presented as the damped random walk(DRW) process, which shows the red noise at high frequency and the white noise at low frequency. On the other hand, when $Q>1/2$ (underdamped), the kernel exhibits quasiperiodic oscillatory behavior, accompanied by a distinct peak in the corresponding PSD, with the peak becoming more pronounced as $Q$ increases. When $Q$ $\gg$ 1/2, the model exhibits high-amplitude QPO. The significance 
can be assessed by the magnitude of the parameter $Q$. \cite{2020ApJ...895..122C} employed this kernel to investigate QPOs in the optical/gamma-ray light curves of blazars. They employed both periodic and nonperiodic kernel function models 
to fit the data. By utilizing the Bayesian information criterion (BIC) for model comparison, they examined whether the periodic kernel model was favored over the nonperiodic model. In this work, we estimate the posterior distribution of the celerite model parameters using the Markov Chain Monte Carlo (MCMC) method, followed by computing the quality factor $Q$ for the quasiperiodic kernel. Thus, the presence and strength of periodic signals 
are determined based on the quality factor $Q$.

There is a kernel with a PSD similar to that of $k_{\rm QPO}$, named the damped simple harmonic oscillator (SHO) kernel. This kernel is also applied in our subsequent analyses. The expression for the SHO kernel is \citep{celerite}
\begin{equation}
    k_{\rm SHO}(\tau_{nm}) = S_0\omega_0 Q \exp(-\frac{\omega_0\tau_{nm}}{2Q})\left\{
    \begin{aligned}
      &\cosh(\eta\omega_0\tau)+\frac{1}{2\eta Q}\sinh(\eta \omega_0 \tau),&0\leq Q\leq1/2 \\ 
       &2(1+\omega_0\tau),&Q=1/2\\
       &\cos(\eta \omega_0\tau)+\frac{1}{2\eta Q}\sin(\eta \omega_0\tau),&1/2\leq Q
       \end{aligned}\ ,
        \right.
        \label{eq: SHO kerenl}
 \end{equation} 
where $\eta = |1-(4Q^2)^{-1}|^{1/2}$, $\omega_0$ is the angular frequency corresponding to the QPO signal, $S_0$ is proportional to the power in the PSD when $\omega$ equals $\omega_0$, and $Q$ is the quality factor, representing the strength of the QPO. The expression for the PSD function of the SHO model is given as \citep{celerite}
\textbf{
\begin{equation}
    S_{\rm SHO}(\omega)=\sqrt{\frac{2}{\pi}}\frac{S_0\omega_0^4}{(\omega^2-\omega_0^2)^2+\omega_0^2\omega^2/Q^2}\ .
\label{eq:psd sho}    
\end{equation}}

A QPO kernel and SHO kernel can be applied in the search for a QPO. We can also employ the DRW kernel for temporal analysis. The PSD of the DRW kernel resembles the PSD of red noise. Consequently, the DRW model cannot capture potential periodicity in light curves. The expression for this kernel is \citep{celerite}
\begin{equation}
    k_{\rm DRW}(\tau_{nm})=a\exp(-c\tau_{nm})\ ,
    \label{eq:DRW kernel}
\end{equation}
where $a$ is the amplitude, $c$ is the reciprocal of the damping timescale, and $\tau_{nm}$ = $|t_n-t_m|$ is the time lag between data points $t_n$ and $t_m$.
The expression for the PSD of the DRW model is given as \citep{celerite}
\begin{equation}
S_{\rm DRW}(\omega)=\sqrt{\frac{2}{\pi}}\frac{a}{c}\frac{1}{1+(\frac{\omega}{c})^2}\ .    
\label{eq:psd drw}
\end{equation}

\subsubsection{Mean Function} 

\begin{figure}
    \centering
    \includegraphics[width=0.85\textwidth]{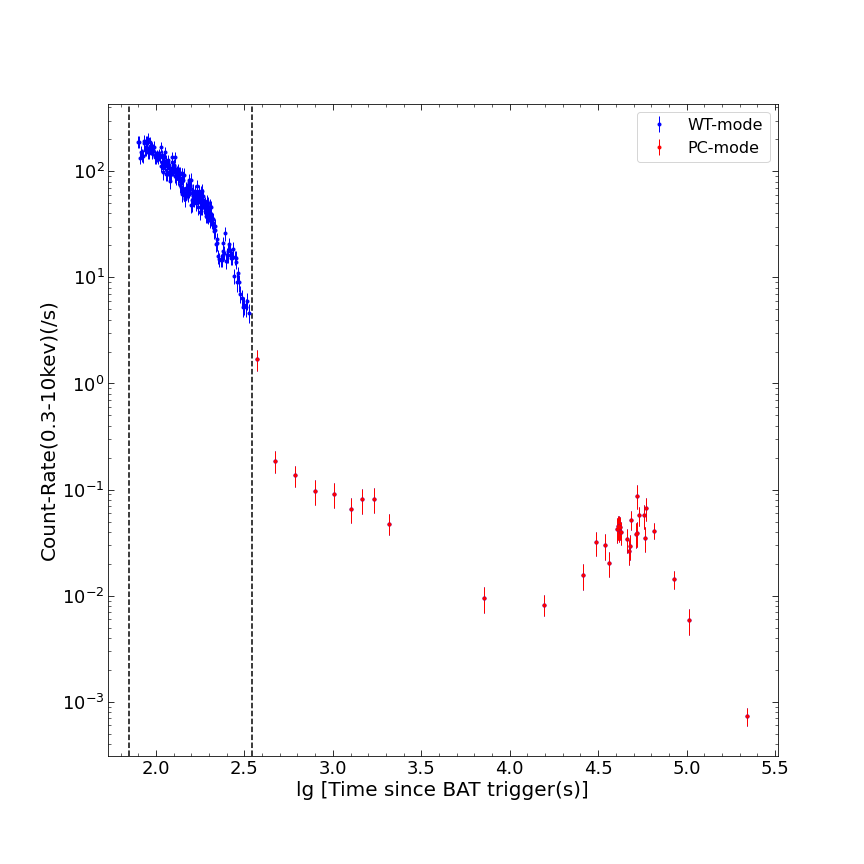}
    \caption{X-ray light curve of GRB 050724. The blue and red dots denote the data points obtained from the WT (windowed timing) mode and the PC (photon counting) mode, respectively. We perform a periodic analysis of the data within the time interval between two vertical dashed lines.}
    \label{fig: lcGRB050724}
\end{figure}

In general, the X-ray light curve of GRBs can be modeled as power-law functions with varying indices in different segments \citep{2015PhR...561....1K}. The power-law functions as the trend in the X-ray light curve of GRBs can be used as the mean function in the celerite model. For example, the X-ray light curve of GRB 050724 within the time interval of [70,350] s exhibits a break at 175 s (see Figure \ref{fig: lcGRB050724}). We thus employ a power-law function with the index $\alpha$ to model the X-ray light curve before 175 s, and a power-law function with the index $\beta$ to model the X-ray light curve after 175 s as
\begin{equation}\label{meanfunction}
M(t)= \left\{\begin{aligned}
&A(\frac{t}{100 s})^{-\alpha}, \ \ t \textless 175 s, \\  
&B(\frac{t}{100 s})^{-\beta},\ \  t \geq 175 s. 
\end{aligned}
\right.
\end{equation}
We ensure the continuity of the mean function at $t$ = 175 s by setting $A(175/100)^{-\alpha}$ = $B(175/100)^{-\beta}$. 
Similar to the mean function of the GRB 050724 X-ray light curve, the mean functions of other GRB X-ray light curves can also be presented as a piecewise power-law function. 

In order to compare the effect of the mean function in the celerite analysis, We can utilize a polynomial function as the mean function in the celerite model. The expression for the polynomial can be expressed as
\begin{equation}
M(t) = p_0+\frac{p_1}{10^{3}}t+\frac{p_2}{10^{6}}t^{2}.   
\end{equation}

In addition to the above two time-varying mean functions, the following analysis also incorporates a time-invariant mean function, where the mean function is equal to a constant as
\begin{equation}
    M(t)=\rm const.
\end{equation}
We choose this mean function when we use the celerite model to study the detrended light curves.

We can select different kernels and different mean functions. 
For simplicity, we label the \textbf{celerite model as GP(Term$_1$ and Term$_2$)}, where the first term represents the mean function used in the celerite model, and the second term represents the kernel used in the celerite model. 
This paper covers various approaches to detrended light curves by subtracting the best-fit mean function of different celerite models. For simplicity, these are labeled as detrended(Term$_1$ and Term$_2$) light curves. 


\subsection{LSP Method}
The LSP method is one of the traditional methods researchers use to analyze time series. We use the traditional LSP method to analyze the GRB X-ray light curves and investigate the possible QPO signals. For the same X-ray light curve, we perform both the GP celerite analysis and the LSP analysis. The results can be compared. We show each of results from the GP celerite method and from the LSP method in the same plot. All the related issues are presented in detail in Appendix \ref{append B}.     

\subsection{Weighted Wavelet Z-transform Method}
Wavelet transform is widely used to search the periodic signal in time series. It is a time-frequency domain method that provides information about how different frequency components in time series data vary over time. In addition to the GP celerite and LSP methods, wavelet transform is also used to analyze the GRB X-ray light curves in this paper. Here, we adopt the weighted wavelet Z-transform \citep[WWZ;][]{wwz} to analyze the GRB X-ray light curves. For the WWZ-transform, the Morlet wavelet is utilized. We note that the detrended process should be performed for an X-ray light curve before we utilize the WWZ analysis.  

\subsection{Temporal Analysis}

We analyze the X-ray light curves of the two sGRBs using a celerite model.
The likelihood function for a general GP model is described in Equation \ref{Equation: likelihood}, and we can obtain the likelihood function from the celerite package. The maximum likelihood estimation (MLE) parameters ($\phi$, $\theta$) for the kernel and the mean function are estimated using the L-BFGS-B nonlinear optimization routine. The parameter numbers obtained from the MLE are applied as the initial input parameter numbers of the MCMC procedure. 

We utilize the emcee package to perform MCMC sampling and estimate the posterior probability density of the parameters. Due to the limitations of data length and data sampling frequency, reliable period estimates have upper and lower bounds. For this case, we constrain the prior of parameter $P$ between $P_{\rm min}=2\pi\bar{\delta t}$ and $P_{\rm max}=\frac{2\pi}{5}\times \rm baseline$, where $\bar{\delta t}$ represents the average time interval between adjacent data points, and the $\rm baseline$ is the time interval of the selected light-curve segment. 
The upper and lower limits of the confidence frequency in the PSD are determined by $1/P_{\rm min}$ and $1/P_{\rm max}$, which is consistent with the confidence frequency range set by \cite{Burke2021}.
Uniform distributions are chosen as priors for all parameters. 

We calculate the PSD of the celerite model for each MCMC sample. This allows us to estimate the median PSD of the celerite model and the 1$\sigma$ uncertainties in the PSD. To assess the significance of the QPO for the celerite model, we compute the probability distribution of quality factor $Q$ (= $\pi/Pc$). For comparison, we also utilize the LSP method 
implemented in the astropy package to obtain the PSDs of the GRB X-ray light curves. 

We study the issue of trending, which has effects on the analysis.
Here, we initially compare the PSD obtained from the analysis of the original data using both the GP method and the LSP method. 
Then, upon detrending the original data by subtracting the best-fit mean function in the GP model, we modify the mean function in the GP model to be a constant. Following this adjustment, we proceed to investigate the PSD of the detrended light curve using both the GP and LSP methods. The PSD obtained from the GP model represents variations relative to the mean function. Therefore, selecting a nonconstant mean function is analogous to a detrending approach. 
When we investigate the trending light curve using the LSP method, an additional approach is required to eliminate the trend in the light curve. In theory, the PSD of a GP model with a predetermined 
trend (nonconstant mean function) is expected to be the same as the PSD of a GP model with a zero mean function obtained by subtracting this predetermined trend \citep{GP}. However, in practical analysis, we do not treat the parameters of the GP model as fixed values but as random variables. Sometimes, these random variables (model parameters) are not entirely independent, potentially leading to differences between the two PSDs.

\begin{figure}
    \centering
    \includegraphics[width=0.45\textwidth]{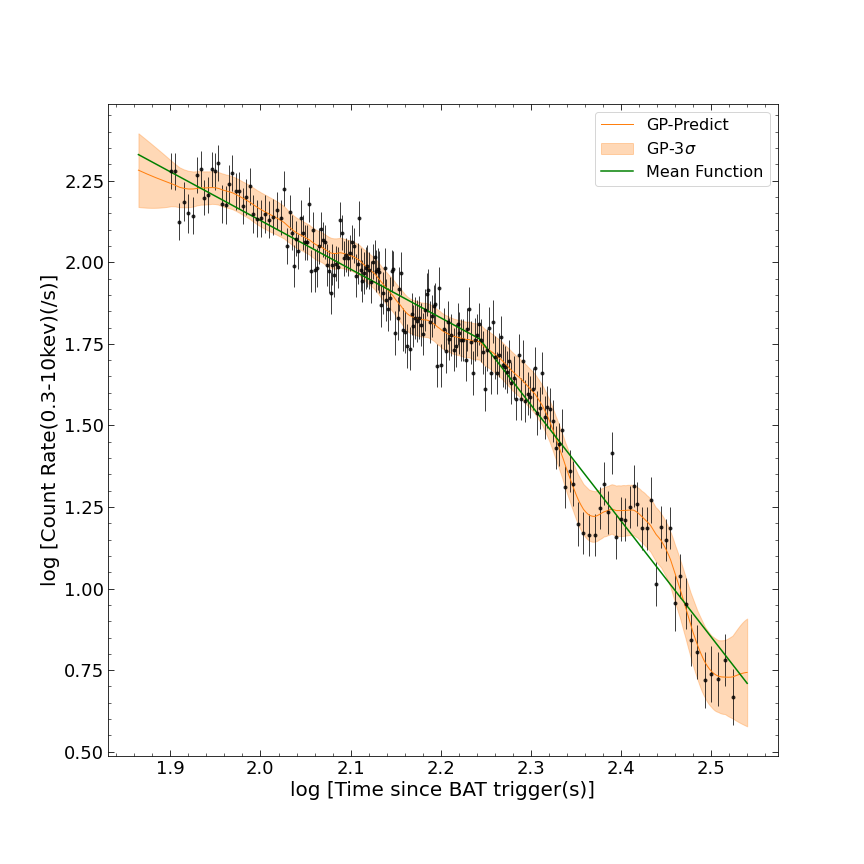}
    {}
    \includegraphics[width=0.45\textwidth]{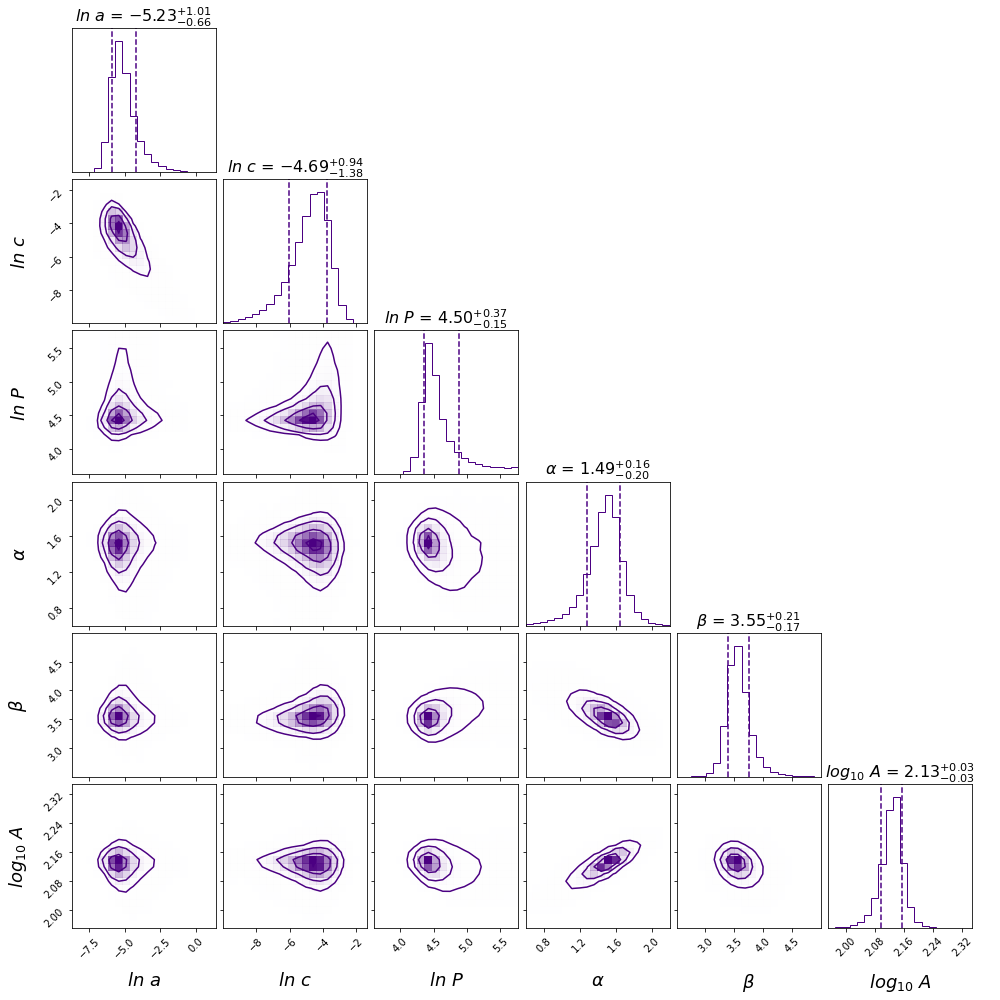}
    \label{fig: original MCMC}
    {} 
    \includegraphics[width=0.45\textwidth]{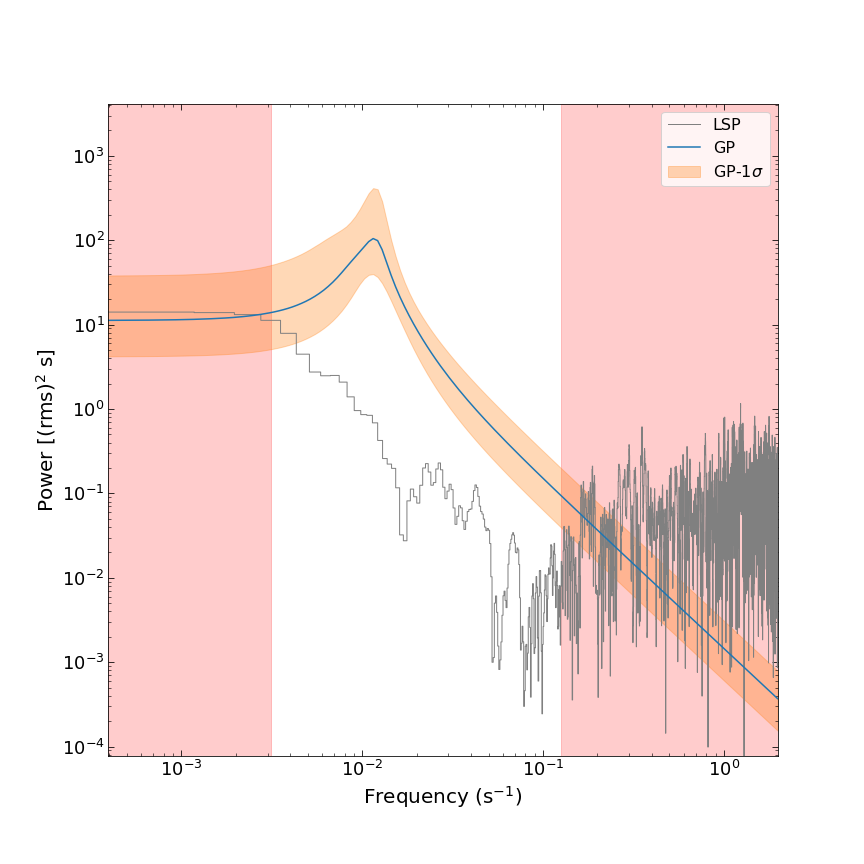}
    \label{fig: original PSD}
    {}
    \includegraphics[width=0.45\textwidth]{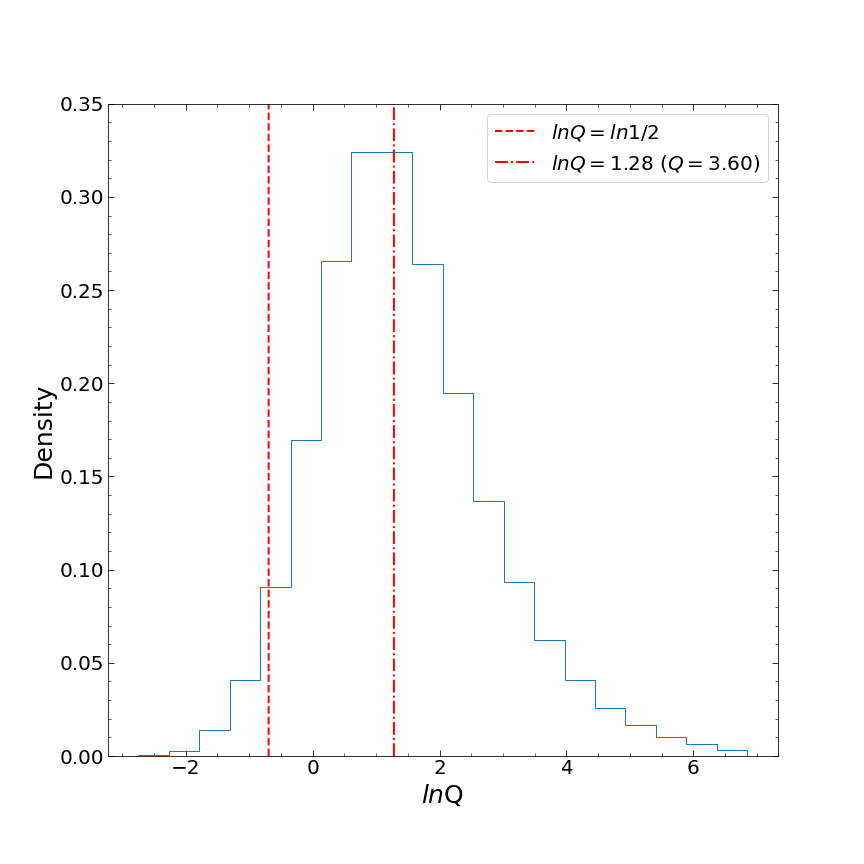}
    \label{fig: original Q}
    \caption{
    The results of the GP(power law, QPO) model analysis for GRB 050724. Top-left panel: the X-ray light curve of GRB 050724 in the time interval of [70-350] s. The black points denote the data observed by Swift, while the orange solid line depicts the prediction of the GP(power law, QPO) model. The orange-shaded region indicates the 3$\sigma$ uncertainties in the model, and the green line represents the best-fit value of the mean function. Top-right panel: The posterior probability density distribution of each parameter in the GP(power law, QPO) model for GRB 050724. The vertical dashed lines indicate the 1$\sigma$ uncertainties. Bottom-left panel: the PSDs of the light curve of GRB 050724. The average PSD of the GP(power law, QPO) model and the PSD obtained from the LSP method are represented by the blue line and the gray line, respectively. The orange-shaded area indicates the 1$\sigma$ uncertainties in the GP(power law, QPO) model PSD. The red-shaded area indicates the unreliable zone, and the white area indicates the confidence zone where the upper and lower limits are 1/$P_{min}$ and 1/$P_{max}$, respectively. The PSD of the GP(power law, QPO) model has an obvious peak at $P$ =90.02s ($f$ = 1.11 $\times$ $10^{-2}$ Hz). Bottom-right panel: the distribution of the quality factor $Q$. The dashed line and the dashed-dotted line indicate the critical quality factor ($Q$ = 0.5) and the median of ln $Q$, respectively.}
\label{fig: result of original GRB050724}
\end{figure}

We show the results of each case of the analysis of the GRB X-ray ligh curves in Figures \ref{fig: result of original GRB050724} - Figure \ref{fig: 060614 dcl GP(poly,drw)}.

\begin{figure}
    \centering
    \includegraphics[width=0.95\textwidth]{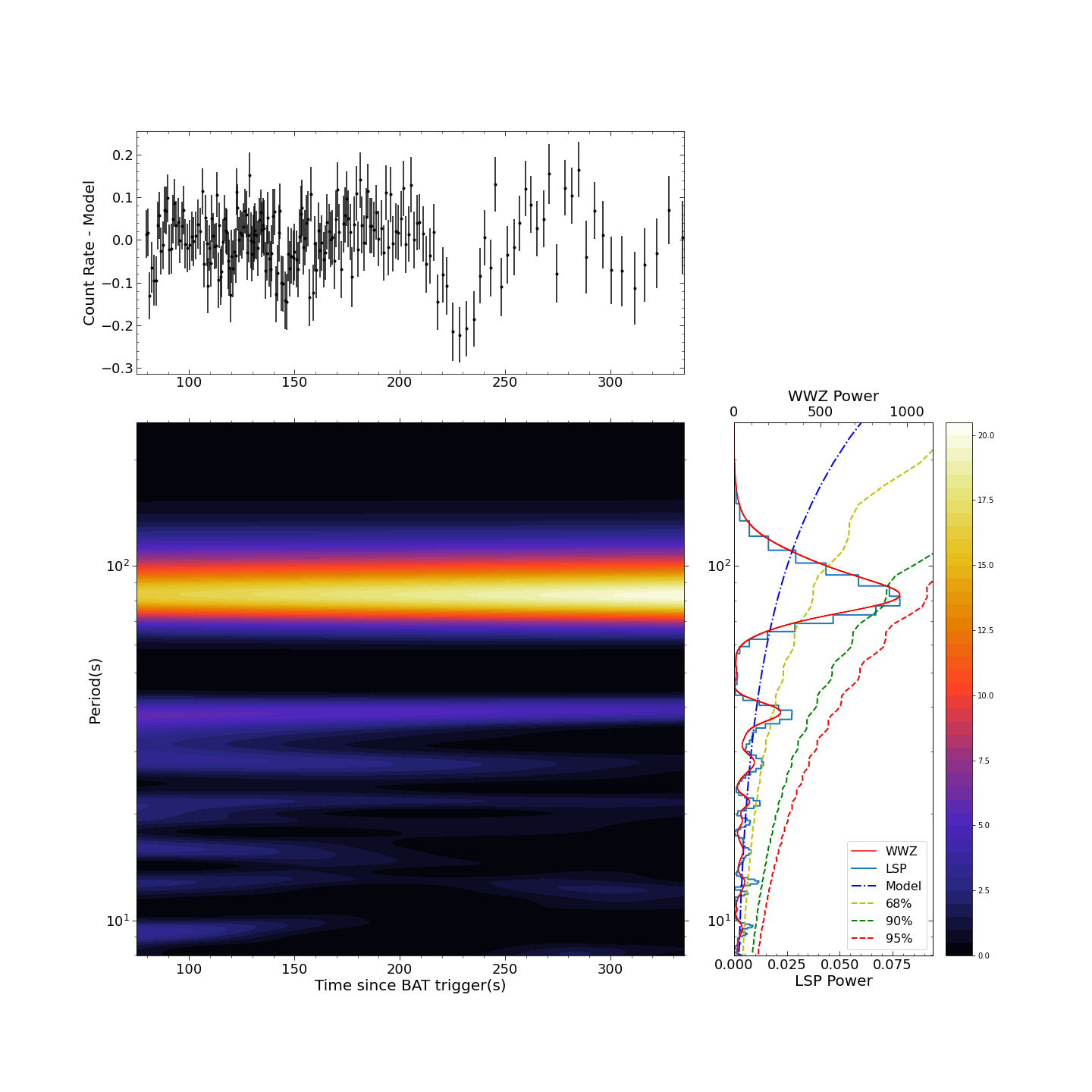}
    \caption{The WWZ analysis of the GRB 050724 X-ray light curve.}
    \label{fig: GRB 050724 WWZ}
\end{figure}

\begin{figure}
    \centering
    \includegraphics[width=0.95\textwidth]{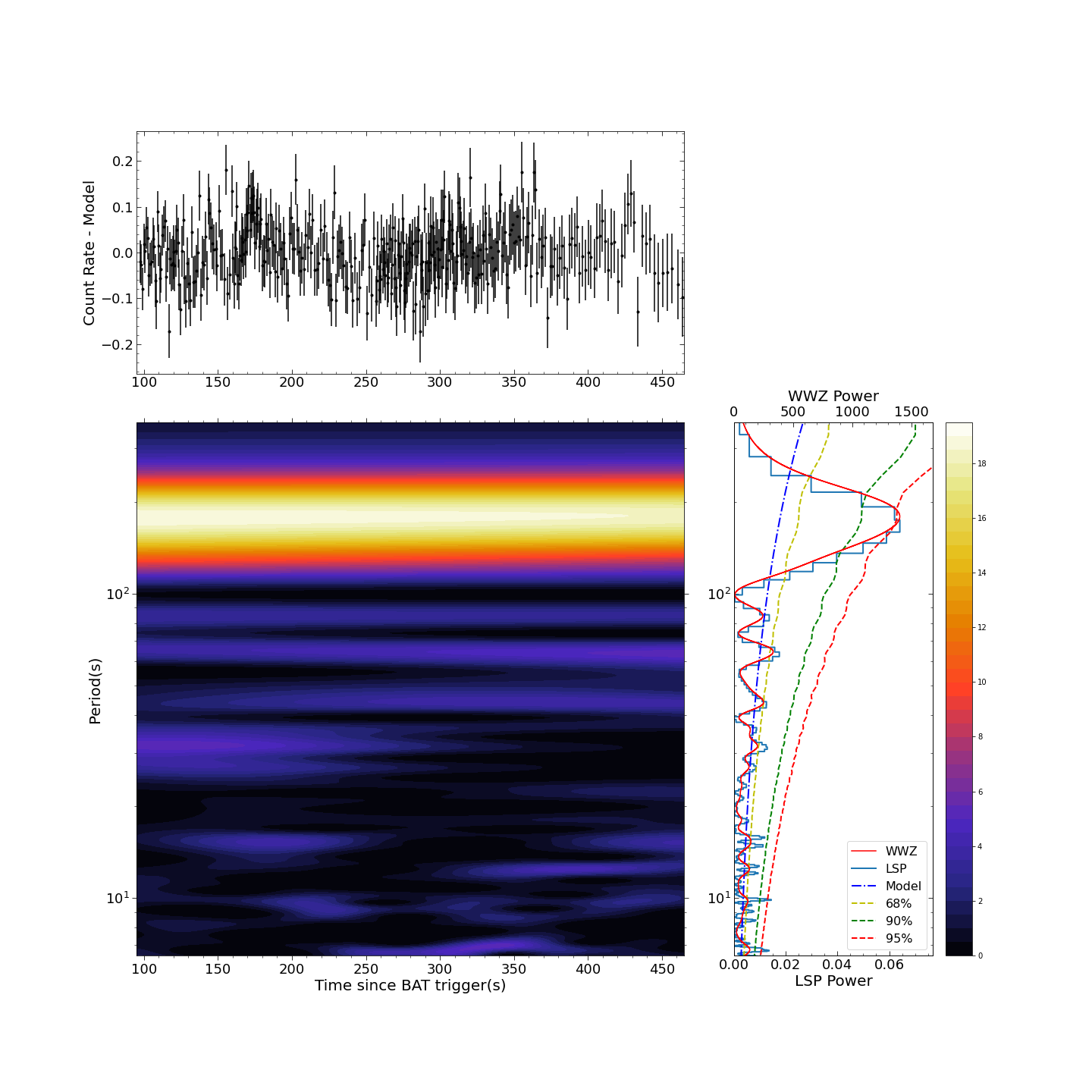}
    \caption{The WWZ analysis of GRB 060614 X-ray light curve.}
    \label{fig: GRB 060614 WWZ}
\end{figure}

We apply WWZ-transform to analyze the X-ray light curve of two GRBs. Prior to the analysis, we performed polynomial fitting to capture the decay trend in the light curves of the two GRBs and subtracted the best-fit polynomial function from the original data. The top panels of Figures \ref{fig: GRB 050724 WWZ} and \ref{fig: GRB 060614 WWZ} show the detrended light curves of GRB 050724 and GRB 060614, respectively.

The results of the WWZ method analysis of the X-ray light curve of GRB 050724 are shown in Figure \ref{fig: GRB 050724 WWZ}.
The bottom-left panel of Figure \ref{fig: GRB 050724 WWZ} shows the results of the wavelet analysis on the X-ray light curve of GRB 050724. In this panel, a periodicity of 82.9 s is presented throughout the entire light curve. 
The LSP method is also employed to obtain the PSD of this light curve. The significance of the peak around 80 s in the PSD is greater than 90\% (see the bottom-right panel of Figure \ref{fig: GRB 050724 WWZ}). The magnitude of this period is consistent with the magnitude of the period identified through the celerite method. 
In the bottom-right panel of Figure \ref{fig: GRB 060614 WWZ}, we present the results of the time-integrated WWZ power (shown by the red solid line), which is quite similar to the PSD obtained by the LSP method.

The results of the WWZ method analysis of the X-ray light curve of GRB 060614 are shown in Figure \ref{fig: GRB 060614 WWZ}. In the bottom-left panel, a persistent periodic signal throughout the entire light curve is observed. The PSD obtained by the LSP method also reveals a peak with a significance greater than 95\% near this periodic signal (see the bottom-right panel of Figure \ref{fig: GRB 060614 WWZ}). The period is approximately 180 s, roughly half the length of the light curve, making this period somewhat speculative. However, the celerite method does not detect this period, which may be due to the choice of the mean function. 

\subsection{Results}

We analyze the X-ray light curves of GRB 050724 and GRB 060614 by the GP celerite, LSP and WWZ methods. The results are summarized in Table 1. All the issues related to the GP results are presented in detail in Appendix \ref{append B}. In addition, we also examine the X-ray light curve of GRB 101225A by the three analysis methods mentioned above. The issues related to the X-ray light curve of GRB 101225A and the analysis process are discussed further in Section \ref{GRB 101225A} and Appendix \ref{appendix: result of GRB 101225A}.

The treatment performed in this paper is a simple test. We further note that, in principle, three parts are included in the time series. One is $S(t)$, which is the true, most likely stochastic, source flux variation, e.g. colored noise with a power-law power spectrum. Another is $T(t)$, which is also referred to as an $external\ trend$. It may be an artifact of the observational instrument or some other component of the source that is neither stochastic nor the QPO of interest. The last one is QPO$(t)$, which is the quasiperiodic signal of interest. It could be a sinusoidal signal whose frequency varies (randomly or deterministically) or a random process that inherently has a QPO power spectrum. The general observational noise in X-ray and gamma-ray observations is not additive noise, white or otherwise, and not Gaussian.  It is more like a process that obeys the a Poisson distribution with a variable event rate parameter, even if this is an approximation. Here, we further discuss the issue of the trend $T(t)$ in a general way. Usually, it is difficult to identify and remove an external trend. $S(t)$ can have what appears to be a trend in the sense that a particular realization of the random process may $accidentally$ have e.g. a linear behavior, even though, on average, there is no trend. Often there is only one realization of the underlying random source variability process, so this $apparent\ trend$ is hard to identify, and it is not even clear if it should be removed. To complicate things further, if $T(t)$ and QPO$(t)$ are missing -- if the data from pure colored noise are detrended, the result can be a spurious peak in the power spectrum -- a false QPO signal.

\section{Discussion}\label{disscu}
\subsection{Discussion of the GP Results Compared to the LSP Results}
\label{sec:discuss}
GP is not a unique method used for temporal analysis. The LSP method, in comparison to the GP method, is widely applied to analyze astrophysical time series. The results from the GP method and those from the LSP method may differ. For example,
\cite{2019MNRAS.482.1270C} employed a frequency domain periodogram method to study the light curves of PG 1553+113 and PKS 2155–304. However, they did not find any significant periodic signals in the optical and gamma-ray light curves. Nevertheless, \cite{2020ApJ...895..122C} reanalyzed the optical and gamma-ray light curves of these blazars using the GP method. They discovered a periodic signal with a period of approximately 2.2 yr in the optical and gamma-ray light curves of PG 1553+113. In the gamma-ray light curve of PKS 2155–304, they also found a weak significance of a periodic signal. 

We have investigated the effect of trending in our work. For the X-ray light curves of GRB 050724 and GRB 060614, it seems that we cannot identify the effect of the detrending using celerite when we choose a certain kernel. For the original light curves analyzed by the LSP method, there is no peak shown in the PSD, while the peak signature is usually shown in the PSD for the detrended light curves analyzed by the LSP method.
It seems that the LSP method requires the light curve to be detrended before application, and its result might be affected by the method of detrending \citep{Auchère_2016}. 

Different kernels may yield different results. 
The kernel is critical as it characterizes the relationship between two data points. The details of the kernel selection can be found in \cite{GP}. 
In this paper, we attempt to use the QPO, SHO, and DRW kernels, respectively.
However, the physical mechanisms of the time variability in the light curves are complex, and the selected kernel with a certain mean function may not completely describe the corresponding physical mechanisms. Although the celerite method allows for increasing the number of exponentials in a kernel to accurately match a light curve, the use of too many exponentials may sacrifice the interpretability of the kernel and may cause degeneracies among some of the parameters. 

We should further mention one issue concerning the LSP method. The model underlying it is a single sinusoidal signal. If there are two or more sinusoids of very different frequencies, the periodogram might work well, and one can find the corresponding peaks in the LSP. However, we may consider the nonsinusoidal signals. The damped sinusoids can be introduced into the LSP method\footnote{https://bayes.wustl.edu/glb/lomb.pdf}. Simulation studies can explore how well the LSP deals with such nonsinusoidal signals in the future.

\subsection{Analysis of the X-Ray Light Curve of GRB 101225A}
\label{GRB 101225A}
It is interesting to use our procedures to examine the periodicity of the GRB X-ray light curve detected by other reference works. \cite{2021ApJ...921L...1Z} employed the LSP method to analyze the XRT data of GRB 101225A and discovered two significant periodic signals with the confidence levels exceeding 90\% ($P_1$ = 460 s, $P_2$ = 228 s) within the time interval of [4900, 7500] s. In order to examine the possible QPOs mentioned above, we performed a celerite analysis 
on the XRT data of GRB 101225A in the time interval of [5050, 7150] s. The complex results are presented in Appendix \ref{appendix: result of GRB 101225A}. 



\section{Summary}
\label{summary}
We investigate the detection of QPO in astronomical time series in general. The GRB X-ray light curves provide a simple example of this kind of exploration. 
We select two GRBs in an sGRB sample. Their X-ray light curves are carefully examined by the GP celerite model. Diversities in the results are shown as we adopt different kernels and different mean functions.
The LSP method is also used for the light-curve analysis. We also utilize the WWZ method for further examination. In particular, some artifacts could be introduced in the analysis of a QPO, resulting in a false QPO signal when one performs the detrending process.
We expect that the comprehensive investigation of the GRB X-ray light curves in this paper can be a reference when we reveal GRB physics by temporal analysis in the future.
From the results provided in this paper, we further suggest a few general comments on the analysis of a QPO in astronomical time series. First, sometimes, a trend that is a long-term flux decreasing or increasing is shown in a light curve. In principle, the effect of trending on the analysis of a QPO should be considered. Second, when one performs an analysis of a QPO, several different methods can be utilized. It is helpful to compare the results obtained by the different methods before a solid conclusion is drawn. Third, when applying a certain method to the light curve, one should pay attention to the conditions of the method for the application.
The exploration in this paper can be helpful for future studies of variability in GRBs, AGNs, and other astronomical sources.

\section*{acknowledgments}
We appreciate the detailed suggestion made by the referee.
This work is supported by the National Key R\&D program (2023YFE0101200), the National Natural Science Foundation of China (NSFC 12393813), and the Yunnan Revitalization Talent Support Program (YunLing Scholar Project). This work made use of data supplied by the UK Swift Science Data Center at the University of Leicester. We use the packages of Astropy \citep{2022ApJ...935..167A}, celerite \citep{celerite}, corner \citep{corner}, emcee \citep{2013PASP..125..306F}, Matplotlib \citep{2007CSE.....9...90H}, NumPy \citep{2020Natur.585..357H}, SciPy \citep{2020NatMe..17..261V}, Stingray \citep{2019ApJ...881...39H}, taufit\citep{Burke2021}, and WWZ\footnote{http://doi.org/10.5281/zenodo.375648}.





\appendix
\section{Simple Description of the GP} \label{appendix: GP}
GP is a type of stochastic process \citep{2022arXiv220908940A}. If the stochastic dataset $y$ = $(y_1,\ y_2,\ y_3,\ \cdots)^T$ at the coordinates $x$ = $(x_1,\ x_2,\ x_3,\ \cdots)^T$ (in this paper, the coordinate is time) is multivariate gaussian, we refer to y as a GP. A GP is characterized by the mean function $(M(x))$ and the kernel function $(k(x,\ x^{\prime}))$, where the kernel can be understood as a covariance or autocorrelation function \citep{GP}. The mean function describes the deterministic trend of the dataset, and the kernel describes the relationships among data points. Periodicity is a special relationship among the data points. Thus, in principle, we may use the GP method with a certain kernel to search the periodicity in a dataset. 


In a GP model, either a kernel or a mean function can be parameterized by a set of parameters. The parameters are referred to as the hyperparameters. Specifically, $\phi$ and $\theta$ are used to label the parameters of the kernel and the mean functions, respectively. According to the definition of GP, the log-likelihood can be written as \citep{celerite}:
\begin{equation}
 \ln L(\phi, \theta)= \ln p(y | x, \phi, \theta)\\ = -\frac{1}{2}(y-M_\theta(x))^T K_\phi^{-1}(y-M_\theta(x))\\-\frac{1}{2}\ln|K_\phi|-\frac{N}{2}\ln(2\pi),\label{Equation: likelihood}
\end{equation}
where $M_\theta(x)$ is the mean function, and the elements of the matrix $K_\phi$ are given by $[K_\phi]_{nm}$ = $k_\phi(x_n, x_m)$. The celerite method reduces the computational cost of a general GP by constraining the covariance function to be a sum of arbitrary exponentials \citep{celerite}.
We then consider Bayes' theorem as 
\begin{equation}
        p(\phi, \theta|y)=\frac{L(y|\phi, \theta)p(\phi, \theta)}{p(y)},
\end{equation}
where $p(\phi,\theta|y)$ is the posterior probability of the parameters, $L(y|\phi,\theta)$ is the likelihood of the data given the parameters, $p(\phi,\theta)$ is the prior of the parameters, and $p(y)$ is a constant known as the model evidence, which is related to the model selection. However, it is difficult to obtain the joint posterior distribution by analytical methods. Therefore, the MCMC method is generally used to obtain samples from the posterior distribution \citep{2022arXiv220908940A}.

\section{Detailed Results of the X-Ray Light Curves of GRB 050724 and GRB 060614}\label{append B}
\subsection{Analysis of the X-Ray Light Curve of GRB 050724}
We analyze the original X-ray light curve of GRB 050724 by the GP(power law, QPO) model.
We find that the PSD of the GP(power law, QPO) model has an obvious peak at $f$ = $1.11\times 10^{-2}$ Hz ($P$ = 90.02 s), which means that celerite method has found a possible QPO signal in the X-ray light curve of GRB 050724. The median of ln $Q$ is 1.28 ($Q$ = 3.6 $\textgreater$ 0.5), and 96.3\% of MCMC samples have $Q$ greater than 0.5. However, we note that the PSD obtained from the LSP method does not show a significant peak near $P$ = 90.02 s. The PSD shows red noise-like behavior in the reliable frequency interval.
The results of the analysis by the GP(power law, QPO) model on the original X-ray light curve of GRB 050724 are presented in Figure \ref{fig: result of original GRB050724}.

\begin{figure}
    \centering
    \includegraphics[width=0.45\textwidth]{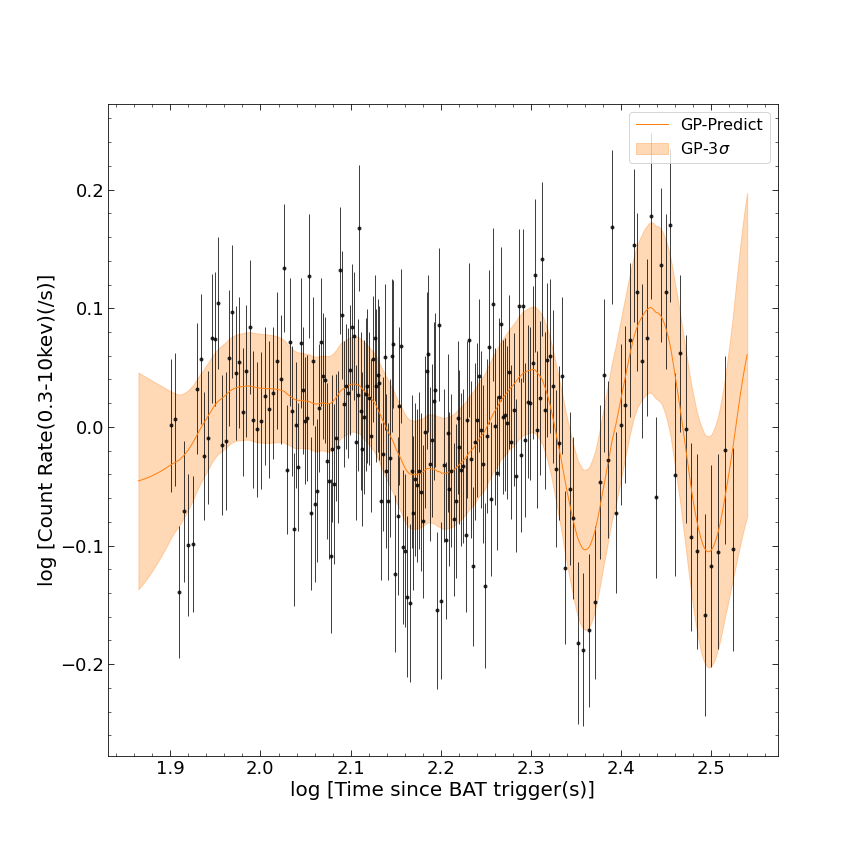}
    \label{fig: detrend MCMC}
    {}
    \includegraphics[width=0.45\textwidth]{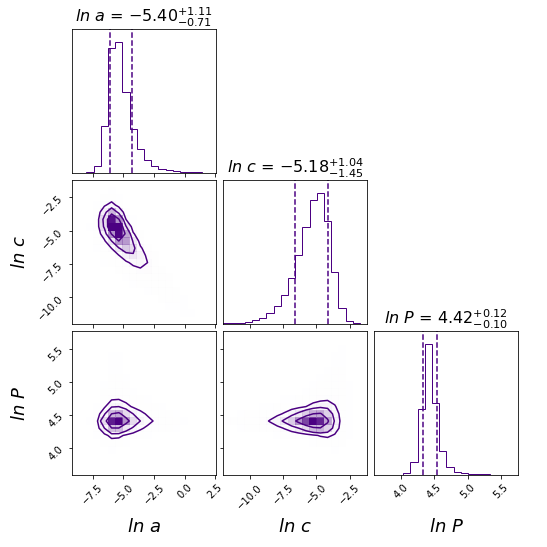}
    \label{fig: detrend MCMC}
    {}
    \includegraphics[width=0.45\textwidth]{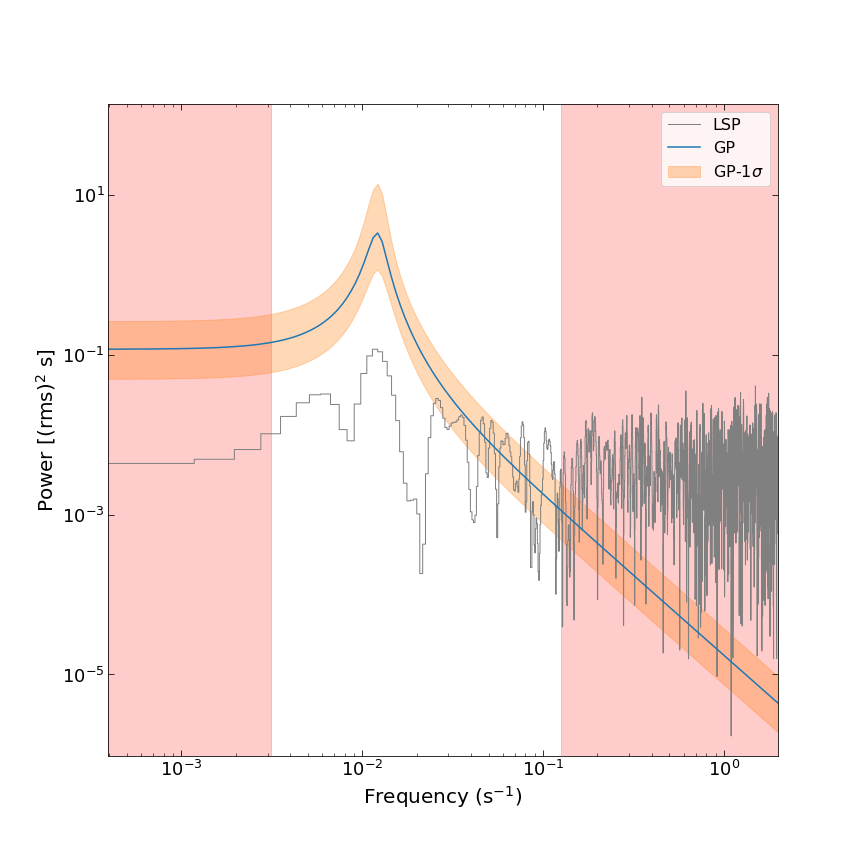}
    \label{fig: detrend PSD}
    {}
    \includegraphics[width=0.45\textwidth]{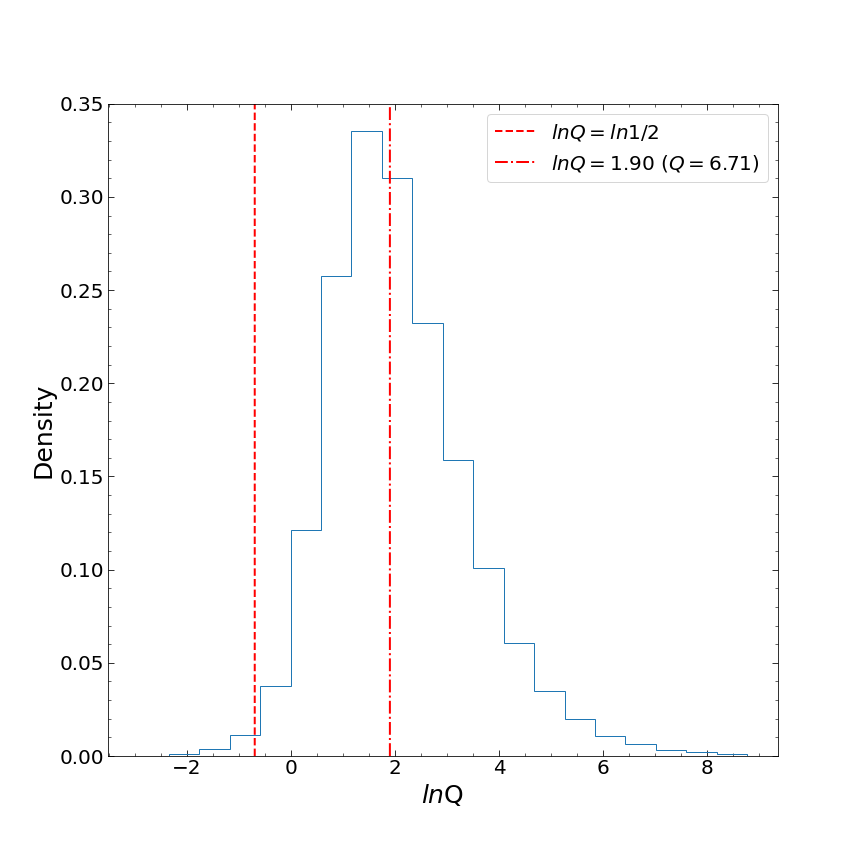}
    \label{fig: detrend Q}
    \caption{
    The results of the GP(constant, QPO) model analysis for the detrended GRB 050724 light curve. The detrended light curve is obtained by subtracting the best-fit mean function in the GP(power law, QPO) model from the original light curve. Top-left panel: the detrended X-ray light curve of GRB 050724 in the time interval of [70-350] s. The black points denote the detrended data, while the orange solid line depicts the prediction of the GP(constant, QPO) model. The orange-shaded region indicates the 3$\sigma$ uncertainties in the model. Top-right panel: the posterior probability density distribution of each parameter in the GP(constant, QPO) model for GRB 050724. The vertical dashed lines indicate the 1$\sigma$ uncertainties. Bottom-left panel: the PSDs of the light curve of GRB 050724. The average PSD of the GP(constant, QPO) model and the PSD obtained from the LSP method are represented by the blue line and the gray line, respectively. The orange-shaded area indicates the 1$\sigma$ uncertainties in the GP(constant, QPO) model PSD. The red-shaded area represents the unreliable zone, and the white area represents the confidence zone where the upper and lower limits are 1/$P_{min}$ and 1/$P_{max}$, respectively. The PSD of the GP(constant, QPO) model has an obvious peak at $P$ = 83.10s ($f$ = 1.20 $\times$ $10^{-2}$ Hz), and the PSD obtained from the LSP method has a suspected peak at $P$ =82.12s ($f$ = 1.22 $\times$ $10^{-2}$ Hz).  Bottom-right panel: the distribution of the quality factor $Q$. The dashed line and the dashed-dotted line indicate the critical quality factor ($Q$ = 0.5) and the median of ln $Q$, respectively.}
    \label{fig: result of detrend}
\end{figure}

We then consider whether the rapid declining trend in the time interval of [70, 350] s 
can have an effect on 
the detection of the QPO using the celerite method. We apply the GP(constant, QPO) model to analyze the detrended(power law, QPO) light curve. 
The PSD of the GP(power law, QPO)model has an obvious peak at $f$ = 1.20 $\times$ $10^{-2}$ Hz\ ($P$ = 83.10 s). The median of ln $Q$ is 1.90 ($Q$ = 6.71 $>$ 0.5), and 99.2\% of MCMC samples have $Q$ greater than 0.5. The PSD obtained from the LSP method has a suspected peak at 1.22 $\times$ $10^{-2}$ Hz\ ($P$ = 82.12 s). To estimate the significance of the QPO obtained from the LSP method, we first simulate the PSD of the LSP method by the red noise model as $S(f)$ = $mf^{k}$. 
Then, the ratio of the power ($I(f_j)$) obtained by the LSP method from the data to the power ($S(f_j)$) from the simulation at a given frequency $f_j$ follows an exponential distribution \citep{Hubner2022ApJS..259...32H}. We thus adopt the Whittle likelihood for the whole dataset, which is calculated as the product of the likelihoods for each frequency \citep{Hubner2022ApJS..259...32H} as $L(I|S)=\prod_{j=1}^{N}\frac{1}{S(f_j)}exp(-I(f_j)/S(f_j))$, where $N$ is the number of data points in the confidence zone. 
We finally estimate the parameters of the PSD model ($S(f_j)$) by the maximum likelihood method. Using the best-fit model $S(f)$, we generate 2 $\times$ $10^5$ light curves implemented in the Stingray package and obtain the PSD of each light curve by the LSP method. We use the exponential distribution of $I(f_j)/S(f_j)$ to calculate the significance of the suspected peak. We find that this suspected peak exceeds the red noise model by a significance level of only 95\%. In this detrended light curve, the QPO signal is detected by both the celerite and the LSP methods.
The results of the analysis of the detrended light curve are presented in Figure \ref{fig: result of detrend}.

For GRB 050724, we can use different mean functions and different kernels to further study its X-ray light curve.
Using the GP(polynomial, QPO) model, we see a signal in the original X-ray light curve of GRB 050724 with a frequency of $f$ = 1.13 $\times$ 10$^{-2}$ Hz ( $P$ = 88.23 s). The median of the quality factor ln $Q$ is 1.21 ($Q$ = 3.35 $\textgreater$ 0.5), and 96.30\% of the MCMC samples have $Q$ greater than 0.5. However, in the vicinity of the signal detected by this model, the LSP method does not detect any significant signals. The results of the analysis of the original light curve are presented in Figure \ref{fig: 050724 ocl GP(poly,qpo)}.
The GP(constant, QPO) model identifies a signal in the detrended (polynomial, QPO) light curve of GRB 050724 with a frequency of $f$ = 1.20 $\times$ 10$^{-2}$ Hz ($P$ = 83.10 s). The median  of the quality factor ln $Q$ is 1.79 ($Q$ = 5.97), and 99.30\% of the MCMC samples have $Q$ greater than 0.5. The LSP method identifies a peak in the PSD of the detrended(polynomial, QPO) light curve at $f$ =1.22 $\times$ 10$^{-2}$ Hz ($P$ = 82.13 s). The results of the analysis of the detrended(polynomial, QPO) light curve are presented in Figure \ref{fig:050724 dcl GP(poly,qpo)}.

When we analyze the original X-ray light curve of GRB 050724 by the GP(polynomial, SHO) model, the median of the quality factor ln $Q$ is -0.99 ($Q$ =0.37 $\textless$ 0.5), and only 44.80\% of MCMC samples have $Q$ values greater than 0.5. This implies that the model does not identify a significant QPO signal. The LSP method, likewise, does not reveal any significant QPQ signals. The results of the analysis of the original light curve are presented in Figure \ref{fig: 050724 ocl GP(poly,sho)}.
We then apply the GP(constant, SHO) model to the detrended(polynomial, SHO) light curve. The model reveals a QPO signal at 75.78s, the median of quality factor ln $Q$ is 1.23 ($Q$ =3.42), and 93.20\% of MCMC samples have $Q$ greater than 0.5.
The LSP method detects a peak at $f$ =1.22 $\times$ 10$^{-2}$ Hz ($P$ = 82.13 s). The results of the analysis of the detrended(polynomial, SHO) light curve using this model are presented in Figure \ref{fig: 050724 dcl GP(poly,sho)}. 

We examine the original X-ray light curve of GRB 050724 by the GP(power law, DRW) model. The PSD of the GP(power law, DRW) model and the PSD obtained by the LSP method both exhibit a red noise-like behavior within the credible frequency range. The results of the analysis of the original light curve using this model are presented in Figure \ref{fig: 050724 ocl GP(power,drw)}.
We then examine the detrended(power law, DRW) light curve by the GP(constant, DRW) model. The PSD of the GP(constant, DRW) model and the PSD obtained by the LSP method exhibit similar trends at low frequencies. However, as the frequency approaches the upper limit of the credible frequency range, the decreasing trend in the PSD obtained by the LSP method gradually becomes slower, while the PSD of the GP model continues to exhibit a single power-law decrease. The results of the analysis of the detrended(power law, DRW) light curve using this model are presented in Figure \ref{fig: 050724 dcl GP(power,drw)}.

We apply the original X-ray light curve of GRB 050724 by the GP(polynomial, DRW) model. The PSD of the GP(polynomial, DRW) model and the PSD obtained by the LSP method both exhibit a behavior similar to red noise in the credible frequency range. The results of the analysis of the original light curve using this model are presented in Figure \ref{fig: 050724 ocl GP(poly,drw)}.
We then apply the detrended(polynomial, DRW) light curve by the GP(constant, DRW) model. The PSD of the GP(constant, DRW) model and the PSD obtained by the LSP method exhibit similar trends at low frequencies. However, as the frequency approaches the upper limit of the credible frequency range, the decreasing trend in the PSD obtained by the LSP method gradually becomes slower, while the PSD of the GP model continues to exhibit a single power-law decrease. The results of the analysis of the detrended(polynomial, DRW) light curve using this model are presented in Figure \ref{fig: 050724 dcl GP(poly,drw)}.

Finally, we calculate the Bayesian Information Criterion (BIC) for the GP(power law, QPO) model and the GP(power law, DRW) model. 
It is shown that the BIC value for the GP(power law, QPO) model is smaller than that for the GP(power law, DRW) model. Here, the BIC is presented as
\begin{equation}
\rm{BIC}=-2\log(\rm likelihood)+K\log(N),
\end{equation}
where $N$ is the number of data points, and $K$ is the number of parameters. However, the difference between the BIC of GP(power law, QPO) model and GP(power law, DRW) model is only 5.59 ($\textless$ 10), indicating that the advantage of the GP(power law, QPO) model is not statistically significant. We note that the detrended cases are not suitable for BIC comparison, as the fitting parameters of the mean function in the GP(constant, QPO) model and those in the GP(constant, DRW) model are different in practice.

\subsection{Analysis of the X-Ray Light Curve of GRB 060614}
\begin{figure}
    \centering
    \includegraphics[width=0.85\textwidth]{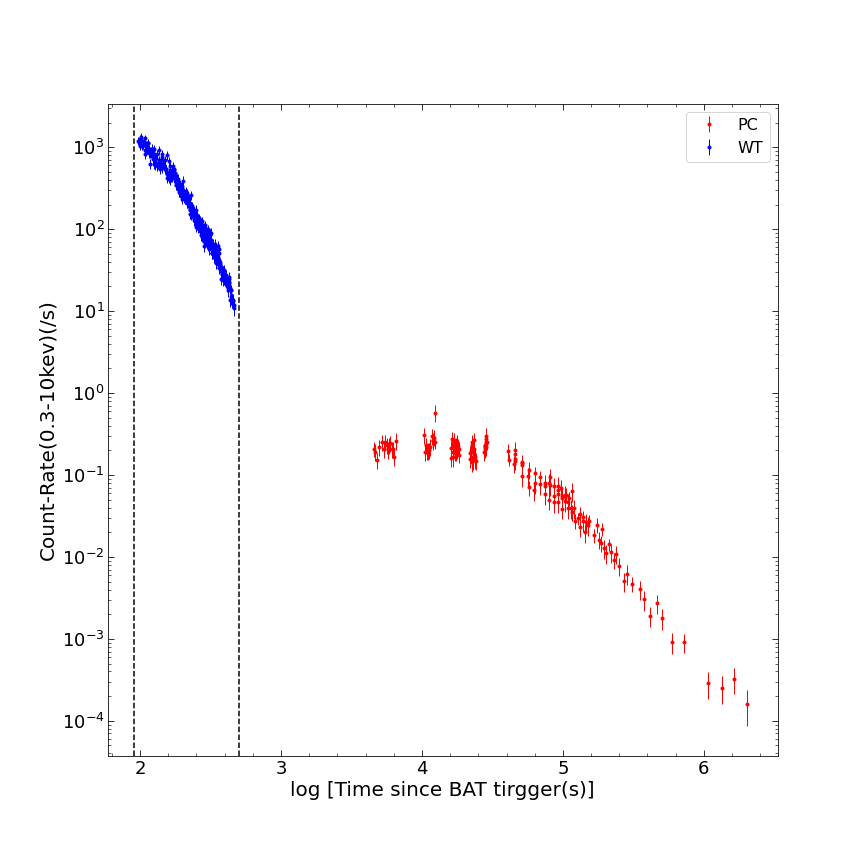}
    \caption{X-ray light curve of GRB 060614. The blue and red dots denote the data points obtained from the WT mode and the PC mode, respectively. We perform periodic analysis of the data within the time interval between two vertical dashed lines.}
    \label{fig: lc060614}
\end{figure}

The GP models performed for the X-ray light curve of GRB 050724 can also be applied to analyze the X-ray light curve of GRB 060614 in the time interval of [90-500] s. The complete X-ray light curve of GRB 060614 is shown in Figure \ref{fig: lc060614}, and we select the data points between the two vertical dotted lines in the figure for analysis.

We fit the original X-ray light curve of GRB 060614 with the GP (power law, QPO) model. Although the PSD of this model shows a significant QPO signal, the median of quality factor ln $Q$ is 3.39 ($Q$ = 29.72), and 99.69\% of the MCMC samples have $Q$ greater than 0.5. However, the period ($P$ = 367.95 s) is comparable to the total length of the light curve (366.61 s), indicating that this is a false QPO signal. The LSP method does not detect any significant QPO signals. The results of the analysis of the original light curve by this model are presented in Figure \ref{fig: 060614 ocl GP(pl,qpo)}.
When we study the detrended(power law, QPO) X-ray light curve of GRB 060614 by the GP(constant, QPO) model, a similar situation arises. This model also detects a significant QPO signal, and the median of the quality factor ln $Q$ is 3.91 ($Q$ =50.15), and 99.97\% of MCMC samples have a $Q$ greater than 0.5. The period ($P$ =357.81s) is also comparable to the total length of the light curve (366.61 s), indicating that this QPO signal is unreliable. The power in the PSD obtained by the LSP method near the position of this false QPO signal is higher compared to the power at other frequencies. The results of analysis of the detrended(power law, QPO) light curve by this model are presented in Figure \ref{fig: 060614 dcl GP(pl,qpo)}.

When we study the original X-ray light curve of GRB 060614 with the GP(polynomial, QPO) model, the model finds a QPO signal with $f$ = 4.47 $\times$ 10$^{-3}$ Hz ($P$ = 223.63 s), the median of quality factor ln $Q$ is 0.38 ($Q$ =1.46 $\textgreater$ 0.5), and only 78.29\% of the MCMC samples have $Q$ greater than 0.5. The value of this QPO signal is greater than half of the length of the light curve. This means that the QPO signal repeats itself less than twice in the duration of the light curve, and this detected QPO signal is not reliable. The PSD obtained by the LSP method does not reveal any significant QPO signals. The results of the analysis of the original light curve by this model are presented in Figure \ref{fig: 060614 ocl GP(poly,qpo)}.
We then apply the GP(constant, QPO) model to study the detrended(polynomial, QPO) light curve of GRB 060614, and this model identifies a QPO signal with a period ($P$ = 183.09 s) approximately half the length of the light curve. The median of the quality factor ln $Q$ is 0.39 ($Q$ =1.48), and 75.12\% of MCMC samples have $Q$ greater than 0.5. It implies that the QPO signal is not significant. The PSD obtained by the LSP method exhibits a broad peak in the vicinity where the QPO signal is identified by the model. The results of the analysis of the detrended(polynomial, QPO) light curve by this model are presented in Figure \ref{fig: 0606014 dcl GP(poly,qpo)}.

We use the GP(polynomial, SHO) model to study the original X-ray light curve of GRB 060614. The PSDs of the model do not show any significant QPO signals. The median of the quality factor ln $Q$ for the model is -4.39 ($Q$ = 0.01), and 8.80\% of the MCMC samples have $Q$ greater than 0.5. Similarly, the LSP method does not detect any QPO signals. The PSD of the GP(polynomial, SHO) model and the PSD obtained from the LSP method in the credible frequency range both exhibit characteristics similar to red noise PSD. The results of the analysis the original light curve by this model are presented in Figure \ref{fig: 060614 ocl GP(poly,sho)}.
We then use the GP(constant, SHO) model to study the detrended(polynomial, SHO) light curve. The PSD of the model does not exhibit any prominent peaks, indicating that the model has not detected any QPO signals. The median of the quality factor ln $Q$ is -1.32 ($Q$ = 0.27) for the mode, and 36.83\% of the MCMC samples have $Q$ greater than 0.5. The LSP method also fails to detect any QPO signals. The results of the analysis the detrended(polynomial, SHO) light curve by this model are presented in Figure \ref{fig: 060614 dcl GP(poly,sho)}.


We employ the GP(power law, DRW) model to analyze the original X-ray light curve of GRB 060614. The PSD of the GP(power law, DRW) model and the PSD obtained from the LSP method both exhibit a trend of single power-law decrease (red noise) within the credible frequency range. The posterior distribution of parameters for this model shows improved convergence compared to the convergence of parameters when fitting the original X-ray light curve with the GP(power law, QPO) model. The results of the analysis of the original light curve by this model are presented in Figure \ref{fig: 060614 ocl GP(pl,drw)}. 
We then employ the GP(constant, DRW) model to analyze the detrended(power law, DRW) light curve of GRB 060614. The PSD of the GP(power law, DRW) model and the PSD obtained by the LSP method exhibit inconsistent behavior in the credible frequency range. The PSD of the GP model displays a trend of a single power-law decrease within the credible frequency range, while the PSD obtained by the LSP method appears like white noise at high frequencies. The posterior distribution of parameters for this model shows improved convergence compared to the convergence of parameters when fitting the detrended(power law, QPO) light curve with the GP(constant, QPO) model. The results of the analysis the detrended(power law, DRW) light curve by this model are presented in Figure \ref{fig: 060614 dcl GP(pl,drw)}. 

We employ the GP(polynomial, DRW) model to examine the original X-ray light curve of GRB 060614. The PSD of the GP(polynomial, DRW) model and the PSD obtained by the LSP method exhibit similar trends within the credible frequency range. Both PSDs exhibit a trend of single power-law decline (red noise). The posterior distribution of parameters for this model shows improved convergence compared to the convergence of parameters when fitting the original X-ray light curve with GP(polynomial, QPO). The results of the analysis of the original light curve by this model are presented in Figure \ref{fig: 060614 ocl GP(poly,drw)}.
We then employ the GP(constant, DRW) model to analyze the detrended(polynomial, DRW) light curve of GRB 060614. The PSD obtained from the GP(constant, DRW) model and the PSD obtained from the LSP method show similar trends at low frequencies. However, as the frequency increases, the PSD of the GP(constant, DRW) model gradually turns to a single power-law decline, while the PSD obtained from the LSP method turns to be shown as white noise. The posterior distribution of parameters for this model shows improved convergence compared to the convergence of parameters when fitting the detrended(polynomial,QPO) light curve with the GP(constant, QPO) model. The results of the analysis of the detrended(polynomial, DRW) light curve by this model are presented in Figure \ref{fig: 060614 dcl GP(poly,drw)}.

\section{QPOs of the GRB 101225A X-Ray Light Curve}\label{appendix: result of GRB 101225A}

\begin{figure}
    \centering
    \includegraphics[width=0.85\textwidth]{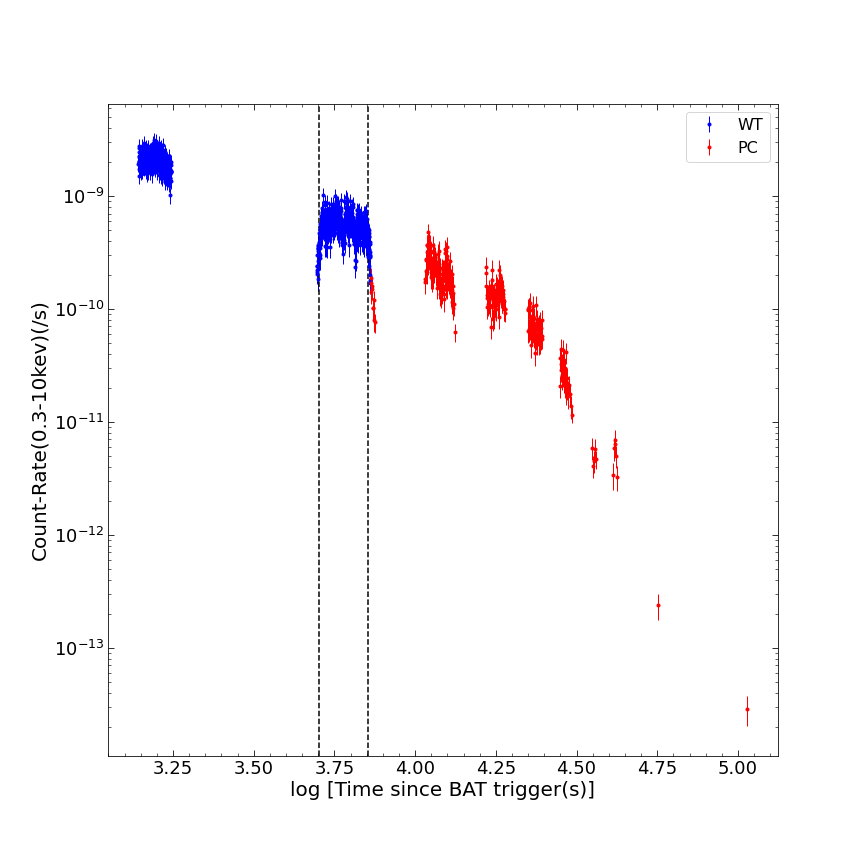}
    \caption{X-ray light curve of GRB 101225A. The blue and red dots denote the data points obtained from the WT mode and the PC mode, respectively. We perform a periodic analysis of the data within the time interval between two vertical dashed lines.}
    \label{fig: lcGRB101225A}
\end{figure}

\begin{figure}
    \centering
    \includegraphics[width=0.45\textwidth]{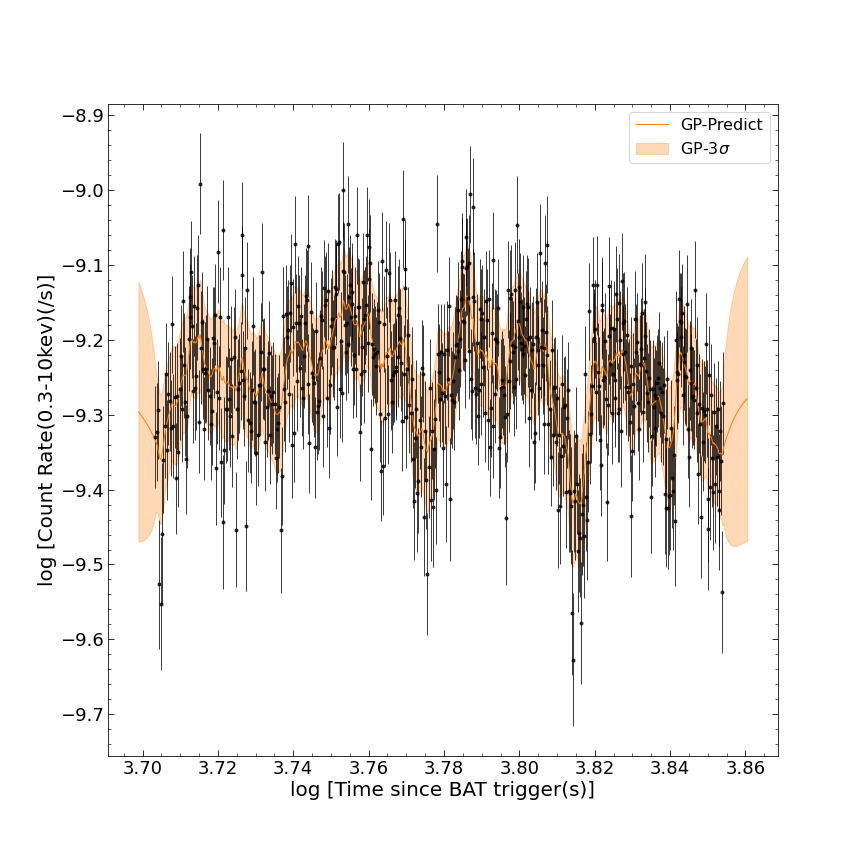}
    \label{fig: detrend MCMC}
    {}
    \includegraphics[width=0.45\textwidth]{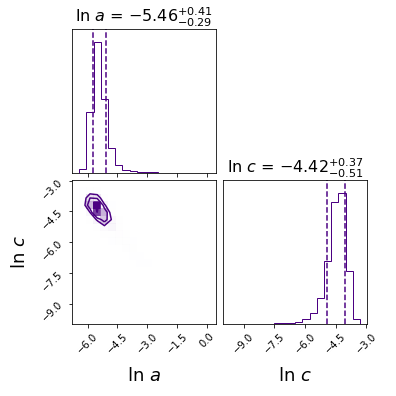}
    \label{fig: detrend MCMC}
    {}
    \includegraphics[width=0.45\textwidth]{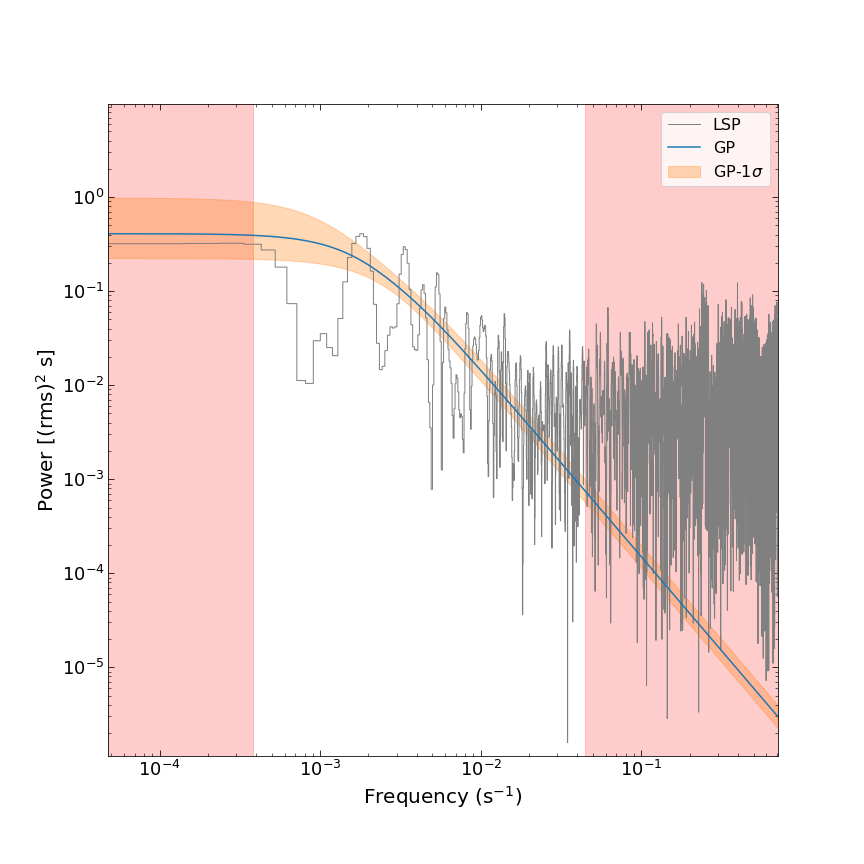}
    \label{fig: detrend PSD}
    \caption{The results of the GP(constant, DRW) model analysis for GRB 101225A. Top-left panel: the X-ray light curve of GRB 101225A in the time interval of [5050-7150] s. The black points denote the data observed by Swift, while the orange solid line depicts the prediction of the GP(constant, DRW) model. The orange-shaded region indicates the 3$\sigma$ uncertainties in the model, and the green line represents the best-fit value of the mean function. Top-right panel: the posterior probability density distribution of each parameter in the GP(constant, DRW) model for GRB 101225A. The vertical dashed lines indicate the 1$\sigma$ uncertainties. Bottom panel: The PSDs of the light curve of GRB 101225A. The average PSD of the GP(constant, DRW) model and the PSD obtained from the LSP method are represented by the blue line and the gray line, respectively. The orange-shaded area indicates the 1$\sigma$ uncertainties in the GP(constant, DRW) model PSD. The red-shaded area represents the unreliable zone, and the white area represents the confidence zone where the upper and lower limits are 1/$P_{min}$ and 1/$P_{max}$, respectively.}
    \label{fig: 101225A  GP(drw)}
\end{figure}

\cite{2021ApJ...921L...1Z} used the LSP method to analyze the X-ray light curve of GRB 101225A observed by Swift-XRT in the time interval of [4900,7500] s. In the analysis, they discovered the QPO signals with the periods of $P$ = 488 s, 304 s, and 205 s (the confidence level is larger than 90\%). 
We adopt the celerite model to analyze the X-ray light curve of GRB 101225A in the time interval of [5050,7150] s.
The complete light curve of GRB 101225A is displayed in Figure \ref{fig: lcGRB101225A}, with the segment of the light curve used in our analysis represented between two vertical dashed lines. This segment of the light curve shows no apparent trend. 
We first employ the GP(constant, QPO) model to study this segment of the light curve.
The value of ln $P$ from the MLE is 6.12 ($P$ = 457.38 s). 
However, the result from MCMC is not constrained in the upper limit of ln $P$. 
The median of ln $P$ is 6.48 ($P$ = 651.97 s).
The median of quality factor $Q$ is 0.47 ($<$ 0.5). 
We also use the GP(constant, SHO) model to study this segment of the light curve. Similarly, this model does not reveal any QPO signal, with a median ln $Q$ of -5.11 ($Q$ = 0.01), and the model's PSD exhibits characteristics similar to red noise. We employ a DRW kernel with a PSD resembling that of red noise, to fit this segment of the light-curve. This model shows well-converged parameters, indicating that a DRW kernel can effectively model this light-curve segment. The results of the analysis of these three models are presented in Figures \ref{fig: 101225A ocl GP(c,qpo)}-\ref{fig: 101225A  GP(drw)}. 

\begin{figure}
    \centering
    \includegraphics[width=0.95\textwidth]{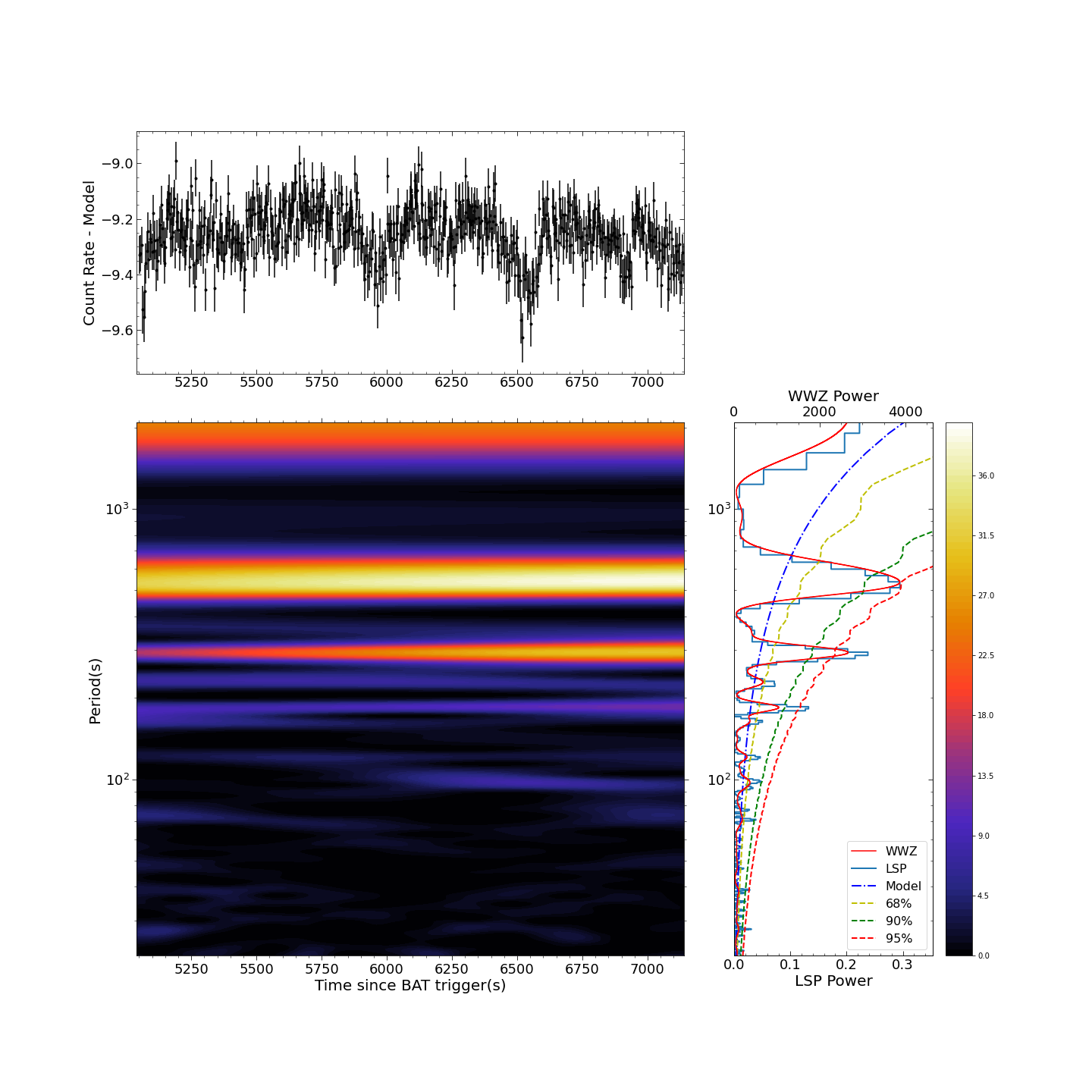}
    \caption{The WWZ analysis of the GRB 101225A X-ray light curve.}
    \label{fig: GRB 101225A WWZ}
\end{figure}

We also employ the WWZ-transform to study this segment of the light curve. The WWZ-transform detects two QPO signals ($P_1$ = 542.84 s, $P_2$ = 295.29 s), as shown in the bottom-left panel of Figure \ref{fig: GRB 101225A WWZ}. We also use the LSP method to obtain the PSD of this light-curve segment. The PSD displays two QPO signals ($P_3$ = 295.12 s, $P_4$ = 185.43 s) with confidence levels exceeding 95\%, as illustrated in the bottom-right panel of Figure \ref{fig: GRB 101225A WWZ}. In this panel, we also present the result of the time-integrated WWZ power.

The analysis examined by the GP mothed simplified to the celerite, LSP, and WWZ methods shows very complex behaviour in the periodicity, although \cite{2021ApJ...921L...1Z} simply found a few period values.

\section{Physical Origin of the Expected Periodicity of GRB 050724}


We discuss the physical origin of the expected QPO signal of the X-ray light curve in GRB 050724 even if the QPO signal is not confirmed. GRB 050724 is a short GRB with the a low-energy prompt X-ray emission lasting for 100 s following the main pulse \citep{2005Natur.438..994B}. 
The host galaxy is an elliptical galaxy at $z$ = 0.258, with a very low star formation rate of 0.02 $M_\odot\rm yr ^{-1}$ \citep{2005GCN..3700....1P}. The stellar population of the galaxy is older than 1 Gyr, and GRB 050724 is offset from the center of the galaxy by approximately 4 kpc. The main burst of GRB 050724 lasted for about 3 s in 15-150 kev energy band, with the energy concentrated primarily in a sharp peak of 0.25 s. 
During the main burst phase, the hardness ratio
of GRB 050724 is 0.91 $\pm$ 0.12, which falls within the lower end of the distribution for short GRB hardness ratios. 
This suggests that GRB 050724 might have originated from the merger of compact binary. \cite{2006A&A...454..113C} discovered four flares in the X-ray afterglow of GRB 050724, with two of them in the time interval of [70,350] s. Due to the ratio of peak time to the width of the flare being less than 1 s, they suggested that the flares are originated from a very dense region. 
In the BAT (15-25 keV) light curve of GRB 050724, a subtle bump is observed at around 100 s, which coincides with early observations by XRT \citep{2005Natur.438..994B,2008arXiv0809.2151C}, suggesting that the central engine of the GRB was still active during the early XRT observations, and the early X-ray flares may be associated with the central engine activity. 
In the scenario of a binary neutron star merger, 
GRB 050724 was identified as a candidate of magnetar-powered merger nova, and 
the magnetar model was further adopted to explain the multiband observation data of GRB 050724 \citep{Gao2017}. 
 
The magnetar oscillations have been used to explain the QPOs in the precursor of sGRBs \citep{Zhang2022apjl}. Considering a binary system composed of a magnetar and a neutron star as the progenitor of the GRB, the tidal forces exerted by the companion neutron star can cause the global breaking of the magnetar crust during the late spiral phase. The breaking magnetar crust undergoes oscillations under the magnetic tension, and the oscillating magnetosphere accompanying the magnetar crust oscillations leads to quasiperiodic variations in the precursor of sGRBs \citep{Gabler2014,Zhang2022apjl}. It is possible to speculate that the QPO observed in the early X-ray afterglow of GRB 050724 can originate from oscillations of a magnetar formed after the binary merger. 
Another possibility for producing QPO in the X-ray light curves of GRBs is the precession of a newborn magnetar, which can generate dipole radiation \citep{2020ApJ...892L..34S}. The precession of a magnetar causes wobbles in the inclination angle between the rotation axis and the magnetic symmetry axis \citep{2015MNRAS.451..695Z,2020ApJ...892L..34S}, leading to oscillations in luminosity \citep{2020ApJ...892L..34S, 2021ApJ...921L...1Z}. In the early phase, 
both the spin frequency and the precession frequency remain 
unchanged. The luminosity exhibits QPOs with two distinct frequency components of the oscillation. One corresponds to the precession frequency, and the other is twice of the precession frequency. The signal of the precession frequency is stronger. 
In the late stage, 
if the spindown of the magnetar does not vary significantly over several precession periods, it is still possible to observe the QPOs in the light curve. We suppose that the oscillations discovered in the X-ray light curve of GRB 050724 using the celerite method are due to the precession effects of the magnetar, and the oscillation frequency corresponds to the low-frequency component of luminosity variation (precession frequency). Assuming no evolution in ellipticity and inclination in the early phase, the ellipticity of the magnetar \citep{2020ApJ...892L..34S} can be estimated as $\epsilon$ $\approx$ $1.11\ \times\ 10^{-5}(\frac{P}{90.02\ s})(\frac{P_{s}}{1\ \rm ms})$, where $P$ and $P_s$ are the precession period and rotation period of the magnetar, respectively. The ellipticity $\epsilon$ exceeds the number of $10^{-6}$ that is obtained from a typical neutron star case 
\citep{Johnson-McDaniel2013PhRvD}. However, this is not unexpected, as a highly elliptical structure was suggested \citep{doneva2015PhRvD..92j4040D,Gao2016PhRvD..93d4065G,sarin2020PhRvD.101f3021S,2020ApJ...892L..34S}. 
Meanwhile, we speculate that a certain accretion is required to trigger the QPO signal, and the fallback accretion could be the required accretion mode.





\bibliography{sample631}{}

\begin{thebibliography}{}
\expandafter\ifx\csname natexlab\endcsname\relax\def\natexlab#1{#1}\fi
\providecommand{\url}[1]{\href{#1}{#1}}
\providecommand{\dodoi}[1]{doi:~\href{http://doi.org/#1}{\nolinkurl{#1}}}
\providecommand{\doeprint}[1]{\href{http://ascl.net/#1}{\nolinkurl{http://ascl.net/#1}}}
\providecommand{\doarXiv}[1]{\href{https://arxiv.org/abs/#1}{\nolinkurl{https://arxiv.org/abs/#1}}}

\bibitem[{{Aigrain} \& {Foreman-Mackey}(2023)}]{2022arXiv220908940A}
{Aigrain}, S., \& {Foreman-Mackey}, D. 2023, \araa, 61, 329, \dodoi{10.1146/annurev-astro-052920-103508}

\bibitem[{{Astropy Collaboration} {et~al.}(2022){Astropy Collaboration}, {Price-Whelan}, {Lim}, {Earl}, {Starkman}, {Bradley}, {Shupe}, {Patil}, {Corrales}, {Brasseur}, {N{\"o}the}, {Donath}, {Tollerud}, {Morris}, {Ginsburg}, {Vaher}, {Weaver}, {Tocknell}, {Jamieson}, {van Kerkwijk}, {Robitaille}, {Merry}, {Bachetti}, {G{\"u}nther}, {Aldcroft}, {Alvarado-Montes}, {Archibald}, {B{\'o}di}, {Bapat}, {Barentsen}, {Baz{\'a}n}, {Biswas}, {Boquien}, {Burke}, {Cara}, {Cara}, {Conroy}, {Conseil}, {Craig}, {Cross}, {Cruz}, {D'Eugenio}, {Dencheva}, {Devillepoix}, {Dietrich}, {Eigenbrot}, {Erben}, {Ferreira}, {Foreman-Mackey}, {Fox}, {Freij}, {Garg}, {Geda}, {Glattly}, {Gondhalekar}, {Gordon}, {Grant}, {Greenfield}, {Groener}, {Guest}, {Gurovich}, {Handberg}, {Hart}, {Hatfield-Dodds}, {Homeier}, {Hosseinzadeh}, {Jenness}, {Jones}, {Joseph}, {Kalmbach}, {Karamehmetoglu}, {Ka{\l}uszy{\'n}ski}, {Kelley}, {Kern}, {Kerzendorf}, {Koch}, {Kulumani}, {Lee}, {Ly}, {Ma}, {MacBride}, {Maljaars}, {Muna}, {Murphy}, {Norman},
  {O'Steen}, {Oman}, {Pacifici}, {Pascual}, {Pascual-Granado}, {Patil}, {Perren}, {Pickering}, {Rastogi}, {Roulston}, {Ryan}, {Rykoff}, {Sabater}, {Sakurikar}, {Salgado}, {Sanghi}, {Saunders}, {Savchenko}, {Schwardt}, {Seifert-Eckert}, {Shih}, {Jain}, {Shukla}, {Sick}, {Simpson}, {Singanamalla}, {Singer}, {Singhal}, {Sinha}, {Sip{\H{o}}cz}, {Spitler}, {Stansby}, {Streicher}, {{\v{S}}umak}, {Swinbank}, {Taranu}, {Tewary}, {Tremblay}, {de Val-Borro}, {Van Kooten}, {Vasovi{\'c}}, {Verma}, {de Miranda Cardoso}, {Williams}, {Wilson}, {Winkel}, {Wood-Vasey}, {Xue}, {Yoachim}, {Zhang}, {Zonca}, \& {Astropy Project Contributors}}]{2022ApJ...935..167A}
{Astropy Collaboration}, {Price-Whelan}, A.~M., {Lim}, P.~L., {et~al.} 2022, \apj, 935, 167, \dodoi{10.3847/1538-4357/ac7c74}

\bibitem[{Auchère {et~al.}(2016)Auchère, Froment, Bocchialini, Buchlin, \& Solomon}]{Auchère_2016}
Auchère, F., Froment, C., Bocchialini, K., Buchlin, E., \& Solomon, J. 2016, The Astrophysical Journal, 825, 110, \dodoi{10.3847/0004-637X/825/2/110}

\bibitem[{{Barthelmy} {et~al.}(2005){Barthelmy}, {Chincarini}, {Burrows}, {Gehrels}, {Covino}, {Moretti}, {Romano}, {O'Brien}, {Sarazin}, {Kouveliotou}, {Goad}, {Vaughan}, {Tagliaferri}, {Zhang}, {Antonelli}, {Campana}, {Cummings}, {D'Avanzo}, {Davies}, {Giommi}, {Grupe}, {Kaneko}, {Kennea}, {King}, {Kobayashi}, {Melandri}, {Meszaros}, {Nousek}, {Patel}, {Sakamoto}, \& {Wijers}}]{2005Natur.438..994B}
{Barthelmy}, S.~D., {Chincarini}, G., {Burrows}, D.~N., {et~al.} 2005, \nat, 438, 994, \dodoi{10.1038/nature04392}

\bibitem[{{Burke} {et~al.}(2021){Burke}, {Shen}, {Blaes}, {Gammie}, {Horne}, {Jiang}, {Liu}, {McHardy}, {Morgan}, {Scaringi}, \& {Yang}}]{Burke2021}
{Burke}, C.~J., {Shen}, Y., {Blaes}, O., {et~al.} 2021, Science, 373, 789, \dodoi{10.1126/science.abg9933}

\bibitem[{{Campana} {et~al.}(2006){Campana}, {Tagliaferri}, {Lazzati}, {Chincarini}, {Covino}, {Page}, {Romano}, {Moretti}, {Cusumano}, {Mangano}, {Mineo}, {La Parola}, {Giommi}, {Perri}, {Capalbi}, {Zhang}, {Barthelmy}, {Cummings}, {Sakamoto}, {Burrows}, {Kennea}, {Nousek}, {Osborne}, {O'Brien}, {Godet}, \& {Gehrels}}]{2006A&A...454..113C}
{Campana}, S., {Tagliaferri}, G., {Lazzati}, D., {et~al.} 2006, \aap, 454, 113, \dodoi{10.1051/0004-6361:20064856}

\bibitem[{{Castro-Tirado} {et~al.}(2021){Castro-Tirado}, {{\O}stgaard}, {G{\"o}{\c{C}}{\textsection}{\"u}{\c{s}}}, {S{\'a}nchez-Gil}, {Pascual-Granado}, {Reglero}, {Mezentsev}, {Gabler}, {Marisaldi}, {Neubert}, {Budtz-J{\o}rgensen}, {Lindanger}, {Sarria}, {Kuvvetli}, {Cerd{\'a}-Dur{\'a}n}, {Navarro-Gonz{\'a}lez}, {Font}, {Zhang}, {Lund}, {Oxborrow}, {Brandt}, {Caballero-Garc{\'\i}a}, {Carrasco-Garc{\'\i}a}, {Castell{\'o}n}, {Castro Tirado}, {Christiansen}, {Eyles}, {Fern{\'a}ndez-Garc{\'\i}a}, {Genov}, {Guziy}, {Hu}, {Nicuesa Guelbenzu}, {Pandey}, {Peng}, {P{\'e}rez del Pulgar}, {Reina Terol}, {Rodr{\'\i}guez}, {S{\'a}nchez-Ram{\'\i}rez}, {Sun}, {Ullaland}, \& {Yang}}]{2021Natur.600..621C}
{Castro-Tirado}, A.~J., {{\O}stgaard}, N., {G{\"o}{\c{C}}{\textsection}{\"u}{\c{s}}}, E., {et~al.} 2021, \nat, 600, 621, \dodoi{10.1038/s41586-021-04101-1}

\bibitem[{{Chincarini} {et~al.}(2008){Chincarini}, {Margutti}, {Mao}, {Pasotti}, {Guidorzi}, {Covino}, \& {D'avanzo}}]{2008arXiv0809.2151C}
{Chincarini}, G., {Margutti}, R., {Mao}, J., {et~al.} 2008, in COSPAR conference, \dodoi{10.48550/arXiv.0809.2151}

\bibitem[{{Covino} {et~al.}(2020){Covino}, {Landoni}, {Sandrinelli}, \& {Treves}}]{2020ApJ...895..122C}
{Covino}, S., {Landoni}, M., {Sandrinelli}, A., \& {Treves}, A. 2020, \apj, 895, 122, \dodoi{10.3847/1538-4357/ab8bd4}

\bibitem[{{Covino} {et~al.}(2019){Covino}, {Sandrinelli}, \& {Treves}}]{2019MNRAS.482.1270C}
{Covino}, S., {Sandrinelli}, A., \& {Treves}, A. 2019, \mnras, 482, 1270, \dodoi{10.1093/mnras/sty2720}

\bibitem[{{{\c{S}}a{\c{s}}maz Mu{\c{s}}} {et~al.}(2019){{\c{S}}a{\c{s}}maz Mu{\c{s}}}, {{\c{C}}{\i}k{\i}nto{\u{g}}lu}, {Ayg{\"u}n}, {Anda{\c{c}}}, \& {Ek{\c{s}}i}}]{2019ApJ...886....5S}
{{\c{S}}a{\c{s}}maz Mu{\c{s}}}, S., {{\c{C}}{\i}k{\i}nto{\u{g}}lu}, S., {Ayg{\"u}n}, U., {Anda{\c{c}}}, I.~C., \& {Ek{\c{s}}i}, K.~Y. 2019, \apj, 886, 5, \dodoi{10.3847/1538-4357/ab498c}

\bibitem[{{Doneva} {et~al.}(2015){Doneva}, {Kokkotas}, \& {Pnigouras}}]{doneva2015PhRvD..92j4040D}
{Doneva}, D.~D., {Kokkotas}, K.~D., \& {Pnigouras}, P. 2015, \prd, 92, 104040, \dodoi{10.1103/PhysRevD.92.104040}

\bibitem[{{Duncan} \& {Thompson}(1992)}]{1992ApJ...392L...9D}
{Duncan}, R.~C., \& {Thompson}, C. 1992, \apjl, 392, L9, \dodoi{10.1086/186413}

\bibitem[{{Evans} {et~al.}(2007){Evans}, {Beardmore}, {Page}, {Tyler}, {Osborne}, {Goad}, {O'Brien}, {Vetere}, {Racusin}, {Morris}, {Burrows}, {Capalbi}, {Perri}, {Gehrels}, \& {Romano}}]{Evans2007A&A...469..379E}
{Evans}, P.~A., {Beardmore}, A.~P., {Page}, K.~L., {et~al.} 2007, \aap, 469, 379, \dodoi{10.1051/0004-6361:20077530}

\bibitem[{{Evans} {et~al.}(2009){Evans}, {Beardmore}, {Page}, {Osborne}, {O'Brien}, {Willingale}, {Starling}, {Burrows}, {Godet}, {Vetere}, {Racusin}, {Goad}, {Wiersema}, {Angelini}, {Capalbi}, {Chincarini}, {Gehrels}, {Kennea}, {Margutti}, {Morris}, {Mountford}, {Pagani}, {Perri}, {Romano}, \& {Tanvir}}]{Evans2009MNRAS}
{Evans}, P.~A., {Beardmore}, A.~P., {Page}, K.~L., {et~al.} 2009, \mnras, 397, 1177, \dodoi{10.1111/j.1365-2966.2009.14913.x}

\bibitem[{Foreman-Mackey(2016)}]{corner}
Foreman-Mackey, D. 2016, The Journal of Open Source Software, 1, 24, \dodoi{10.21105/joss.00024}

\bibitem[{{Foreman-Mackey} {et~al.}(2017){Foreman-Mackey}, {Agol}, {Ambikasaran}, \& {Angus}}]{celerite}
{Foreman-Mackey}, D., {Agol}, E., {Ambikasaran}, S., \& {Angus}, R. 2017, \aj, 154, 220, \dodoi{10.3847/1538-3881/aa9332}

\bibitem[{{Foreman-Mackey} {et~al.}(2013){Foreman-Mackey}, {Hogg}, {Lang}, \& {Goodman}}]{2013PASP..125..306F}
{Foreman-Mackey}, D., {Hogg}, D.~W., {Lang}, D., \& {Goodman}, J. 2013, \pasp, 125, 306, \dodoi{10.1086/670067}

\bibitem[{{Foster}(1996)}]{wwz}
{Foster}, G. 1996, \aj, 112, 1709, \dodoi{10.1086/118137}

\bibitem[{{Gabler} {et~al.}(2014){Gabler}, {Cerd{\'a}-Dur{\'a}n}, {Stergioulas}, {Font}, \& {M{\"u}ller}}]{Gabler2014}
{Gabler}, M., {Cerd{\'a}-Dur{\'a}n}, P., {Stergioulas}, N., {Font}, J.~A., \& {M{\"u}ller}, E. 2014, \mnras, 443, 1416, \dodoi{10.1093/mnras/stu1263}

\bibitem[{{Gao} {et~al.}(2016){Gao}, {Zhang}, \& {L{\"u}}}]{Gao2016PhRvD..93d4065G}
{Gao}, H., {Zhang}, B., \& {L{\"u}}, H.-J. 2016, \prd, 93, 044065, \dodoi{10.1103/PhysRevD.93.044065}

\bibitem[{{Gao} {et~al.}(2017){Gao}, {Zhang}, {L{\"u}}, \& {Li}}]{Gao2017}
{Gao}, H., {Zhang}, B., {L{\"u}}, H.-J., \& {Li}, Y. 2017, \apj, 837, 50, \dodoi{10.3847/1538-4357/aa5be3}

\bibitem[{{Gehrels} {et~al.}(2004){Gehrels}, {Chincarini}, {Giommi}, {Mason}, {Nousek}, {Wells}, {White}, {Barthelmy}, {Burrows}, {Cominsky}, {Hurley}, {Marshall}, {M{\'e}sz{\'a}ros}, {Roming}, {Angelini}, {Barbier}, {Belloni}, {Campana}, {Caraveo}, {Chester}, {Citterio}, {Cline}, {Cropper}, {Cummings}, {Dean}, {Feigelson}, {Fenimore}, {Frail}, {Fruchter}, {Garmire}, {Gendreau}, {Ghisellini}, {Greiner}, {Hill}, {Hunsberger}, {Krimm}, {Kulkarni}, {Kumar}, {Lebrun}, {Lloyd-Ronning}, {Markwardt}, {Mattson}, {Mushotzky}, {Norris}, {Osborne}, {Paczynski}, {Palmer}, {Park}, {Parsons}, {Paul}, {Rees}, {Reynolds}, {Rhoads}, {Sasseen}, {Schaefer}, {Short}, {Smale}, {Smith}, {Stella}, {Tagliaferri}, {Takahashi}, {Tashiro}, {Townsley}, {Tueller}, {Turner}, {Vietri}, {Voges}, {Ward}, {Willingale}, {Zerbi}, \& {Zhang}}]{2004ApJ...611.1005G}
{Gehrels}, N., {Chincarini}, G., {Giommi}, P., {et~al.} 2004, \apj, 611, 1005, \dodoi{10.1086/422091}

\bibitem[{{Harris} {et~al.}(2020){Harris}, {Millman}, {van der Walt}, {Gommers}, {Virtanen}, {Cournapeau}, {Wieser}, {Taylor}, {Berg}, {Smith}, {Kern}, {Picus}, {Hoyer}, {van Kerkwijk}, {Brett}, {Haldane}, {del R{\'\i}o}, {Wiebe}, {Peterson}, {G{\'e}rard-Marchant}, {Sheppard}, {Reddy}, {Weckesser}, {Abbasi}, {Gohlke}, \& {Oliphant}}]{2020Natur.585..357H}
{Harris}, C.~R., {Millman}, K.~J., {van der Walt}, S.~J., {et~al.} 2020, \nat, 585, 357, \dodoi{10.1038/s41586-020-2649-2}

\bibitem[{{Heger} {et~al.}(2007){Heger}, {Cumming}, \& {Woosley}}]{nsacc}
{Heger}, A., {Cumming}, A., \& {Woosley}, S.~E. 2007, \apj, 665, 1311, \dodoi{10.1086/517491}

\bibitem[{{H{\"u}bner} {et~al.}(2022){H{\"u}bner}, {Huppenkothen}, {Lasky}, \& {Inglis}}]{Hubner2022ApJS..259...32H}
{H{\"u}bner}, M., {Huppenkothen}, D., {Lasky}, P.~D., \& {Inglis}, A.~R. 2022, \apjs, 259, 32, \dodoi{10.3847/1538-4365/ac49ec}

\bibitem[{{Hunter}(2007)}]{2007CSE.....9...90H}
{Hunter}, J.~D. 2007, Computing in Science and Engineering, 9, 90, \dodoi{10.1109/MCSE.2007.55}

\bibitem[{{Huppenkothen} {et~al.}(2019){Huppenkothen}, {Bachetti}, {Stevens}, {Migliari}, {Balm}, {Hammad}, {Khan}, {Mishra}, {Rashid}, {Sharma}, {Martinez Ribeiro}, \& {Valles Blanco}}]{2019ApJ...881...39H}
{Huppenkothen}, D., {Bachetti}, M., {Stevens}, A.~L., {et~al.} 2019, \apj, 881, 39, \dodoi{10.3847/1538-4357/ab258d}

\bibitem[{{Johnson-McDaniel} \& {Owen}(2013)}]{Johnson-McDaniel2013PhRvD}
{Johnson-McDaniel}, N.~K., \& {Owen}, B.~J. 2013, \prd, 88, 044004, \dodoi{10.1103/PhysRevD.88.044004}

\bibitem[{{Kumar} \& {Zhang}(2015)}]{2015PhR...561....1K}
{Kumar}, P., \& {Zhang}, B. 2015, \physrep, 561, 1, \dodoi{10.1016/j.physrep.2014.09.008}

\bibitem[{{Li} {et~al.}(2021){Li}, {Zhang}, {Zhang}, \& {Zhen}}]{2021ApJS..252...16L}
{Li}, X.~J., {Zhang}, Z.~B., {Zhang}, X.~L., \& {Zhen}, H.~Y. 2021, \apjs, 252, 16, \dodoi{10.3847/1538-4365/abd3fd}

\bibitem[{{Lomb}(1976)}]{1976Ap&SS..39..447L}
{Lomb}, N.~R. 1976, \apss, 39, 447, \dodoi{10.1007/BF00648343}

\bibitem[{{Oganesyan} {et~al.}(2018){Oganesyan}, {Nava}, {Ghirlanda}, \& {Celotti}}]{Oganesyan2018A&A...616A.138O}
{Oganesyan}, G., {Nava}, L., {Ghirlanda}, G., \& {Celotti}, A. 2018, \aap, 616, A138, \dodoi{10.1051/0004-6361/201732172}

\bibitem[{{Pescalli} {et~al.}(2018){Pescalli}, {Ronchi}, {Ghirlanda}, \& {Ghisellini}}]{2018A&A...615A..80P}
{Pescalli}, A., {Ronchi}, M., {Ghirlanda}, G., \& {Ghisellini}, G. 2018, \aap, 615, A80, \dodoi{10.1051/0004-6361/201732270}

\bibitem[{{Prochaska} {et~al.}(2005){Prochaska}, {Bloom}, {Chen}, {Hansen}, {Kalirai}, {Rich}, \& {Richer}}]{2005GCN..3700....1P}
{Prochaska}, J.~X., {Bloom}, J.~S., {Chen}, H.~W., {et~al.} 2005, GRB Coordinates Network, 3700, 1

\bibitem[{{Rasmussen} \& {Williams}(2006)}]{GP}
{Rasmussen}, C.~E., \& {Williams}, C.~K.~I. 2006, Gaussian Processes for Machine Learning (MIT Press)

\bibitem[{{Rastinejad} {et~al.}(2022){Rastinejad}, {Gompertz}, {Levan}, {Fong}, {Nicholl}, {Lamb}, {Malesani}, {Nugent}, {Oates}, {Tanvir}, {de Ugarte Postigo}, {Kilpatrick}, {Moore}, {Metzger}, {Ravasio}, {Rossi}, {Schroeder}, {Jencson}, {Sand}, {Smith}, {Ag{\"u}{\'\i} Fern{\'a}ndez}, {Berger}, {Blanchard}, {Chornock}, {Cobb}, {De Pasquale}, {Fynbo}, {Izzo}, {Kann}, {Laskar}, {Marini}, {Paterson}, {Escorial}, {Sears}, \& {Th{\"o}ne}}]{2022Natur.612..223R}
{Rastinejad}, J.~C., {Gompertz}, B.~P., {Levan}, A.~J., {et~al.} 2022, \nat, 612, 223, \dodoi{10.1038/s41586-022-05390-w}

\bibitem[{{Sarin} {et~al.}(2020){Sarin}, {Lasky}, \& {Ashton}}]{sarin2020PhRvD.101f3021S}
{Sarin}, N., {Lasky}, P.~D., \& {Ashton}, G. 2020, \prd, 101, 063021, \dodoi{10.1103/PhysRevD.101.063021}

\bibitem[{{Scargle}(1982)}]{1982ApJ...263..835S}
{Scargle}, J.~D. 1982, \apj, 263, 835, \dodoi{10.1086/160554}

\bibitem[{{Suvorov} \& {Kokkotas}(2020)}]{2020ApJ...892L..34S}
{Suvorov}, A.~G., \& {Kokkotas}, K.~D. 2020, \apjl, 892, L34, \dodoi{10.3847/2041-8213/ab8296}

\bibitem[{{Usov}(1992)}]{1992Natur.357..472U}
{Usov}, V.~V. 1992, \nat, 357, 472, \dodoi{10.1038/357472a0}

\bibitem[{{Virtanen} {et~al.}(2020){Virtanen}, {Gommers}, {Oliphant}, {Haberland}, {Reddy}, {Cournapeau}, {Burovski}, {Peterson}, {Weckesser}, {Bright}, {van der Walt}, {Brett}, {Wilson}, {Millman}, {Mayorov}, {Nelson}, {Jones}, {Kern}, {Larson}, {Carey}, {Polat}, {Feng}, {Moore}, {VanderPlas}, {Laxalde}, {Perktold}, {Cimrman}, {Henriksen}, {Quintero}, {Harris}, {Archibald}, {Ribeiro}, {Pedregosa}, {van Mulbregt}, \& {SciPy 1. 0 Contributors}}]{2020NatMe..17..261V}
{Virtanen}, P., {Gommers}, R., {Oliphant}, T.~E., {et~al.} 2020, Nature Methods, 17, 261, \dodoi{10.1038/s41592-019-0686-2}

\bibitem[{{Xiao} {et~al.}(2022){Xiao}, {Zhang}, {Zhu}, {Xiong}, {Gao}, {Xu}, {Zhang}, {Peng}, {Li}, {Zhang}, {Lu}, {Lin}, {Liu}, {Zhang}, {Ge}, {Tuo}, {Xue}, {Fu}, {Liu}, {Li}, {Wang}, {Zheng}, {Wang}, {Jiang}, {Li}, {Liu}, {Cao}, {Cai}, {Yi}, {Zhao}, {Xie}, {Li}, {Luo}, {Liao}, {Song}, {Zhang}, {Qu}, {Liu}, {Li}, {Xu}, \& {Li}}]{2022arXiv220502186X}
{Xiao}, S., {Zhang}, Y.-Q., {Zhu}, Z.-P., {et~al.} 2022, arXiv e-prints, arXiv:2205.02186, \dodoi{10.48550/arXiv.2205.02186}

\bibitem[{{Yang} {et~al.}(2022){Yang}, {Ai}, {Zhang}, {Zhang}, {Liu}, {Wang}, {Yang}, {Yin}, {Li}, \& {L{\"u}}}]{2022Natur.612..232Y}
{Yang}, J., {Ai}, S., {Zhang}, B.-B., {et~al.} 2022, \nat, 612, 232, \dodoi{10.1038/s41586-022-05403-8}

\bibitem[{{Zanazzi} \& {Lai}(2015)}]{2015MNRAS.451..695Z}
{Zanazzi}, J.~J., \& {Lai}, D. 2015, \mnras, 451, 695, \dodoi{10.1093/mnras/stv955}

\bibitem[{{Zhang} {et~al.}(2022){Zhang}, {Yi}, {Zhang}, {Xiong}, \& {Xiao}}]{Zhang2022apjl}
{Zhang}, Z., {Yi}, S.-X., {Zhang}, S.-N., {Xiong}, S.-L., \& {Xiao}, S. 2022, \apjl, 939, L25, \dodoi{10.3847/2041-8213/ac9b55}

\bibitem[{{Zou} {et~al.}(2021){Zou}, {Zheng}, {Yang}, {Zhang}, {Li}, {Ren}, {Lin}, \& {Liang}}]{2021ApJ...921L...1Z}
{Zou}, L., {Zheng}, T.-C., {Yang}, X., {et~al.} 2021, \apjl, 921, L1, \dodoi{10.3847/2041-8213/ac2ee4}

\end{thebibliography}
\bibliographystyle{aasjournal}



\begin{longrotatetable}
\begin{deluxetable}{lcccccccc}\label{Tabel}
\tablecaption{Signal Types of X-ray Light Curves for Three GRBs Obtained Using Different Methods}
\tablewidth{0pt}
\tabletypesize{\scriptsize}

\startdata
{} & 
{Color noise with} & 
{Color noise} & 
{Color noise with} & 
{Color noise} & 
{Color noise with} & 
{Color noise with} & 
{Color noise with} & 
{Color noise with} \\[0.5em]
{GRB 050724} & 
{power law trend} & 
{(detrend power law)} & 
{polynomial trend} & 
{(detrend polynomial)} & 
{power law trend and} & 
{QPO signal} & 
{polynomial trend and} & 
{QPO signal} \\[0.5em]
{} & 
{} & 
{} & 
{} & 
{} & 
{QPO signal} & 
{(detrend power law)} & 
{QPO signal} & 
{(detrend polynomial)}\\[0.5em]
\hline
GP(QPO) & F & F & F & F & S & S & S & S \\[0.5em]
GP(SHO) & - & - & F & F & - & - & S & S \\[0.5em]
GP(DRW) & S & S & S & S & F & F & F & F \\[0.5em]
LSP & S & F & S & F & F & S & F & S \\[0.5em]
WWZ & - & - & - & F & - & - & - & S \\[0.5em]
\hline
{}& Color noise with & Color noise & Color noise with & Color noise  & Color noise with & Color noise with & Color noise with & Color noise with \\[0.5em]
{GRB 060614} & power law trend & (detrend power law) & polynomial trend &(detrend polynomial) & power law trend and & QPO signal & polynomial trend and & QPO signal \\[0.5em]
{} & & & & & QPO signal &(detrend power law) &QPO signal & (detrend polynomial) \\[0.5em] 
\hline
GP(QPO) & F & F & F & F & S(N) & S(N) & S(N) & S \\[0.5em]
GP(SHO) & - & - & S & S & - & - & F & F \\[0.5em]
GP(DRW) & S & S & S & S & F & F & F & F \\[0.5em]
LSP & S & S & S & F & F & F & F & S \\[0.5em]
WWZ & - & - & - & F & - & - & - & S \\[0.5em]
\hline
{} & Color noise & Color noise with & ~ & ~ & ~ & ~ & ~ & ~ \\[0.5em] 
{GRB 101225A} & ~& QPO signal & ~ & ~ & ~ & ~ & ~ & ~ \\[0.5em]
{} & ~& ~ & ~ & ~ & ~ & ~ & ~ & ~ \\[0.5em] 
\hline
GP(QPO) & S & F & ~ & ~ & ~ & ~ & ~ & ~ \\[0.5em]
GP(SHO) & S & F & ~ & ~ & ~ & ~ & ~ & ~ \\[0.5em]
GP(DRW) & S & F & ~ & ~ & ~ & ~ & ~ & ~ \\[0.5em]
LSP & F & S & ~ & ~ & ~ & ~ & ~ & ~ \\[0.5em]
WWZ & F & S & ~ & ~ & ~ & ~ & ~ & ~ \\[0.5em]
\enddata
\tablecomments{If a method obtains the result as shown by the signal type listed, it is denoted as ``S (Success)''. If a method obtains a result that is different from the signal type listed, we denote it as ``F (Failure)''. If a method is not used to study the type of signal shown, it is denoted by ``-''. (Here, we are inclined to use a general polynomial trend rather than a special power-law trend. Thus, some cases with the power-law trend are not tested.) S(N) indicates that the QPO signal detected by the celerite method is false. In these cases, because the period of the QPO signal detected by the celerite method is longer than half of the light-curve duration time, we do not take this detected QPO signal to be real. Although the QPO signal seems to be detected in some cases and labeled by S, we note that the trending of the light curves may produce a false QPO signal (see the main text for detail).}
\end{deluxetable}
\end{longrotatetable}

\begin{figure}
    \centering
    \includegraphics[width=0.45\textwidth]{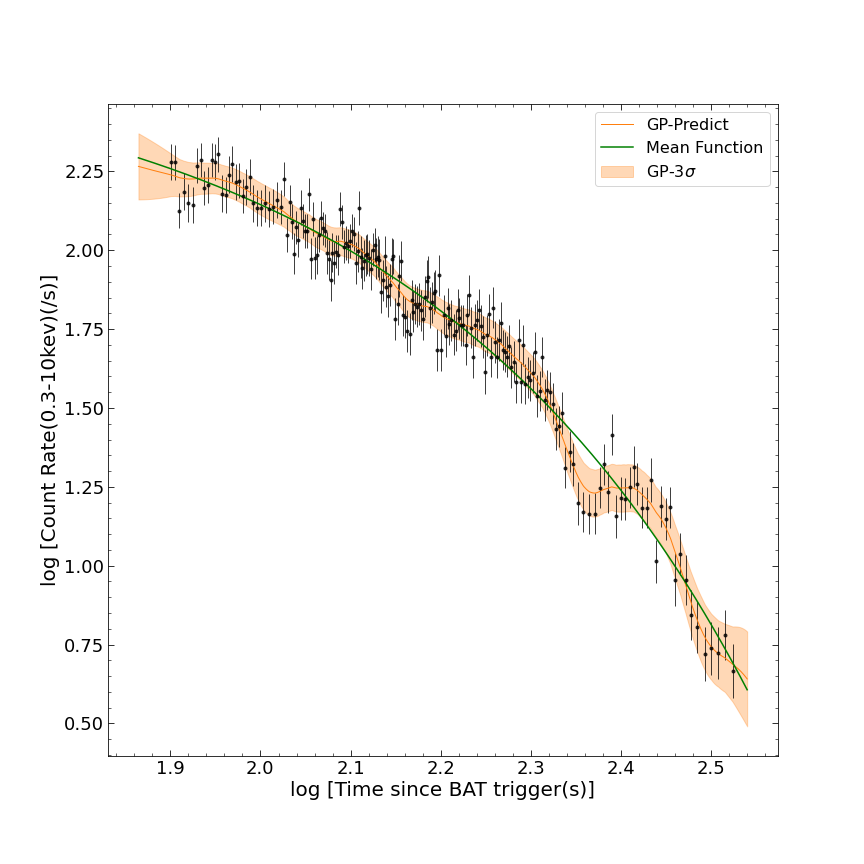}
    \label{fig: detrend MCMC}
    {}
    \includegraphics[width=0.45\textwidth]{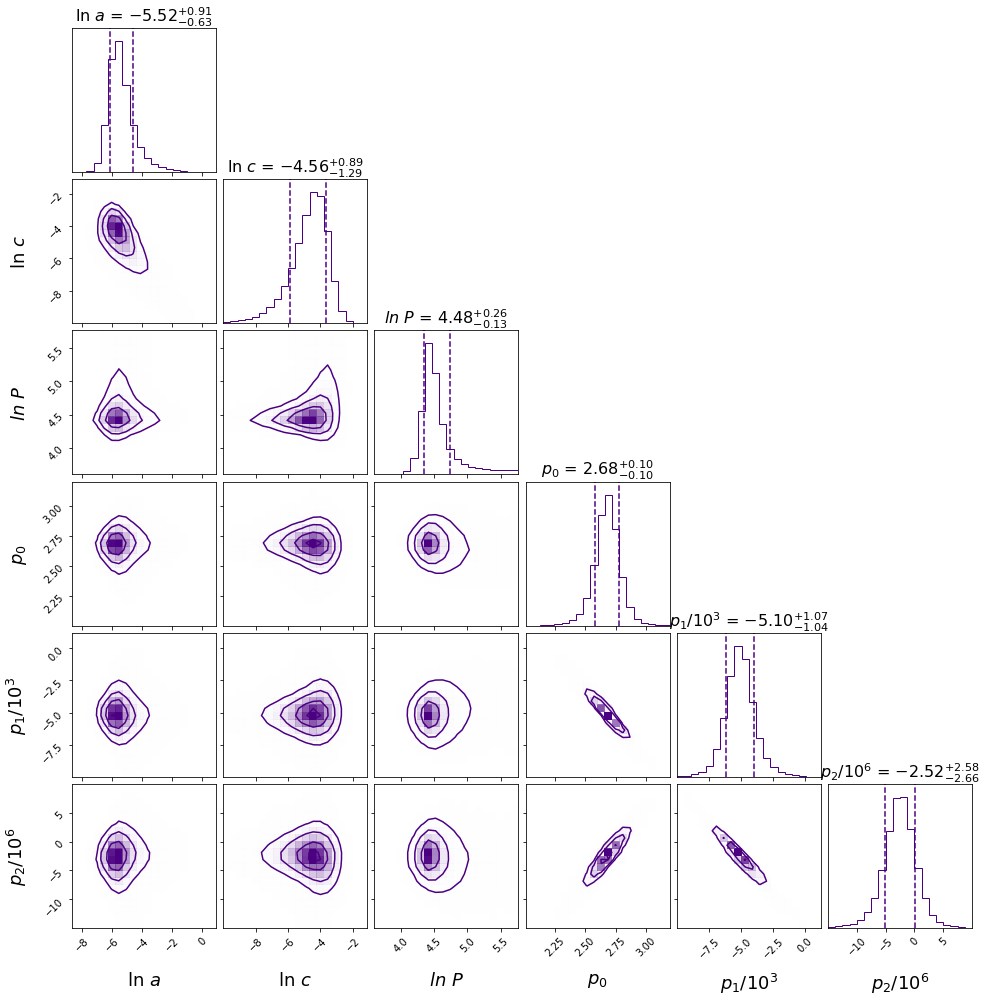}
    \label{fig: detrend MCMC}
    {}
    \includegraphics[width=0.45\textwidth]{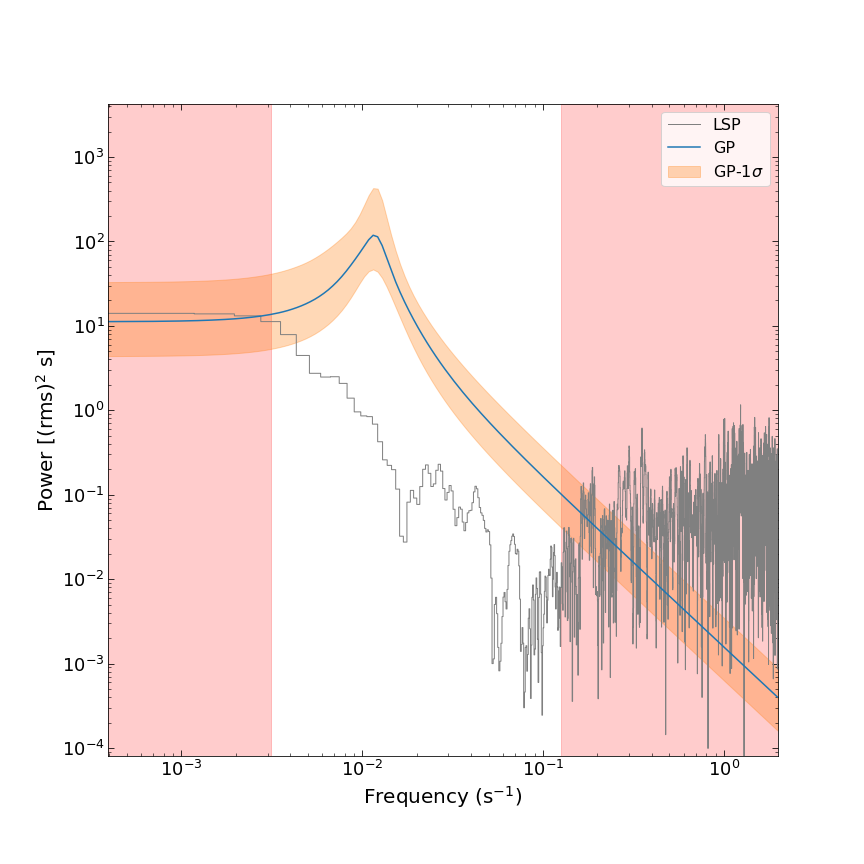}
    \label{fig: detrend PSD}
    {}
    \includegraphics[width=0.45\textwidth]{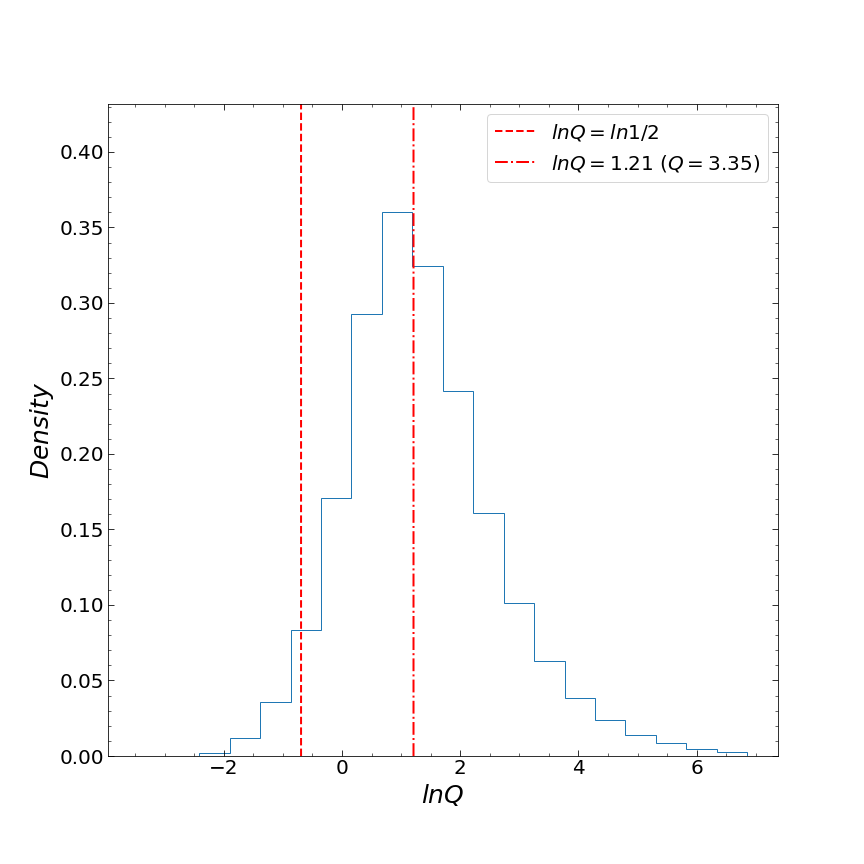}
    \label{fig: detrend Q}
    \caption{The results of the GP(polynomial, QPO) model analysis for GRB 050724. Top-left panel: the X-ray light curve of GRB 050724 in the time interval of [70-350] s. The black points denote the data observed by Swift, while the orange solid line depicts the prediction of the GP(polynomial, QPO) model. The orange-shaded region indicates the 3$\sigma$ uncertainties in the model, and the green line represents the best-fit value of the mean function. Top-right panel: the posterior probability density distribution of each parameter in the GP(polynomial, QPO) model for GRB 050724. The vertical dashed lines indicate 1$\sigma$ uncertainties. Bottom-left panel: the PSDs of the light curve of GRB 050724. The average PSD of the GP(polynomial, QPO) model and the PSD obtained from the LSP method are represented by the blue line and the gray line, respectively. The orange-shaded area indicates the 1$\sigma$ uncertainties in the GP(polynomial, QPO) model PSD. The red-shaded area represents the unreliable zone, and the white area represents the confidence zone where the upper and lower limits are 1/$P_{min}$ and 1/$P_{max}$, respectively. Bottom-right panel: the distribution of the quality factor $Q$. The dashed line and the dashed-dotted line indicate the critical quality factor ($Q$ = 0.5) and the median of ln $Q$, respectively.}
    \label{fig: 050724 ocl GP(poly,qpo)}
\end{figure}
\begin{figure}
    \centering
    \includegraphics[width=0.45\textwidth]{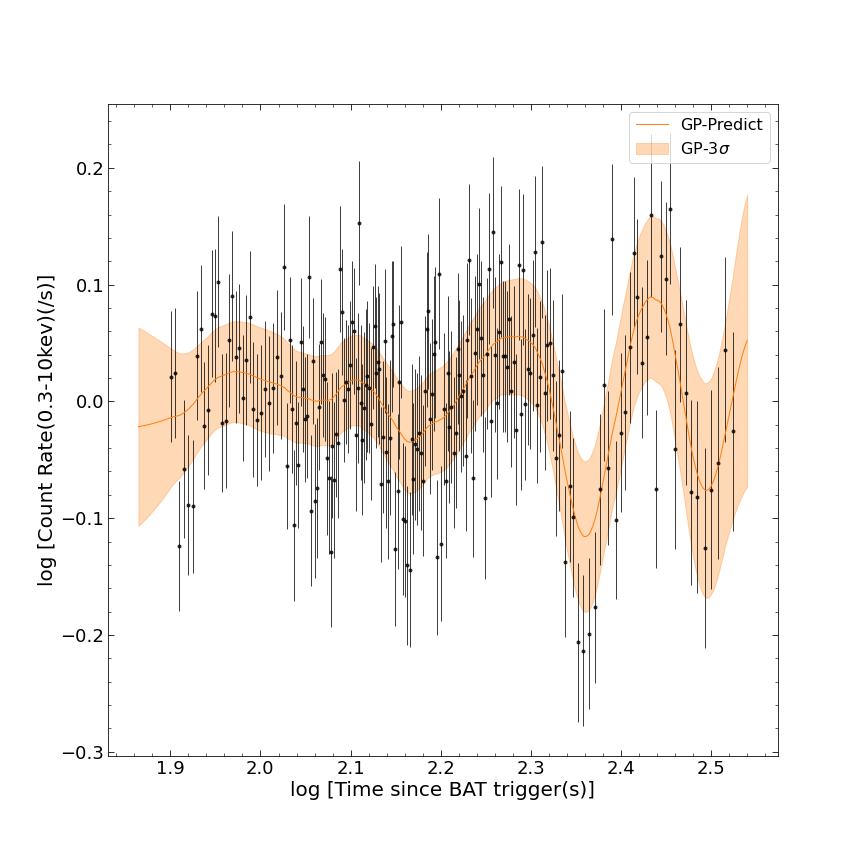}
    \label{fig: detrend MCMC}
    {}
    \includegraphics[width=0.45\textwidth]{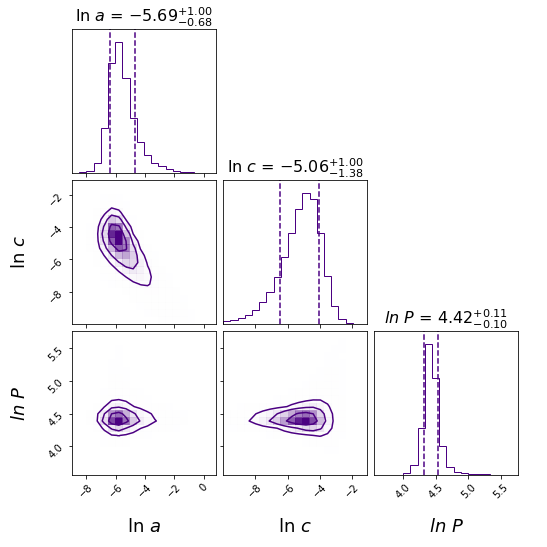}
    \label{fig: detrend MCMC}
    {}
    \includegraphics[width=0.45\textwidth]{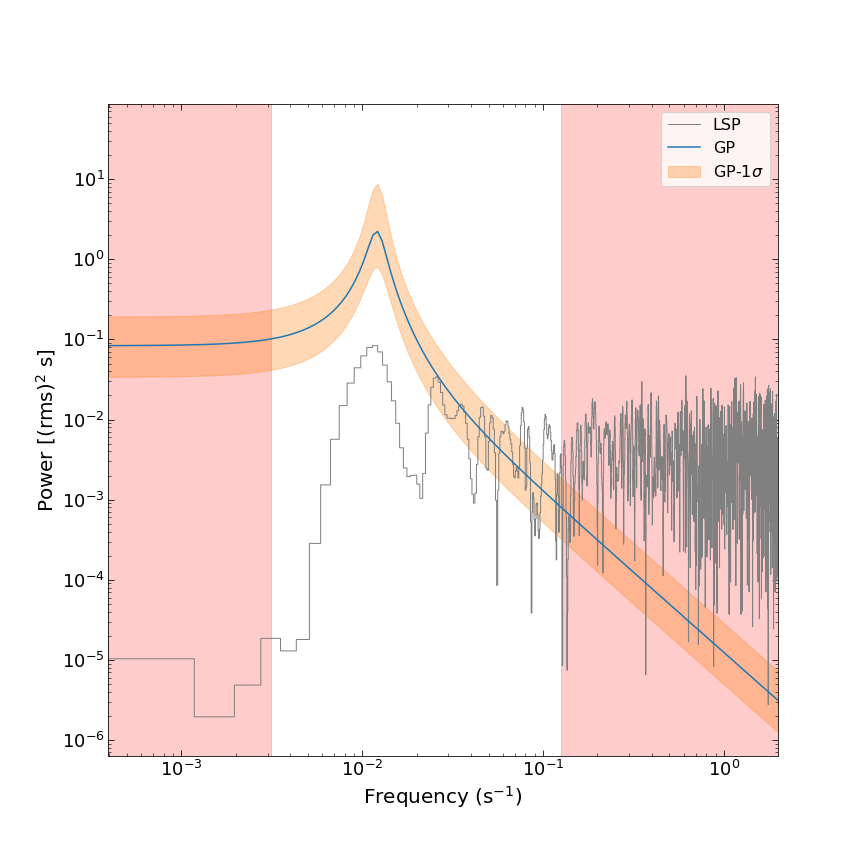}
    \label{fig: detrend PSD}
    {}
    \includegraphics[width=0.45\textwidth]{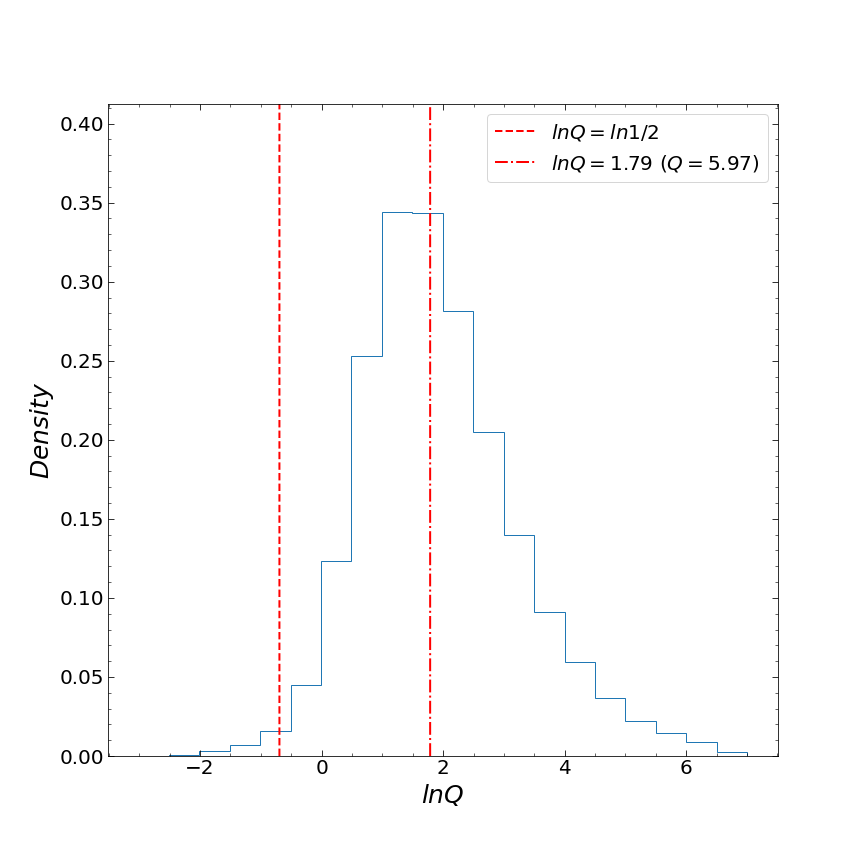}
    \label{fig: detrend Q}
    \caption{The results of the GP(constant, QPO) model analysis for the detrended GRB 050724 light curve. The detrended light curve is obtained by subtracting the best-fit mean function in the GP(polynomial, QPO) model from the original light curve. Top-left panel: the detrended X-ray light curve of GRB 050724 in the time interval of [70-350] s. The black points denote the detrended data, while the orange solid line depicts the prediction of the GP(constant, QPO) model. The orange-shaded region indicates the 3$\sigma$ uncertainties in the model. Top-right panel: the posterior probability density distribution of each parameter in the GP(constant, QPO) model for GRB 050724. The vertical dashed lines indicate the 1$\sigma$ uncertainties. Bottom-left panel: the PSDs of the light curve of GRB 050724. The average PSD of the GP(constant, QPO) model and the PSD obtained from the LSP method are represented by the blue line and the gray line, respectively. The orange-shaded area indicates 1$\sigma$ uncertainties in the GP(constant, QPO) model PSD. The red-shaded area represents the unreliable zone, and the white area represents the confidence zone where the upper and lower limits are 1/$P_{min}$ and 1/$P_{max}$, respectively. Bottom-right panel: the distribution of the quality factor $Q$. The dashed line and the dashed-dotted line indicate the critical quality factor ($Q$ = 0.5) and the median of ln $Q$, respectively.}
    \label{fig:050724 dcl GP(poly,qpo)}
\end{figure}
\begin{figure}
    \centering
    \includegraphics[width=0.45\textwidth]{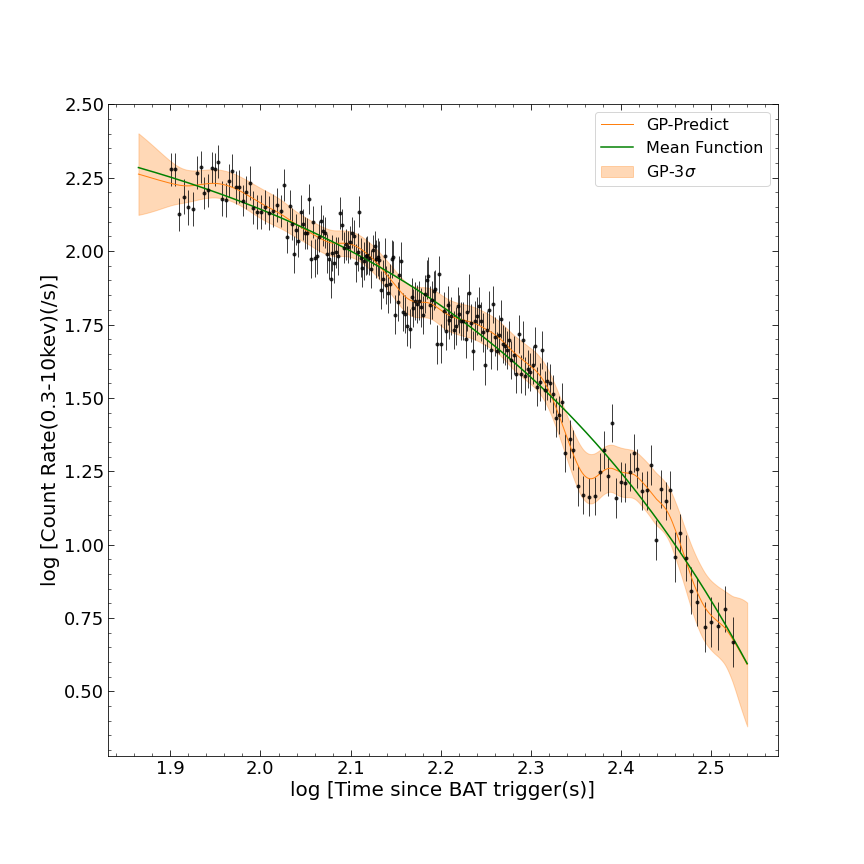}
    \label{fig: detrend MCMC}
    {}
    \includegraphics[width=0.45\textwidth]{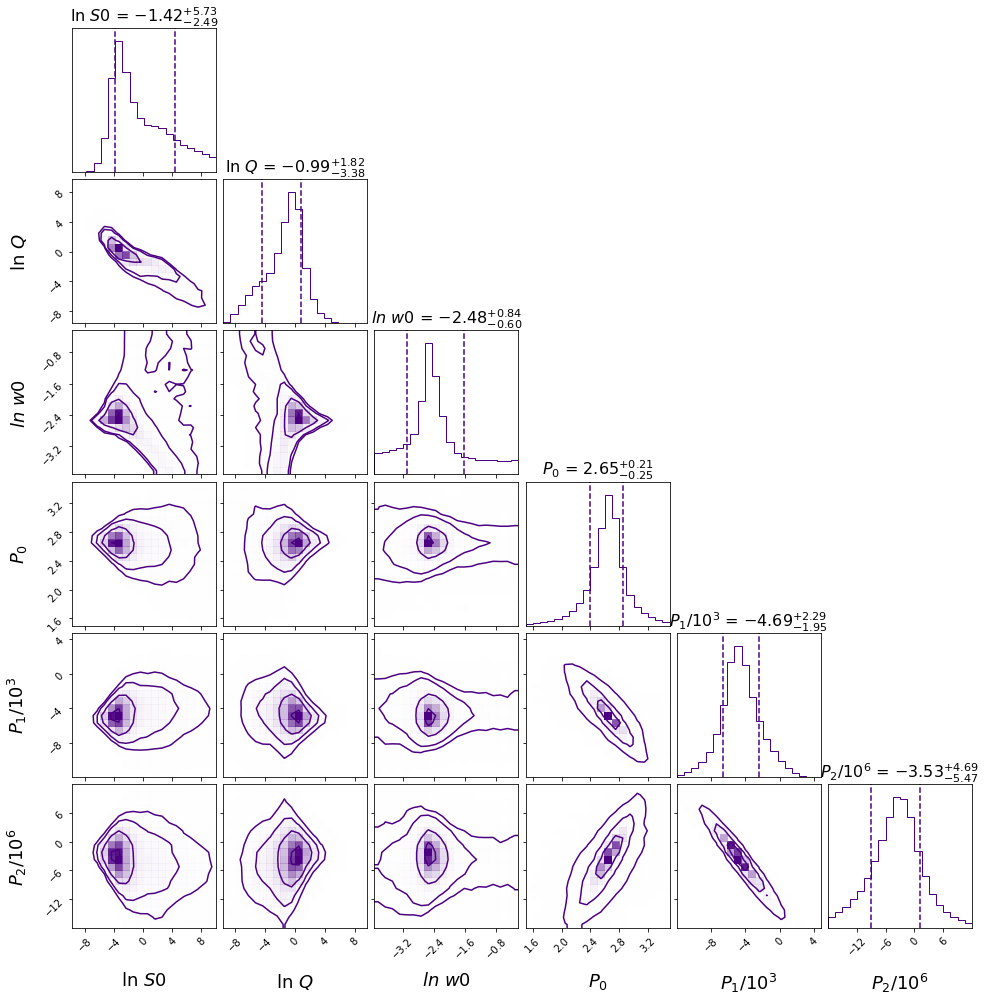}
    \label{fig: detrend MCMC}
    {}
    \includegraphics[width=0.45\textwidth]{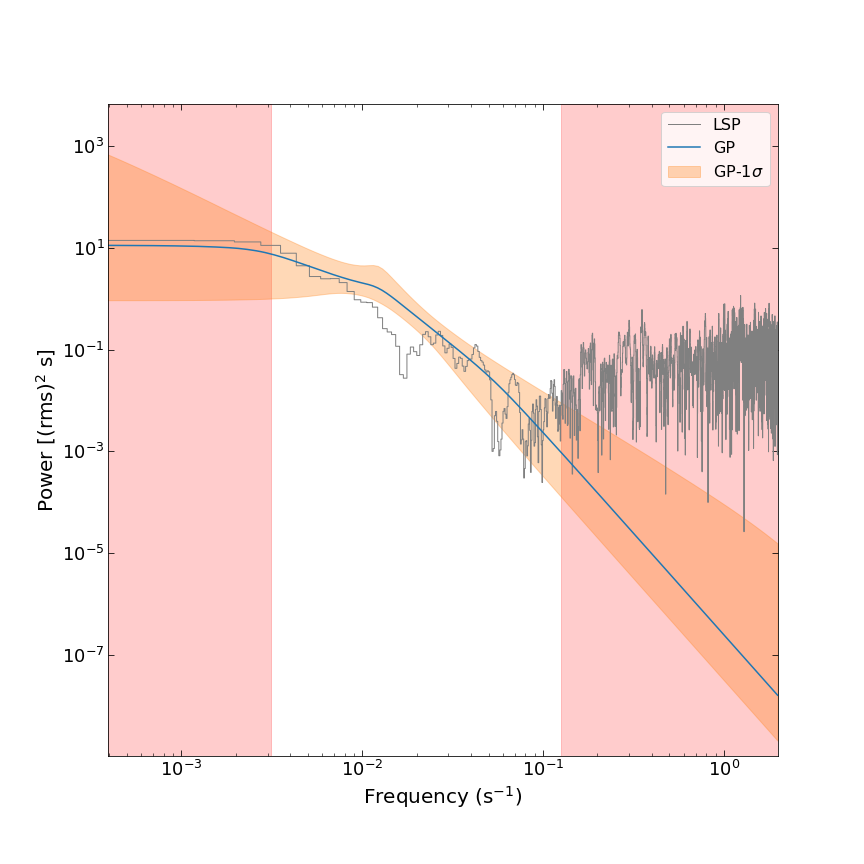}
    \label{fig: detrend PSD}
    {}
    \includegraphics[width=0.45\textwidth]{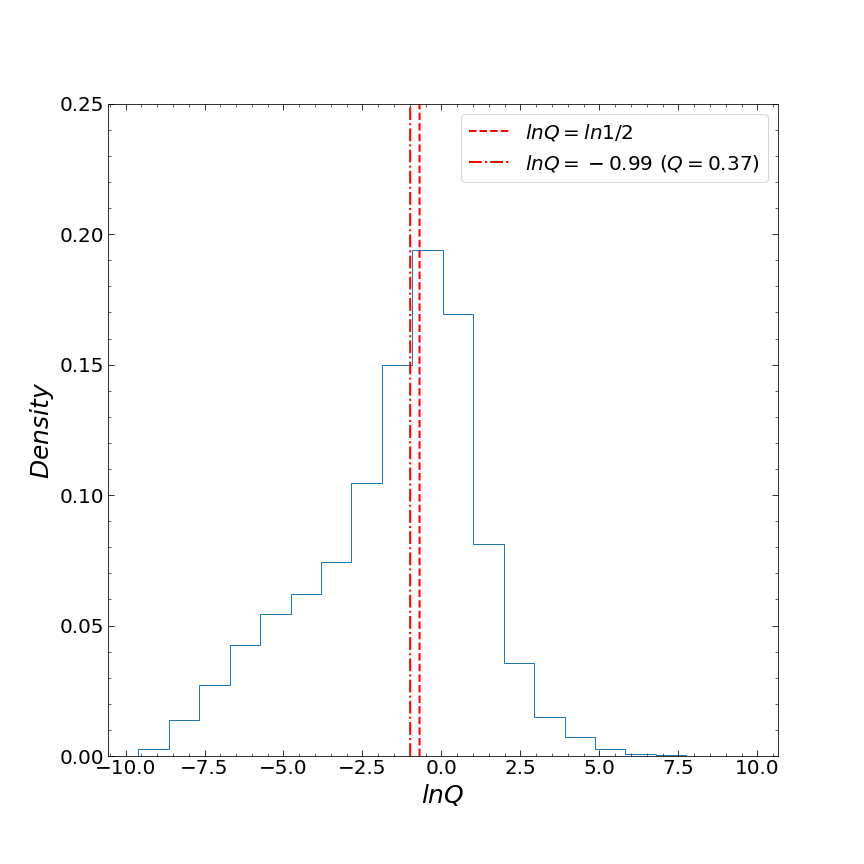}
    \label{fig: detrend Q}
    \caption{The results of the GP(polynomial, SHO) model analysis for GRB 050724. Top-left panel: the X-ray light curve of GRB 050724 in the time interval of [70-350] s. The black points dnote the data observed by Swift, while the orange solid line depicts the prediction of the GP(polynomial, SHO) model. The orange-shaded region indicates the 3$\sigma$ uncertainties in the model, and the green line represents the best-fit value of the mean function. Top-right panel: the posterior probability density distribution of each parameter in the GP(polynomial, SHO) model for GRB 050724. The vertical dashed lines indicate the 1$\sigma$ uncertainties. Bottom-left panel: the PSDs of the light curve of GRB 050724. The average PSD of the GP(polynomial, SHO) model and the PSD obtained from the LSP method are represented by the blue line and the gray line, respectively. The orange-shaded area indicates the 1$\sigma$ uncertainties in the GP(polynomial, SHO) model PSD. The red-shaded area represents the unreliable zone, and the white area represents the confidence zone where the upper and lower limits are 1/$P_{min}$ and 1/$P_{max}$, respectively. Bottom-right panel: the distribution of the quality factor $Q$. The dashed line and the dashed-dotted line indicate the critical quality factor ($Q$ = 0.5) and the median of ln $Q$, respectively.}
    \label{fig: 050724 ocl GP(poly,sho)}
\end{figure}
\begin{figure}
    \centering
    \includegraphics[width=0.45\textwidth]{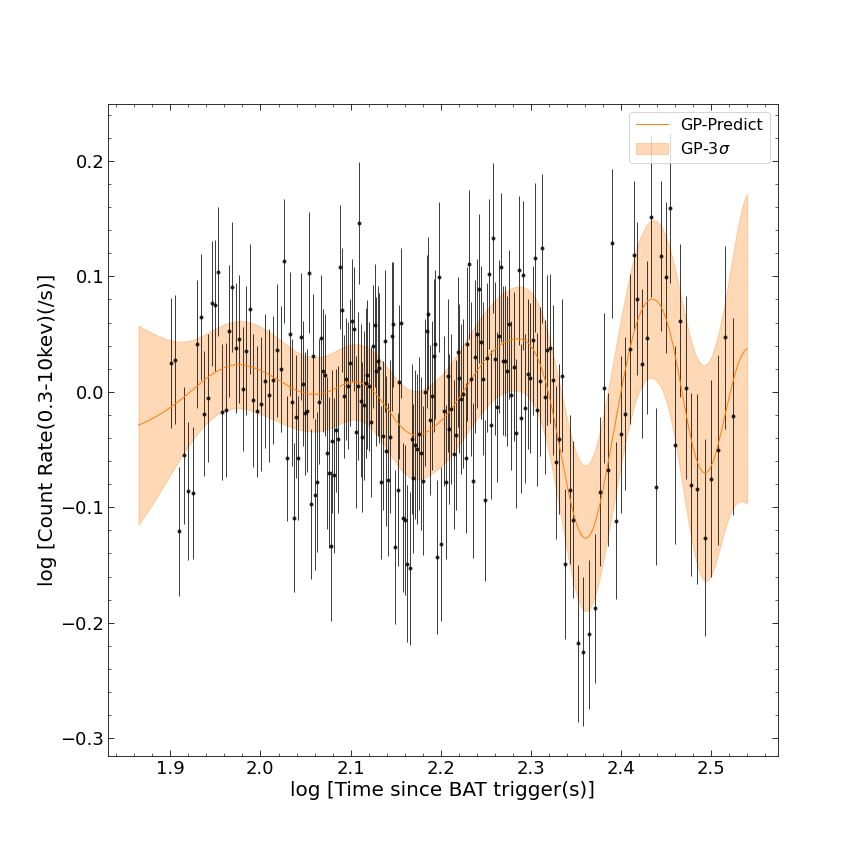}
    \label{fig: detrend MCMC}
    {}
    \includegraphics[width=0.45\textwidth]{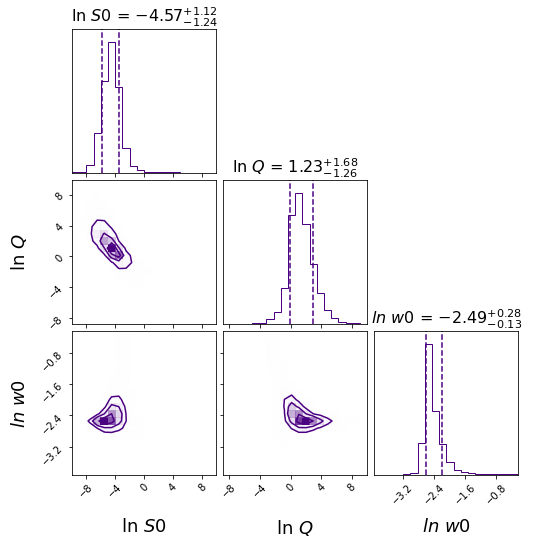}
    \label{fig: detrend MCMC}
    {}
    \includegraphics[width=0.45\textwidth]{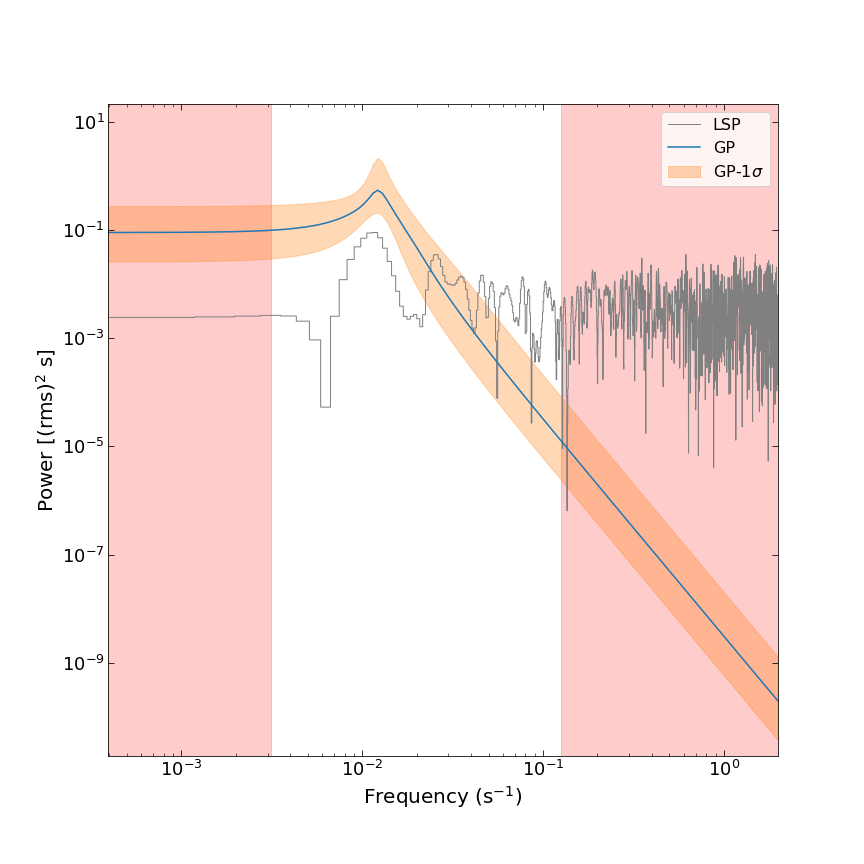}
    \label{fig: detrend PSD}
    {}
    \includegraphics[width=0.45\textwidth]{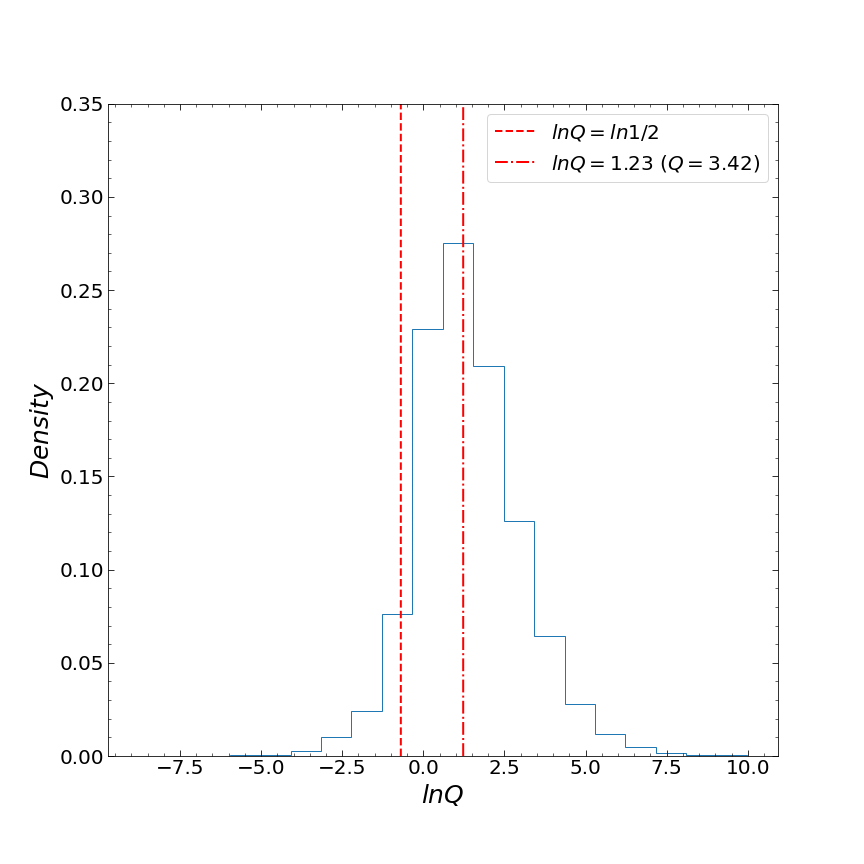}
    \label{fig: detrend Q}
    \caption{The results of the GP(constant, SHO) model analysis for the detrended GRB 050724 light curve. The detrended light curve is obtained by subtracting the best-fit mean function in the GP(polynomial, SHO) from the original light curve. Top-left panel: the detrended X-ray light curve of GRB 050724 in the time interval of [70-350] s. The black points denote the detrended data, while the orange solid line depicts the prediction of the GP(constant, SHO) model. The orange-shaded region indicates the 3$\sigma$ uncertainties in the model. Top-right panel: the posterior probability density distribution of each parameter in the GP(constant, SHO) model for GRB 050724. The vertical dashed lines indicate the 1$\sigma$ uncertainties. Bottom-left panel: the PSDs of the light curve of GRB 050724. The average PSD of the GP(constant, SHO) model and the PSD obtained from the LSP method are represented by the blue line and the gray line, respectively. The orange-shaded area indicates the 1$\sigma$ uncertainties in the GP(constant, SHO) model PSD. The red-shaded area represents the unreliable zone, and the white area represents the confidence zone where the upper and lower limits are 1/$P_{min}$ and 1/$P_{max}$, respectively. Bottom-right panel: the distribution of the quality factor $Q$. The dashed line and the dashed-dotted line indicate the critical quality factor ($Q$ = 0.5) and the median of ln $Q$, respectively.}
    \label{fig: 050724 dcl GP(poly,sho)}
\end{figure}
\begin{figure}
    \centering
    \includegraphics[width=0.45\textwidth]{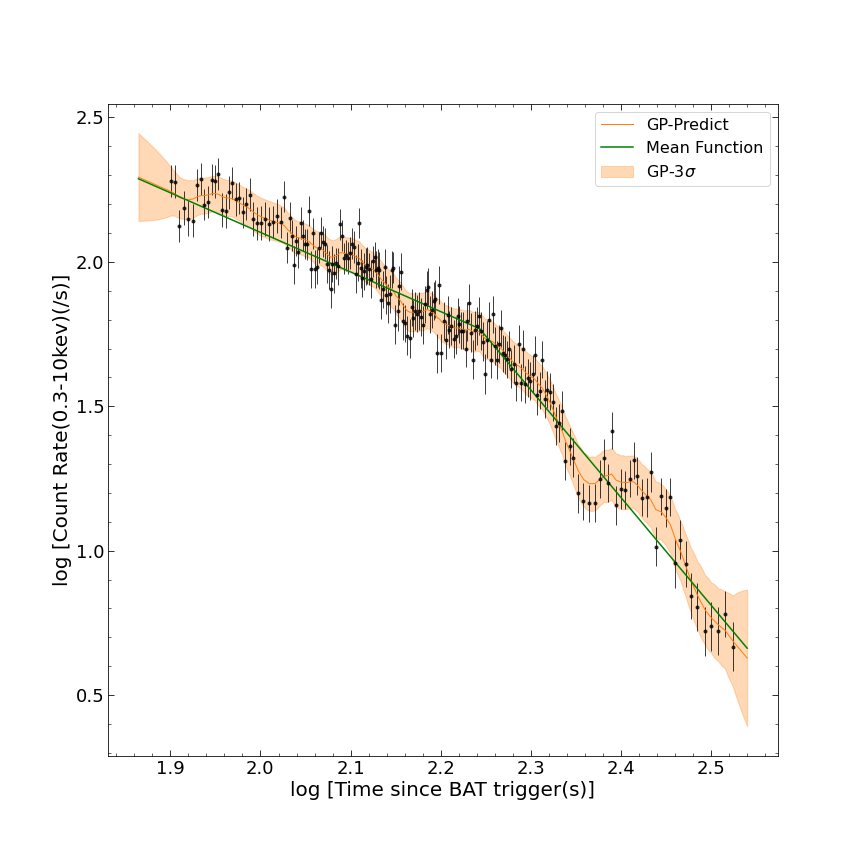}
    \label{fig: detrend MCMC}
    {}
    \includegraphics[width=0.45\textwidth]{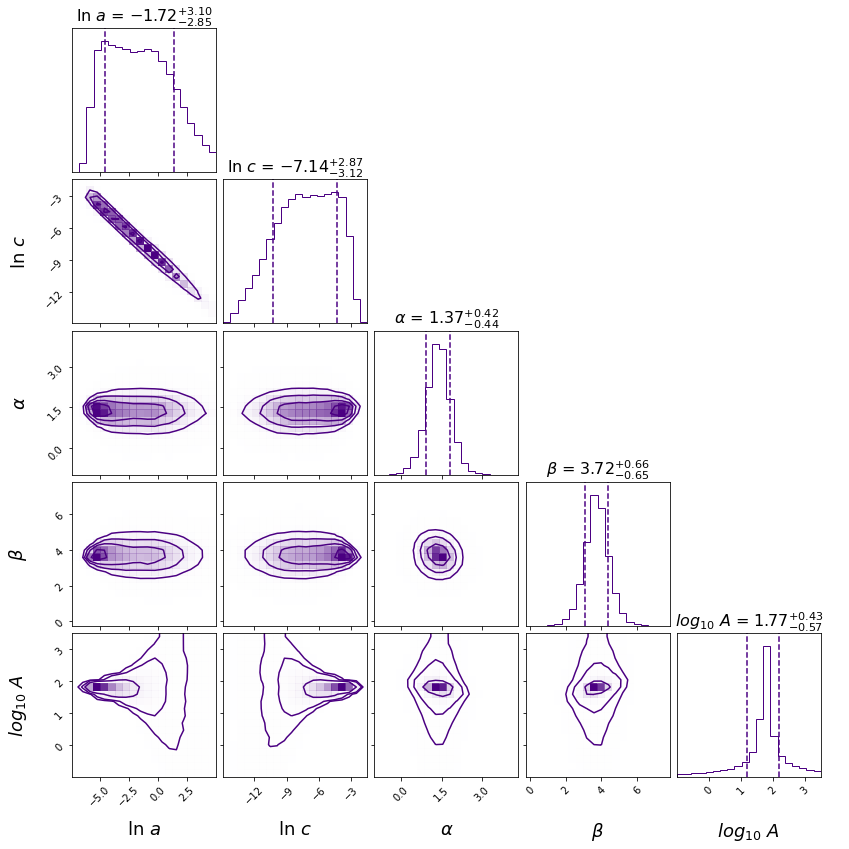}
    \label{fig: detrend MCMC}
    {}
    \includegraphics[width=0.45\textwidth]{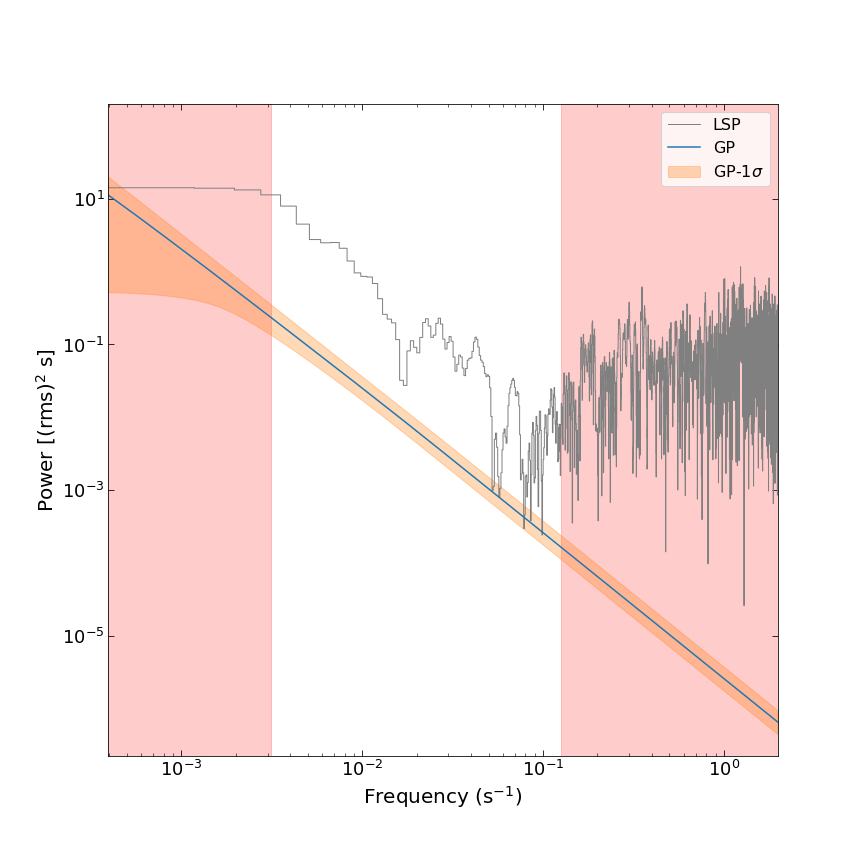}
    \label{fig: detrend PSD}
    {}
    \caption{The results of the GP(power law, DRW) model analysis for GRB 050724. Top-left panel: the X-ray light curve of GRB 050724 in the time interval of [70-350] s. The black points denote the data observed by Swift, while the orange solid line depicts the prediction of the GP(power law, DRW) model. The orange-shaded region indicates the 3$\sigma$ uncertainties in the model, and the green line represents the best-fit value of the mean function. Top-right panel: the posterior probability density distribution of each parameter in the GP(power law, DRW) model for GRB 050724. The vertical dashed lines indicate the 1$\sigma$ uncertainties. Bottom panel: the PSDs of the light curve of GRB 050724. The average PSD of the GP(power law, DRW) model and the PSD obtained from the LSP method are represented by the blue line and the gray line, respectively. The orange-shaded area indicates the 1$\sigma$ uncertainties in the GP(power law, DRW) model PSD. The red-shaded area represents the unreliable zone, and the white area represents the confidence zone where the upper and lower limits are 1/$P_{min}$ and 1/$P_{max}$, respectively. }
    \label{fig: 050724 ocl GP(power,drw)}
\end{figure}
\begin{figure}
    \centering
    \includegraphics[width=0.45\textwidth]{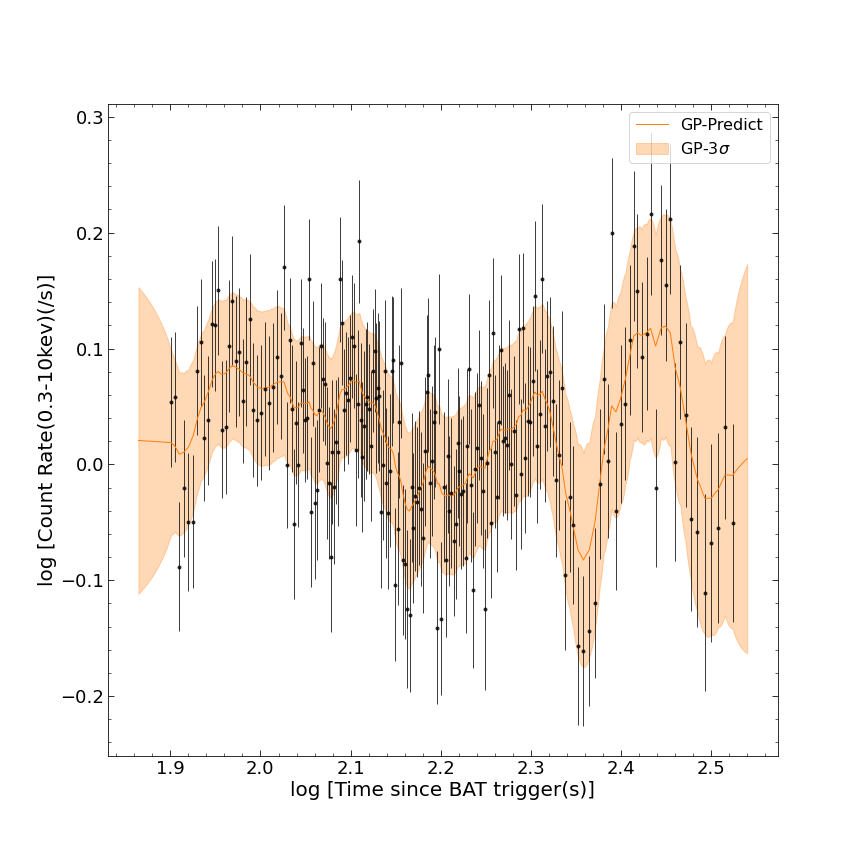}
    \label{fig: detrend MCMC}
    {}
    \includegraphics[width=0.45\textwidth]{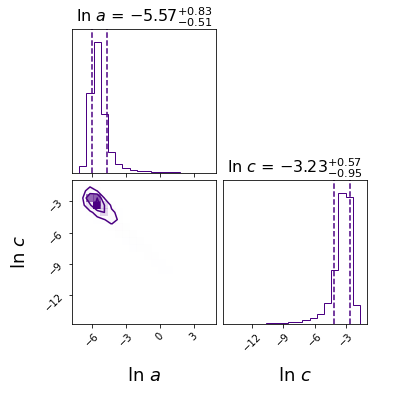}
    \label{fig: detrend MCMC}
    {}
    \includegraphics[width=0.45\textwidth]{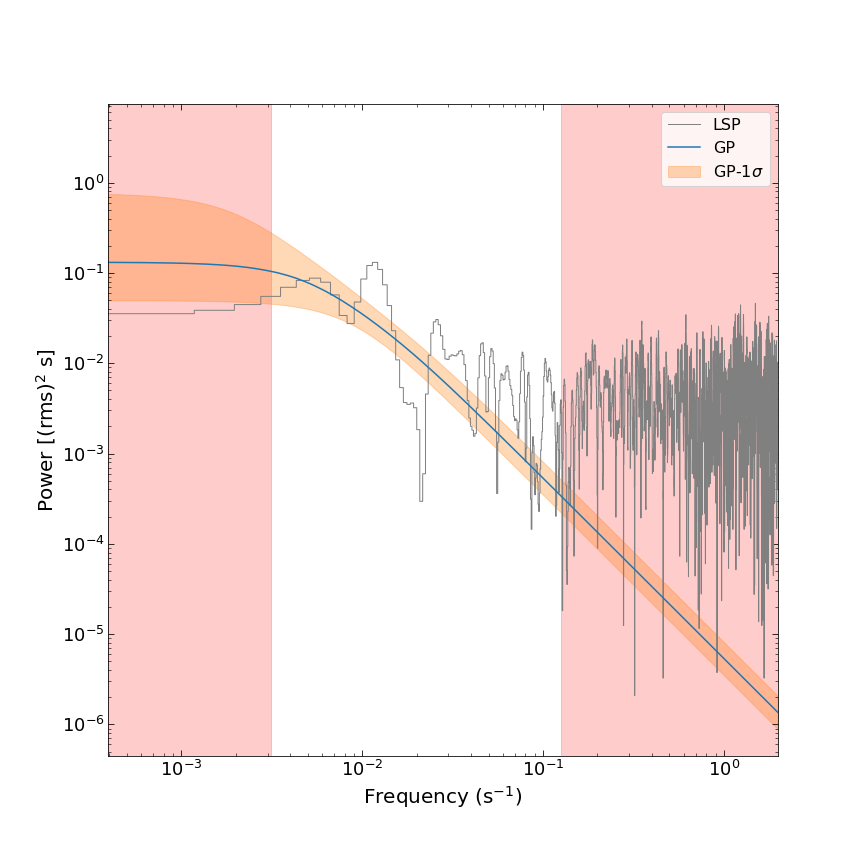}
    \label{fig: detrend PSD}
    {}
    \caption{The results of the GP(constant, DRW) model analysis for the detrended GRB 050724 light curve. The detrended light curve is obtained by subtracting the best-fit mean function in the GP(power law, DRW) model from the original light curve. Top-left panel: the detrended X-ray light curve of GRB 050724 in the time interval of [70-350] s. The black points denote the detrended data, while the orange solid line depicts the prediction of the GP(constant, DRW) model. The orange-shaded region indicates the 3$\sigma$ uncertainties in the model. Top-right panel: the posterior probability density distribution of each parameter in the GP(constant, DRW) model for GRB 050724. The vertical dashed lines indicate the 1$\sigma$ uncertainties. Bottom-left panel: the PSDs of the light curve of GRB 050724. The average PSD of the GP(constant, DRW) model and the PSD obtained from the LSP method are represented by the blue line and the gray line, respectively. The orange-shaded area indicates the 1$\sigma$ uncertainties in the GP(constant, DRW) model PSD. The red-shaded area represents the unreliable zone, and the white area represents the confidence zone where the upper and lower limits are 1/$P_{min}$ and 1/$P_{max}$, respectively. Bottom-right panel: the distribution of the quality factor $Q$. The dashed line and the dashed-dotted line indicate the critical quality factor ($Q$ = 0.5) and the median of ln $Q$, respectively.}
    \label{fig: 050724 dcl GP(power,drw)}
\end{figure}
\begin{figure}
    \centering
    \includegraphics[width=0.45\textwidth]{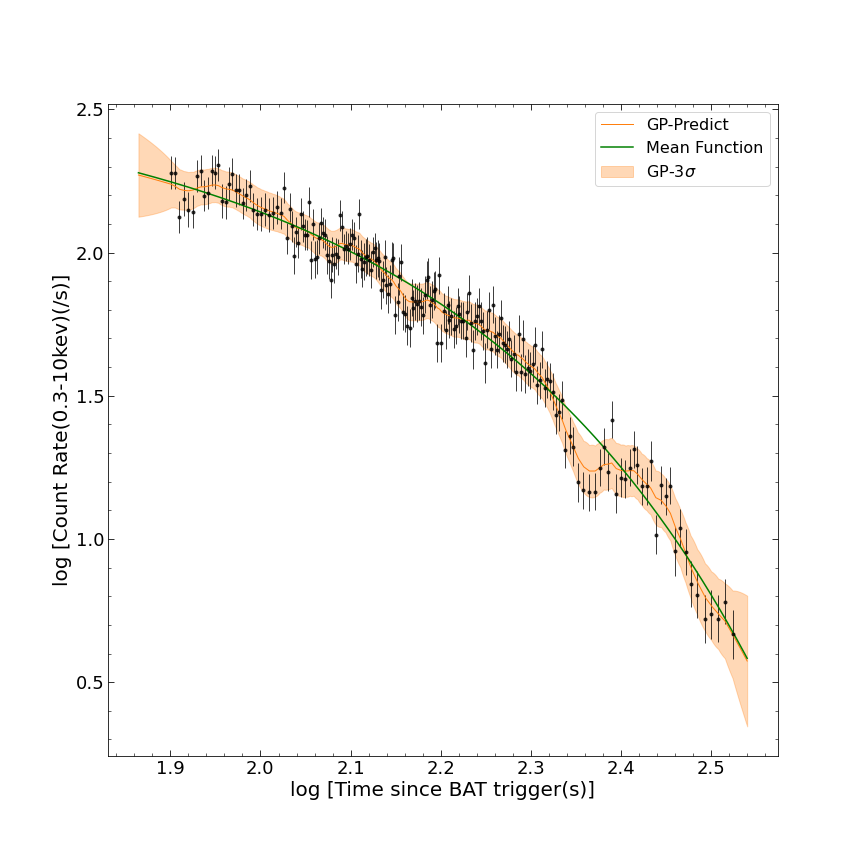}
    \label{fig: detrend MCMC}
    {}
    \includegraphics[width=0.45\textwidth]{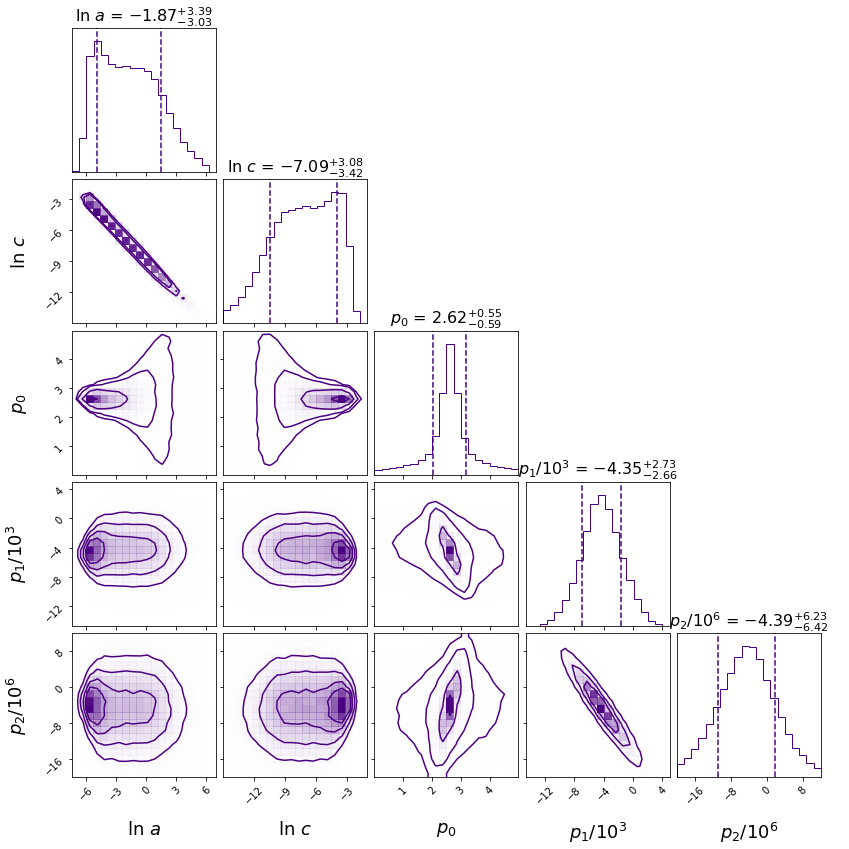}
    \label{fig: detrend MCMC}
    {}
    \includegraphics[width=0.45\textwidth]{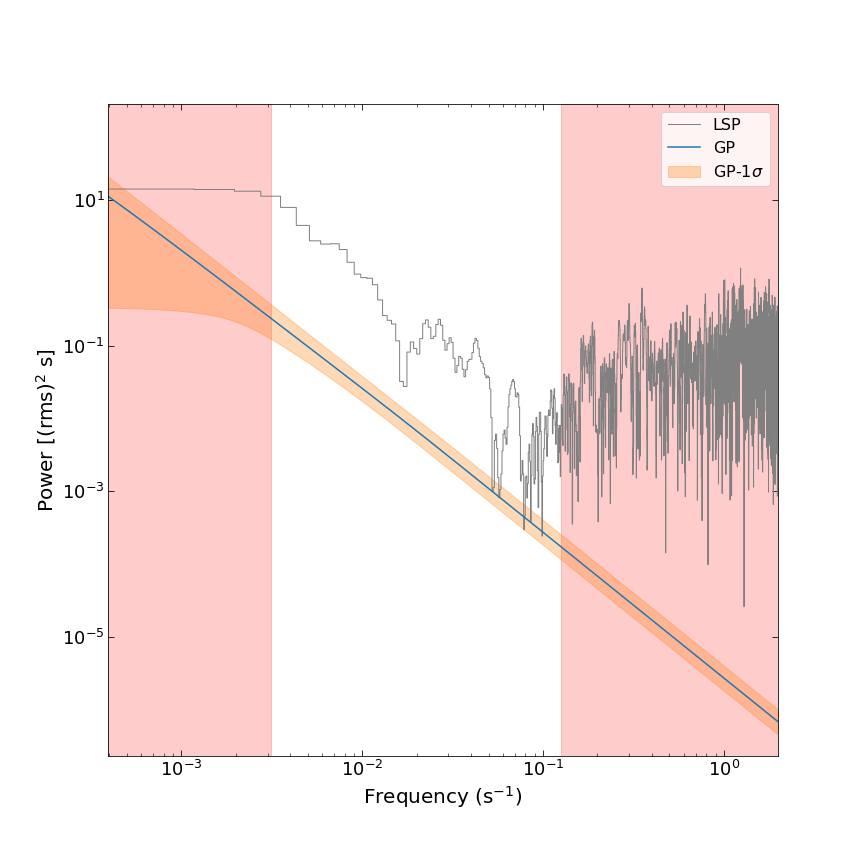}
    \label{fig: detrend PSD}
    {}
    \caption{The results of the GP(polynomial, DRW) model analysis for GRB 050724. Top-left panel: the X-ray light curve of GRB 050724 in the time interval of [70-350] s. The black points denote the data observed by Swift, while the orange solid line depicts the prediction of the GP(polynomial, DRW) model. The orange-shaded region indicates the 3$\sigma$ uncertainties in the model, and the green line represents the best-fit value of the mean function. Top-right panel: the posterior probability density distribution of each parameter in the GP(polynomial, DRW) model for GRB 050724. The vertical dashed lines indicate the 1$\sigma$ uncertainties. Bottom panel: the PSDs of the light curve of GRB 050724. The average PSD of the GP(polynomial, DRW) model and the PSD obtained from the LSP method are represented by the blue line and the gray line, respectively. The orange-shaded area indicates the 1$\sigma$ uncertainties in the GP(polynomial, DRW) model PSD. The red-shaded area represents the unreliable zone, and the white area represents the confidence zone where the upper and lower limits are 1/$P_{min}$ and 1/$P_{max}$, respectively.}
    \label{fig: 050724 ocl GP(poly,drw)}
\end{figure}
\begin{figure}
    \centering
    \includegraphics[width=0.45\textwidth]{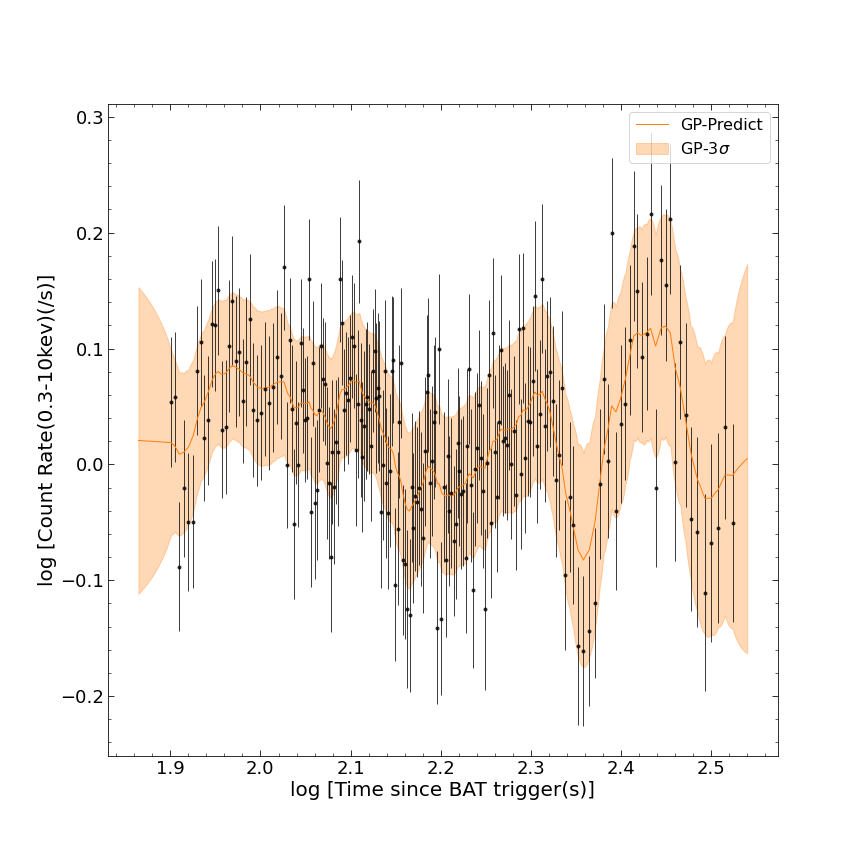}
    \label{fig: detrend MCMC}
    {}
    \includegraphics[width=0.45\textwidth]{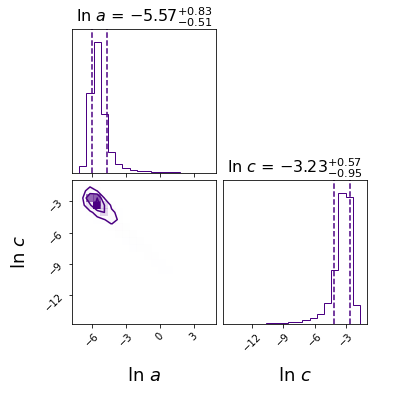}
    \label{fig: detrend MCMC}
    {}
    \includegraphics[width=0.45\textwidth]{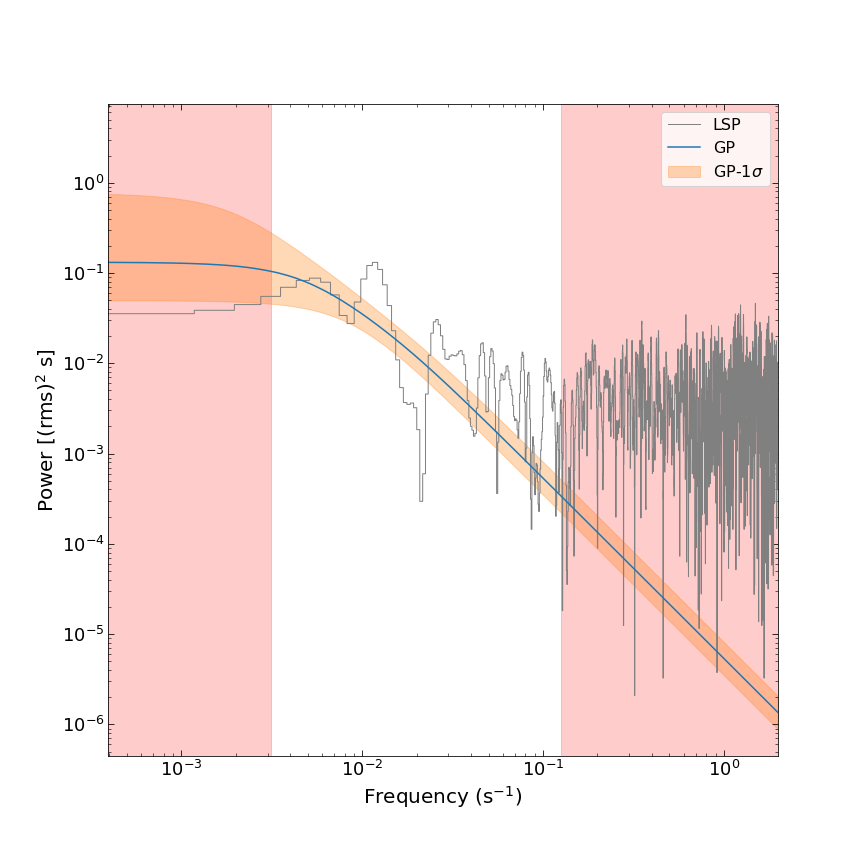}
    \label{fig: detrend PSD}
    {}
    \caption{The results of the GP(constant, DRW) model analysis for the detrended GRB 050724 light curve. The detrended light curve is obtained by subtracting the best-fit mean function in the GP(polynomial, DRW) model from the original light curve. Top-left panel: the detrended X-ray light curve of GRB 050724 in the time interval of [70-350] s. The black points denote the detrended data, while the orange solid line depicts the prediction of the GP(constant, DRW) model. The orange-shaded region indicates the 3$\sigma$ uncertainties in the model. Top-right panel: the posterior probability density distribution of each parameter in the GP(constant, DRW) model for GRB 050724. The vertical dashed lines are 1$\sigma$ uncertainties. Bottom panel: the PSDs of the light curve of GRB 050724. The average PSD of the GP(constant, DRW) model and the PSD obtained from the LSP method are represented by the blue line and the gray line, respectively. The orange-shaded area indicates the 1$\sigma$ uncertainties in the GP(constant, DRW) model PSD. The red-shaded area represents the unreliable zone, and the white area represents the confidence zone where the upper and lower limits are 1/$P_{min}$ and 1/$P_{max}$, respectively.}
    \label{fig: 050724 dcl GP(poly,drw)}
\end{figure}
\begin{figure}
    \centering
    \includegraphics[width=0.45\textwidth]{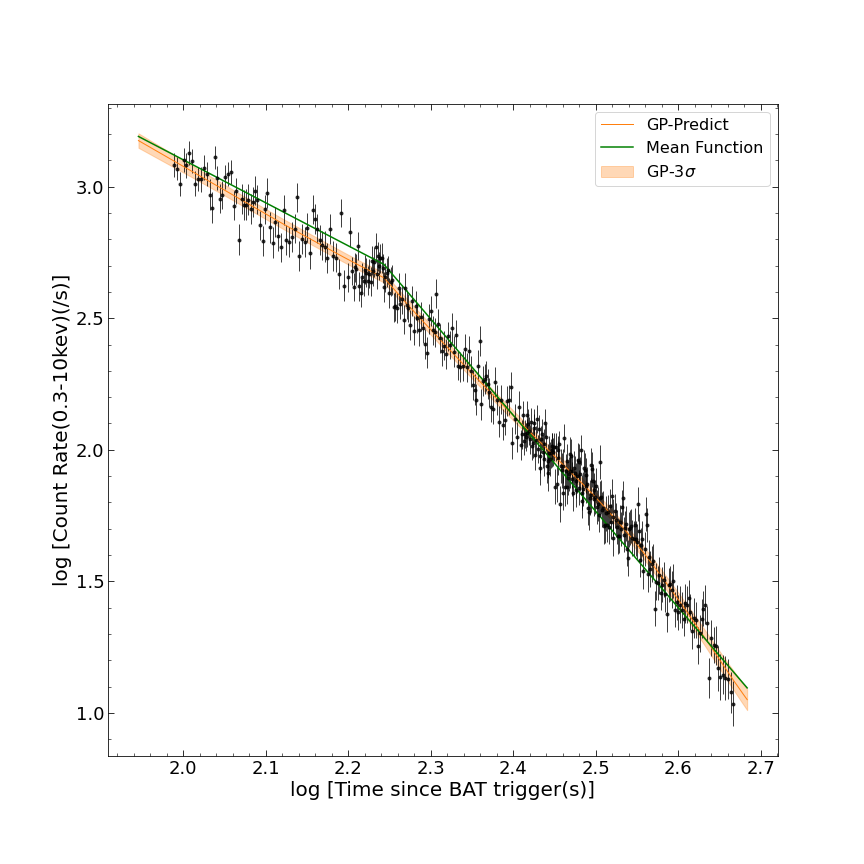}
    \label{fig: detrend MCMC}
    {}
    \includegraphics[width=0.45\textwidth]{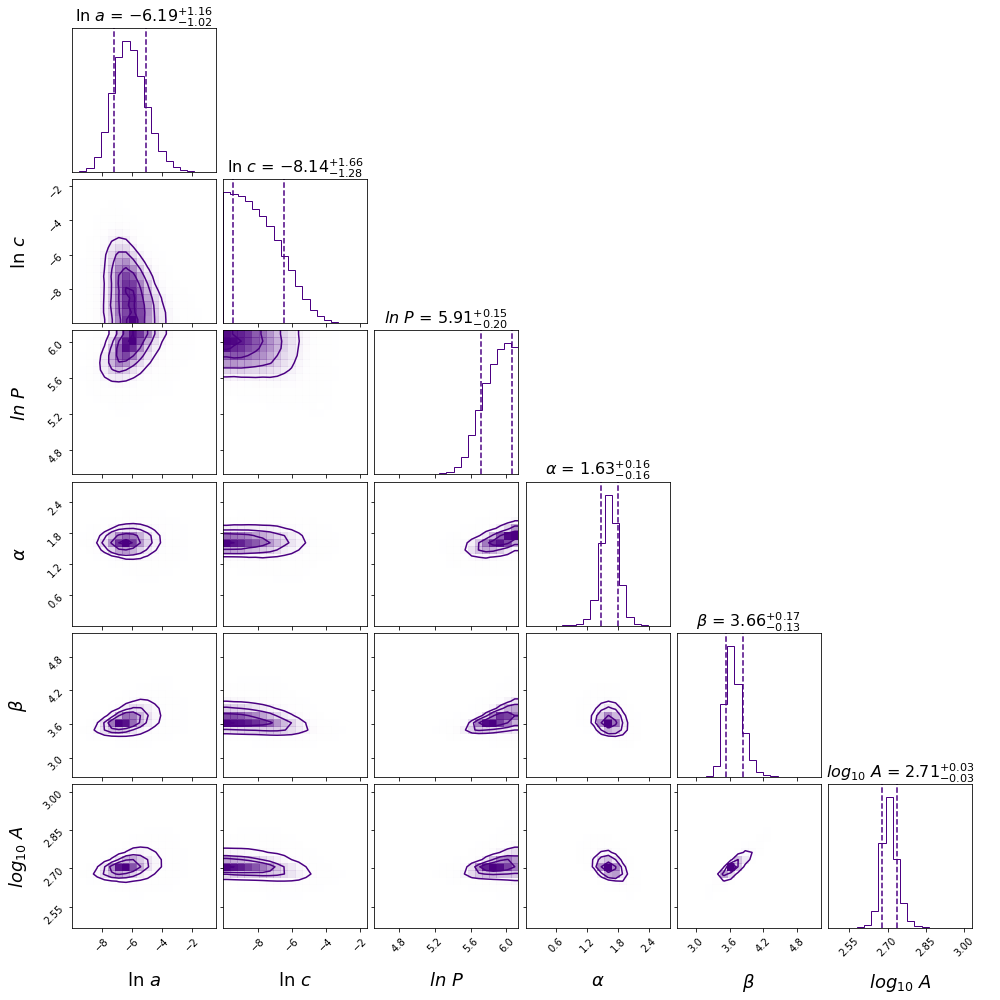}
    \label{fig: detrend MCMC}
    {}
    \includegraphics[width=0.45\textwidth]{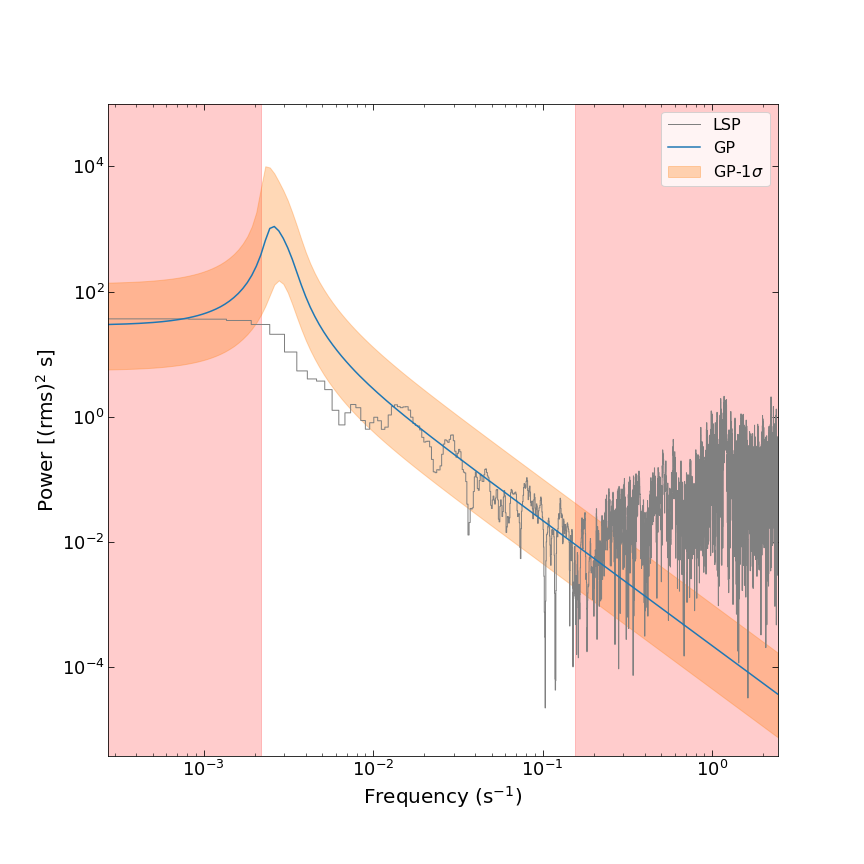}
    \label{fig: detrend PSD}
    {}
    \includegraphics[width=0.45\textwidth]{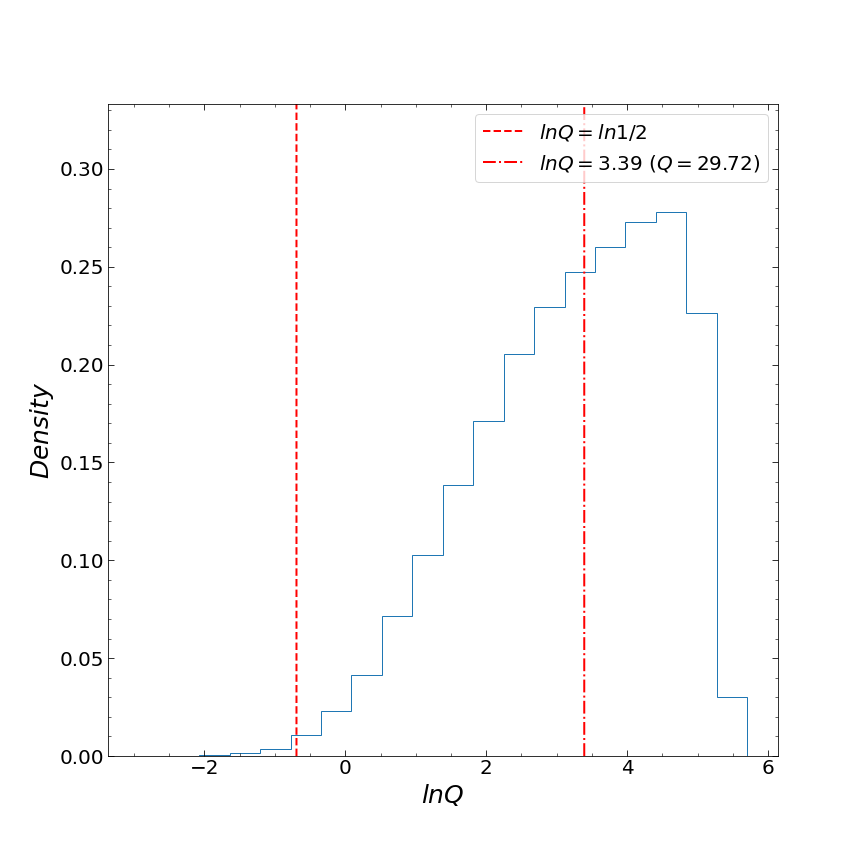}
    \label{fig: detrend Q}
    \caption{The results of the GP(power law, QPO) model analysis for GRB 060614. Top-left panel: the X-ray light curve of GRB 060614 in the time interval of [90-500] s. The black points denote the data observed by Swift, while the orange solid line depicts the prediction of the GP(power law, QPO) model. The orange-shaded region indicates the 3$\sigma$ uncertainties in the model, and the green line represents the best-fit value of the mean function. Top-right panel: the posterior probability density distribution of each parameter in the GP(power law, QPO) model for GRB 060614 The vertical dashed lines indicate the 1$\sigma$ uncertainties. Bottom-left panel: the PSDs of the light curve of GRB 060614. The average PSD of the GP(power law, QPO) model and the PSD obtained from the LSP method are represented by the blue line and the gray line, respectively. The orange-shaded area indicates the 1$\sigma$ uncertainties in the GP(power law, QPO) model PSD. The red-shaded area represents the unreliable zone, and the white area represents the confidence zone where the upper and lower limits are 1/$P_{min}$ and 1/$P_{max}$, respectively. Bottom-right panel: the distribution of the quality factor $Q$. The dashed line and the dashed-dotted line indicate the critical quality factor ($Q$ = 0.5) and the median of ln $Q$, respectively.}
    \label{fig: 060614 ocl GP(pl,qpo)}
\end{figure}
\begin{figure}
    \centering
    \includegraphics[width=0.45\textwidth]{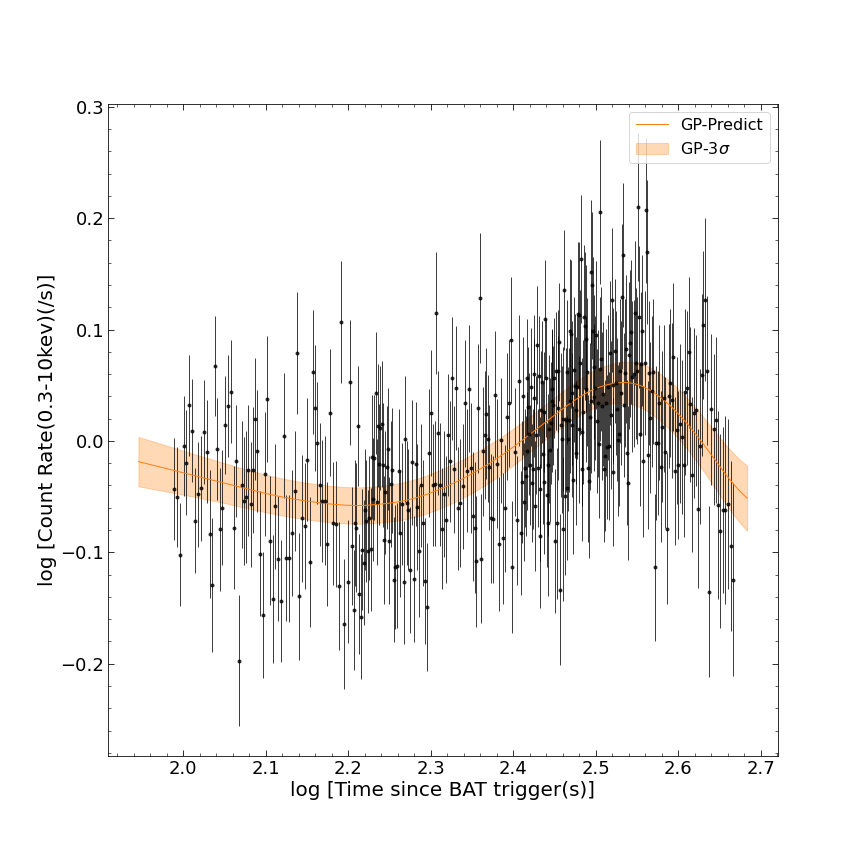}
    \label{fig: detrend MCMC}
    {}
    \includegraphics[width=0.45\textwidth]{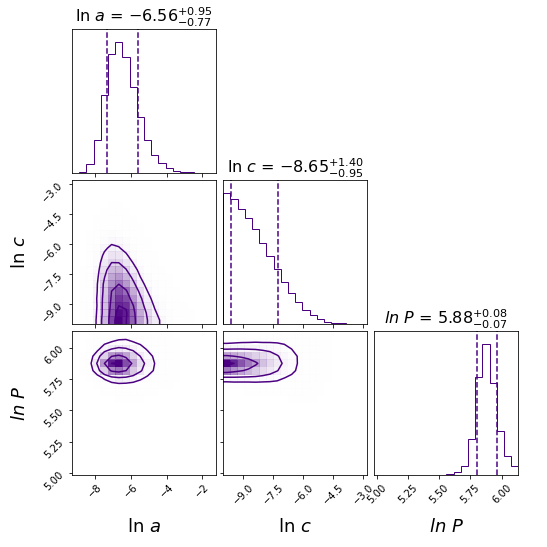}
    \label{fig: detrend MCMC}
    {}
    \includegraphics[width=0.45\textwidth]{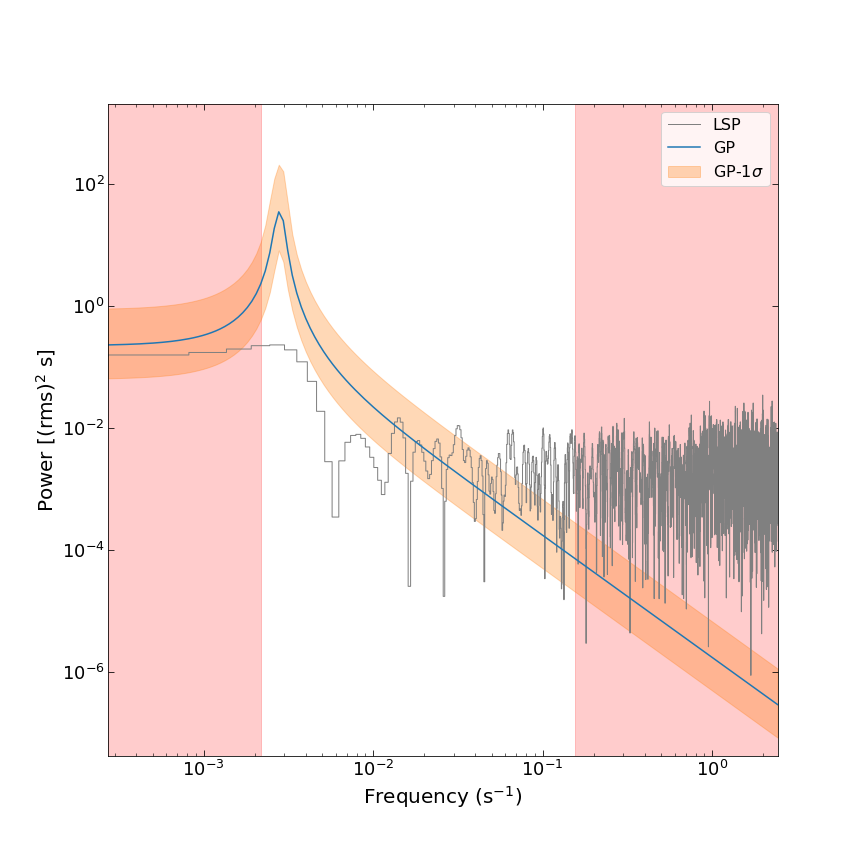}
    \label{fig: detrend PSD}
    {}
    \includegraphics[width=0.45\textwidth]{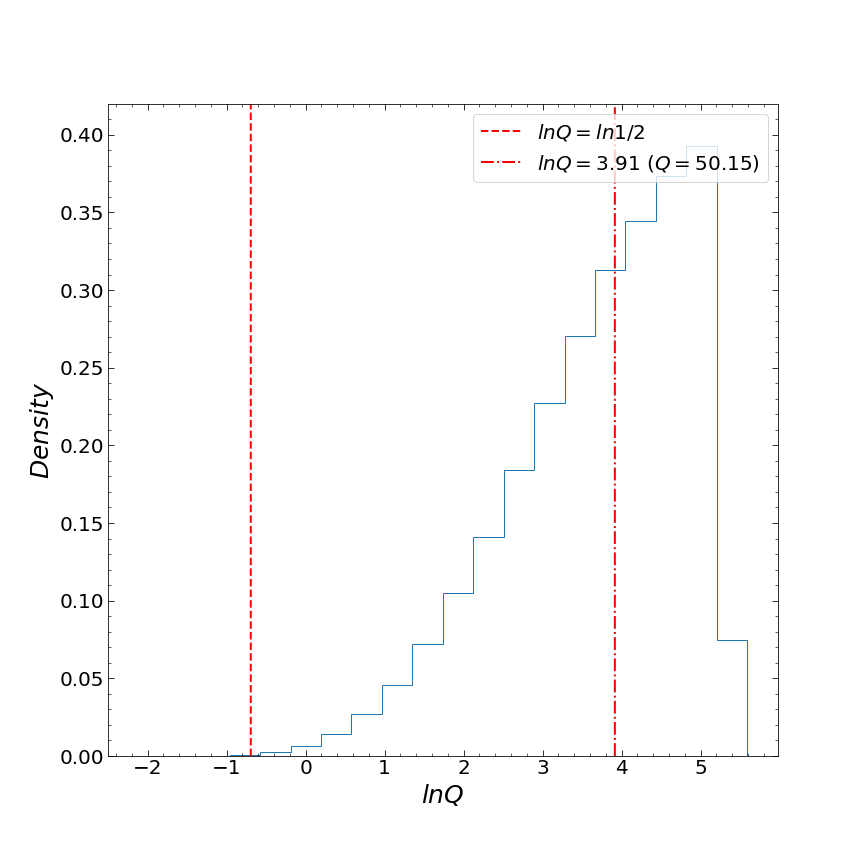}
    \label{fig: detrend Q}
    \caption{The results of the GP(constant, QPO) model analysis for the detrended GRB 060614 light curve. The detrended light curve is obtained by subtracting the best-fit mean function in the GP(power law, QPO) model from the original light curve. Top-left panel: the detrended X-ray light curve of GRB 060614 in the time interval of [90-500] s. The black points denote the detrended data, while the orange solid line depicts the prediction of the GP(constant, QPO) model. The orange-shaded region indicates the 3$\sigma$ uncertainties in the model. Top-right panel: the posterior probability density distribution of each parameter in the GP(constant, QPO) model for GRB 060614. The vertical dashed lines indicate the 1$\sigma$ uncertainties. Bottom-left panel: the PSDs of the light curve of GRB 060614. The average PSD of the GP(constant, QPO) model and the PSD obtained from the LSP method are represented by the blue line and the gray line, respectively. The orange-shaded area indicates the 1$\sigma$ uncertainties in the GP(constant, QPO) model PSD. The red-shaded area represents the unreliable zone, and the white area represents the confidence zone where the upper and lower limits are 1/$P_{min}$ and 1/$P_{max}$, respectively. Bottom-right panel: the distribution of the quality factor $Q$. The dashed line and the dashed-dotted line indicate the critical quality factor ($Q$ = 0.5) and the median of ln $Q$, respectively.}
    \label{fig: 060614 dcl GP(pl,qpo)}
\end{figure}
\begin{figure}
    \centering
    \includegraphics[width=0.45\textwidth]{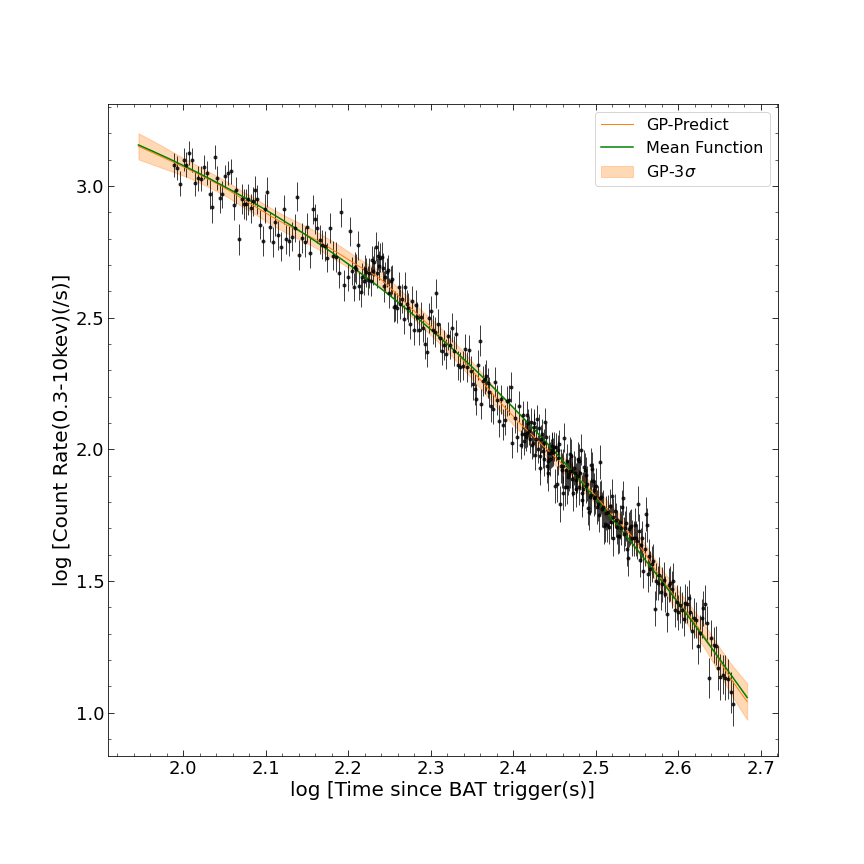}
    \label{fig: detrend MCMC}
    {}
    \includegraphics[width=0.45\textwidth]{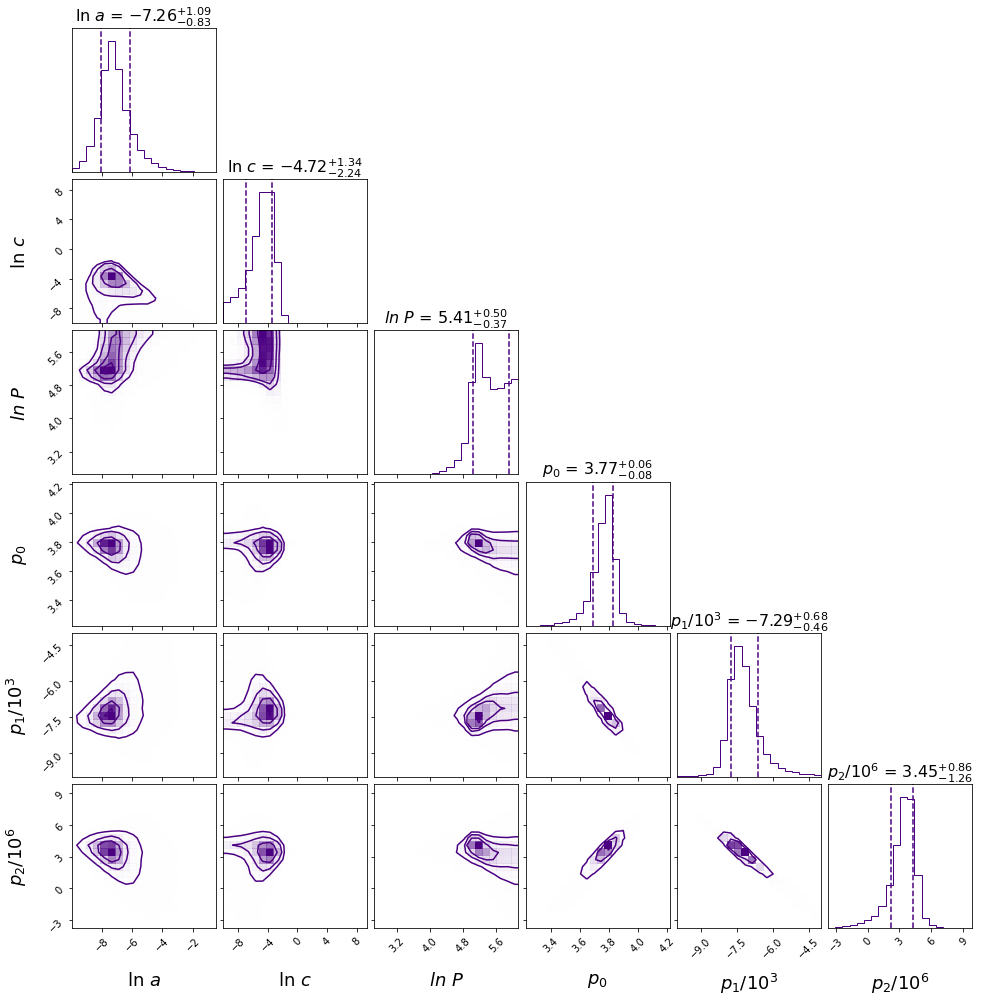}
    \label{fig: detrend MCMC}
    {}
    \includegraphics[width=0.45\textwidth]{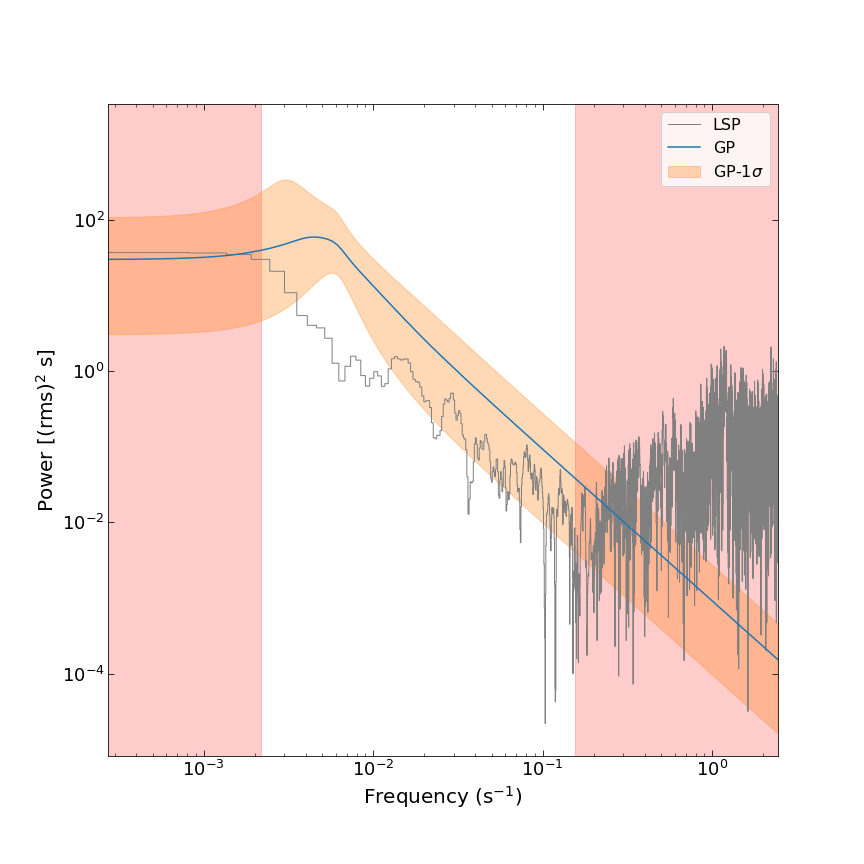}
    \label{fig: detrend PSD}
    {}
    \includegraphics[width=0.45\textwidth]{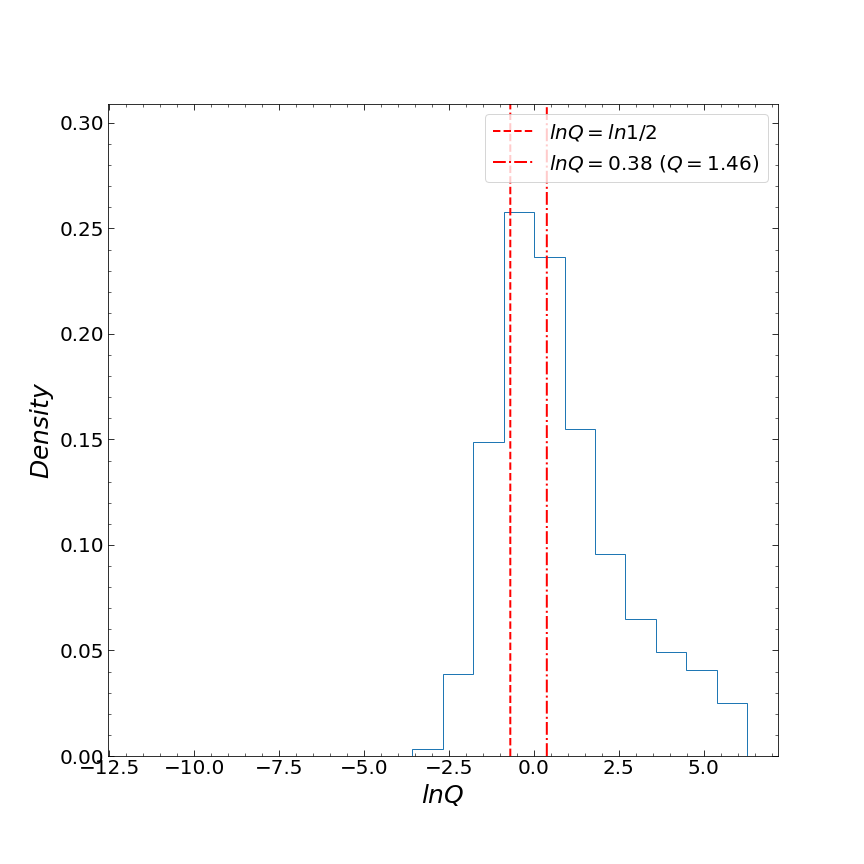}
    \label{fig: detrend Q}
    \caption{The results of the GP(polynomial, QPO) model analysis for GRB 060614. Top-left panel: the X-ray light curve of GRB 060614 in the time interval of [90-500] s. The black points debite the data observed by Swift, while the orange solid line depicts the prediction of the GP(polynomial, QPO) model. The orange-shaded region indicates the 3$\sigma$ uncertainties in the model, and the green line represents the best-fit value of the mean function. Top-right panel: the posterior probability density distribution of each parameter in the GP(polynomial, QPO) model for GRB 060614 The vertical dashed lines indecates the 1$\sigma$ uncertainties. Bottom-left panel: the PSDs of the light curve of GRB 060614. The average PSD of the GP(polynomial, QPO) model and the PSD obtained from the LSP method are represented by the blue line and the gray line, respectively. The orange-shaded area indicates the 1$\sigma$ uncertainties in the GP(polynomial, QPO) model PSD. The red-shaded area represents the unreliable zone, and the white area represents the confidence zone where the upper and lower limits are 1/$P_{min}$ and 1/$P_{max}$, respectively. Bottom right panel: the distribution of the quality factor $Q$. The dashed line and the dashed-dotted line indicate the critical quality factor ($Q$ = 0.5) and the median of ln $Q$, respectively.}
    \label{fig: 060614 ocl GP(poly,qpo)}
\end{figure}
\begin{figure}
    \centering
    \includegraphics[width=0.45\textwidth]{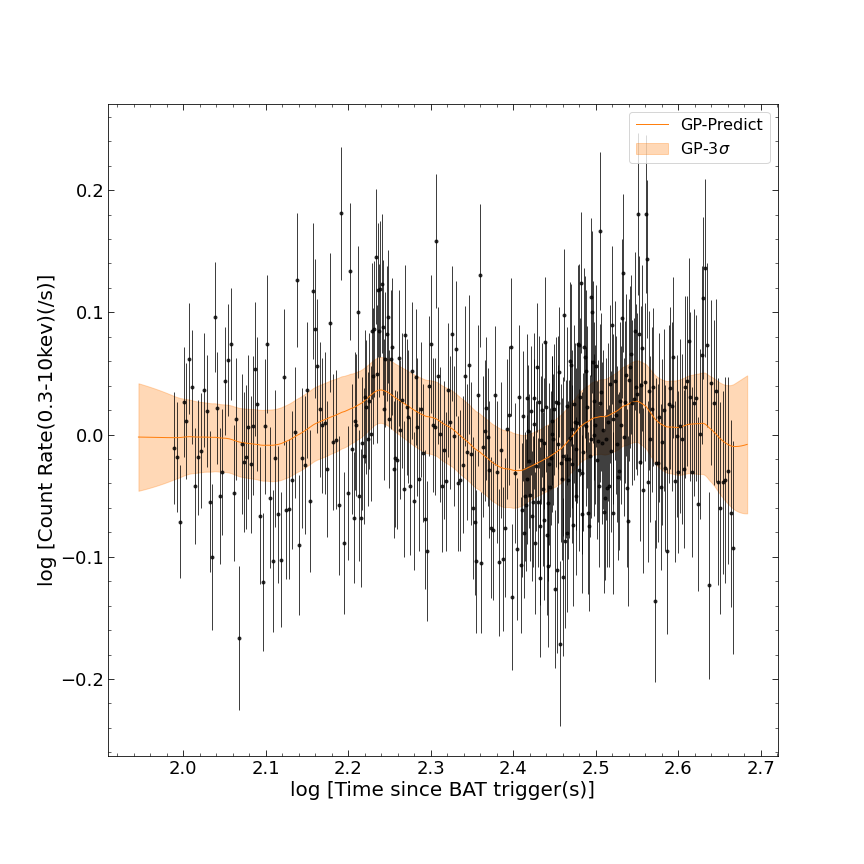}
    \label{fig: detrend MCMC}
    {}
    \includegraphics[width=0.45\textwidth]{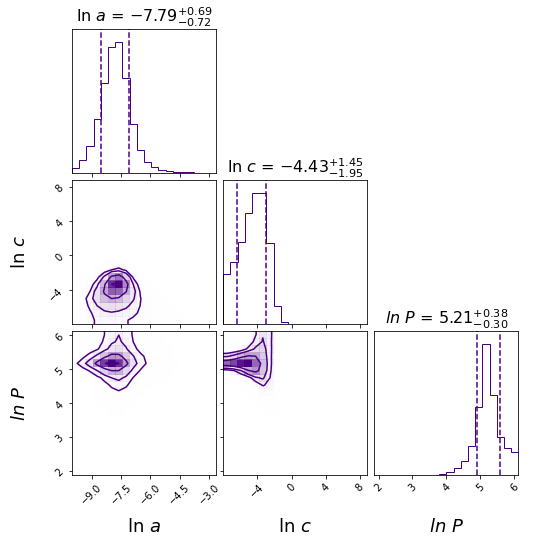}
    \label{fig: detrend MCMC}
    {}
    \includegraphics[width=0.45\textwidth]{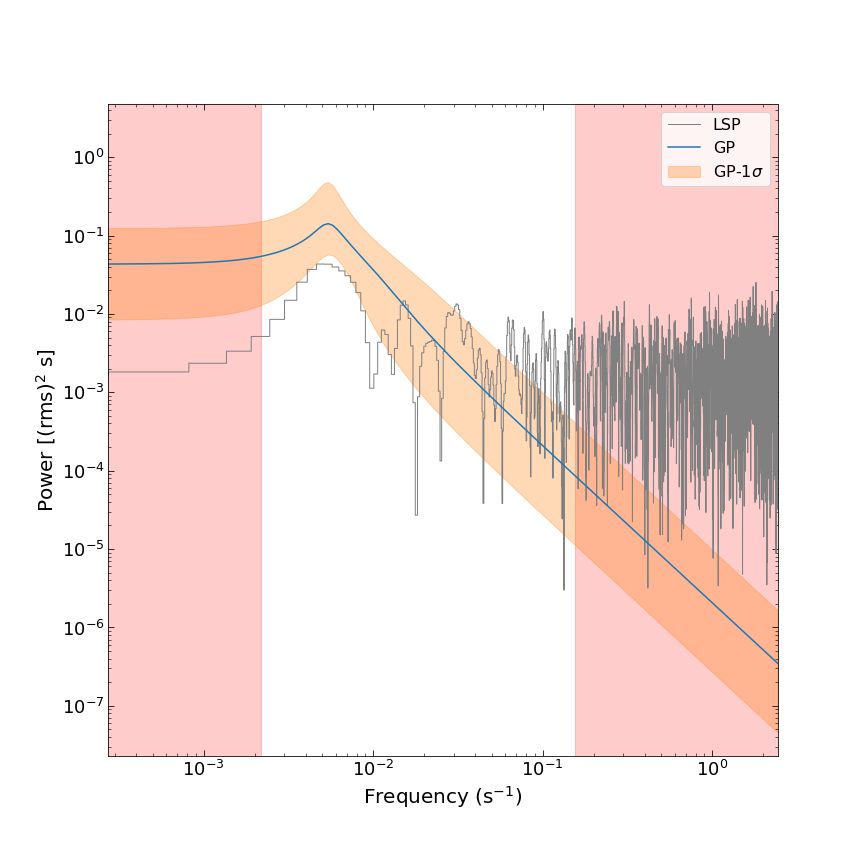}
    \label{fig: detrend PSD}
    {}
    \includegraphics[width=0.45\textwidth]{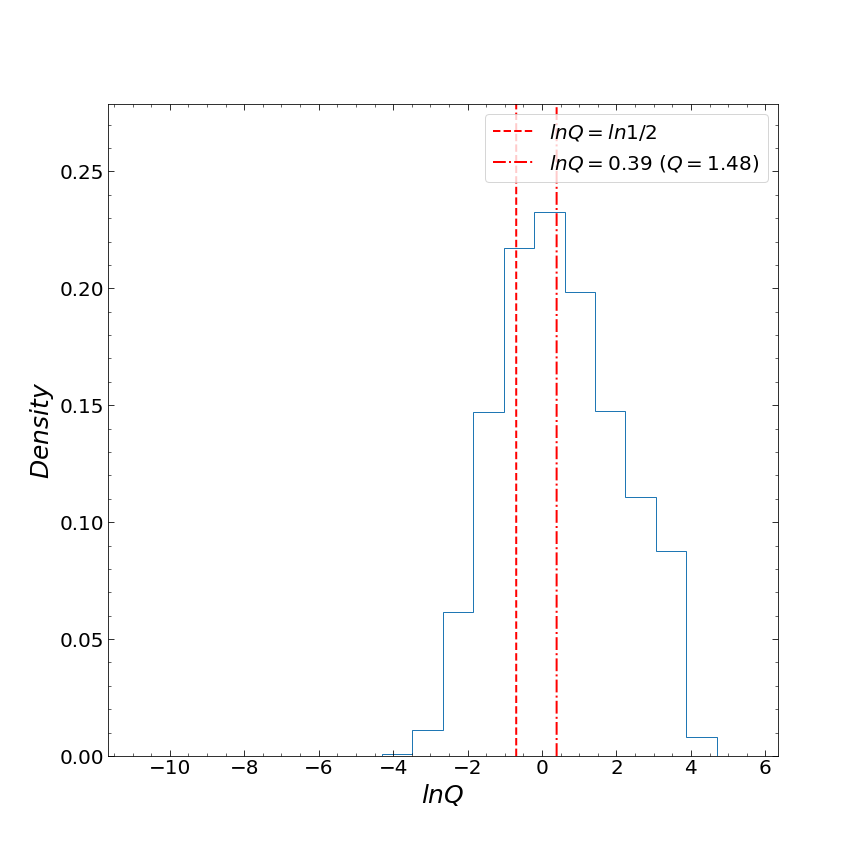}
    \label{fig: detrend Q}
    \caption{The results of the GP(constant, QPO) model analysis for the detrended GRB 060614 light curve. The detrended light curve is obtained by subtracting the best-fit mean function in the GP(polynomial, QPO) model from the original light curve. Top-left panel: the detrended X-ray light curve of GRB 060614 in the time interval of [90-500] s. The black points denote the detrended data, while the orange solid line depicts the prediction of the GP(constant, QPO) model. The orange-shaded region indicates the 3$\sigma$ uncertainties in the model. Top-right panel: the posterior probability density distribution of each parameter in the GP(constant, QPO) model for GRB 060614. The vertical dashed lines indicate the 1$\sigma$ uncertainties. Bottom-left panel: the PSDs of the light curve of GRB 060614. The average PSD of the GP(constant, QPO) model and the PSD obtained from the LSP method are represented by the blue line and the gray line, respectively. The orange-shaded area indicates the 1$\sigma$ uncertainties in the GP(constant, QPO) model PSD. The red-shaded area represents the unreliable zone, and the white area represents the confidence zone where the upper and lower limits are 1/$P_{min}$ and 1/$P_{max}$, respectively. Bottom-right panel: the distribution of the quality factor $Q$. The dashed line and the dashed-dotted line indicate the critical quality factor ($Q$ = 0.5) and the median of ln $Q$, respectively.}
    \label{fig: 0606014 dcl GP(poly,qpo)}
\end{figure}
\begin{figure}
    \centering
    \includegraphics[width=0.45\textwidth]{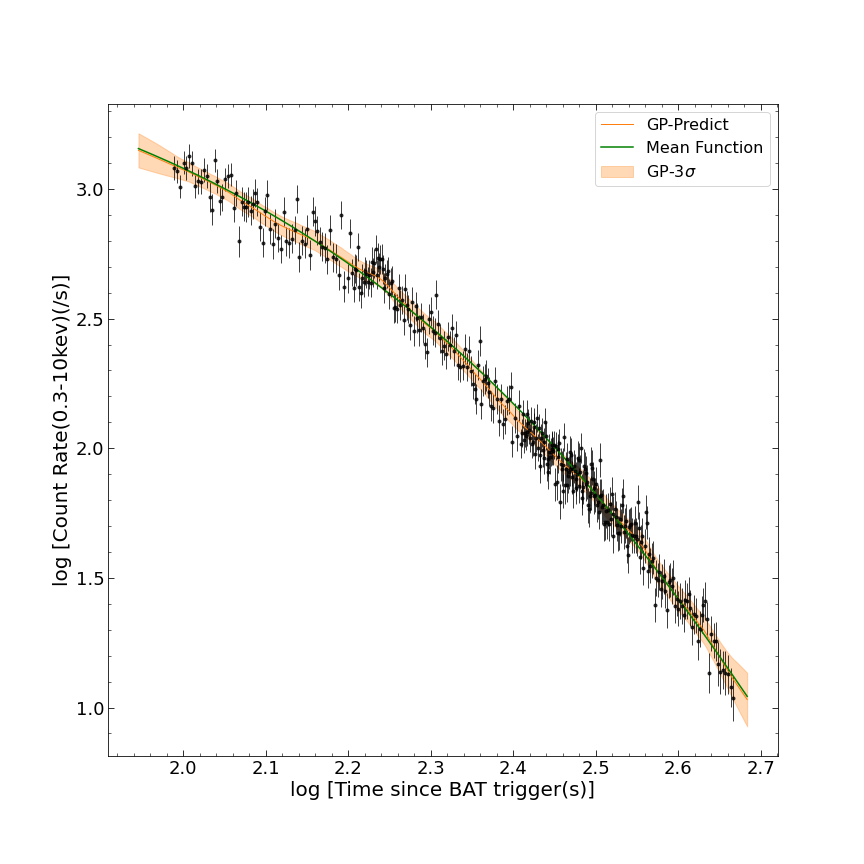}
    \label{fig: detrend MCMC}
    {}
    \includegraphics[width=0.45\textwidth]{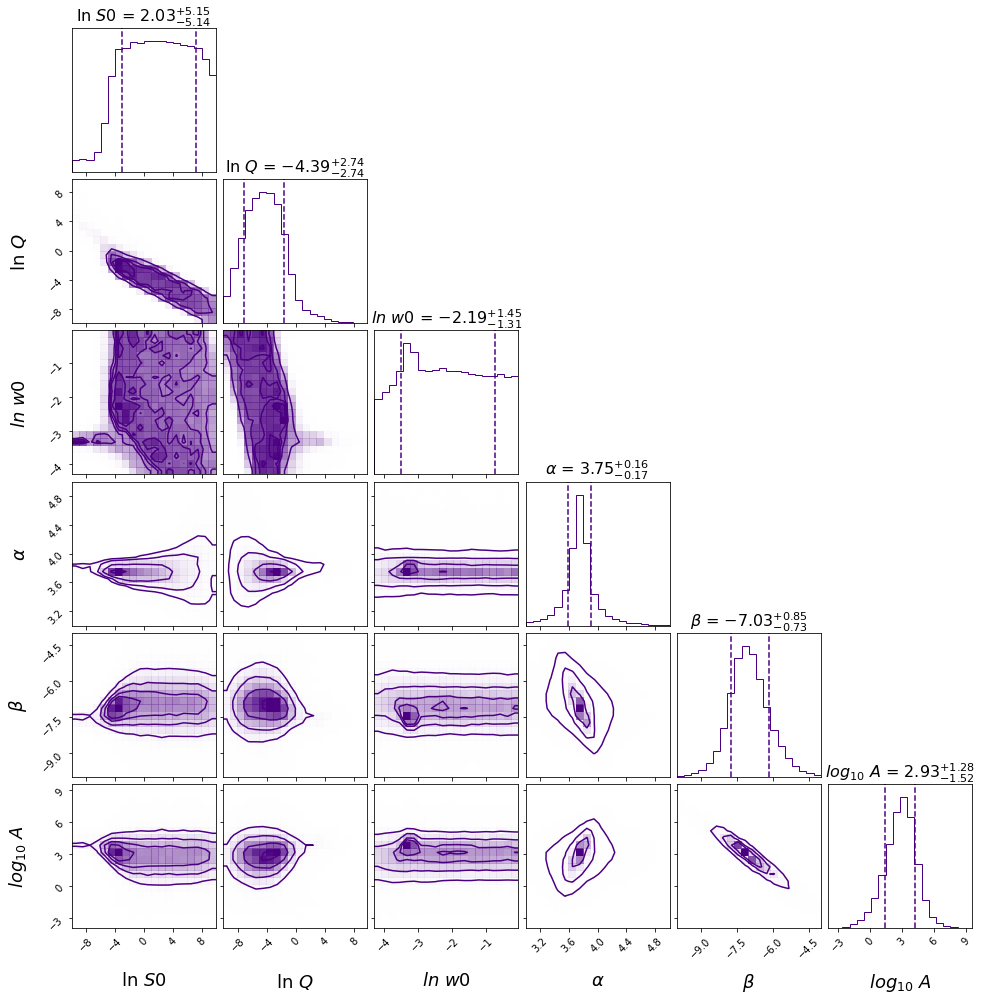}
    \label{fig: detrend MCMC}
    {}
    \includegraphics[width=0.45\textwidth]{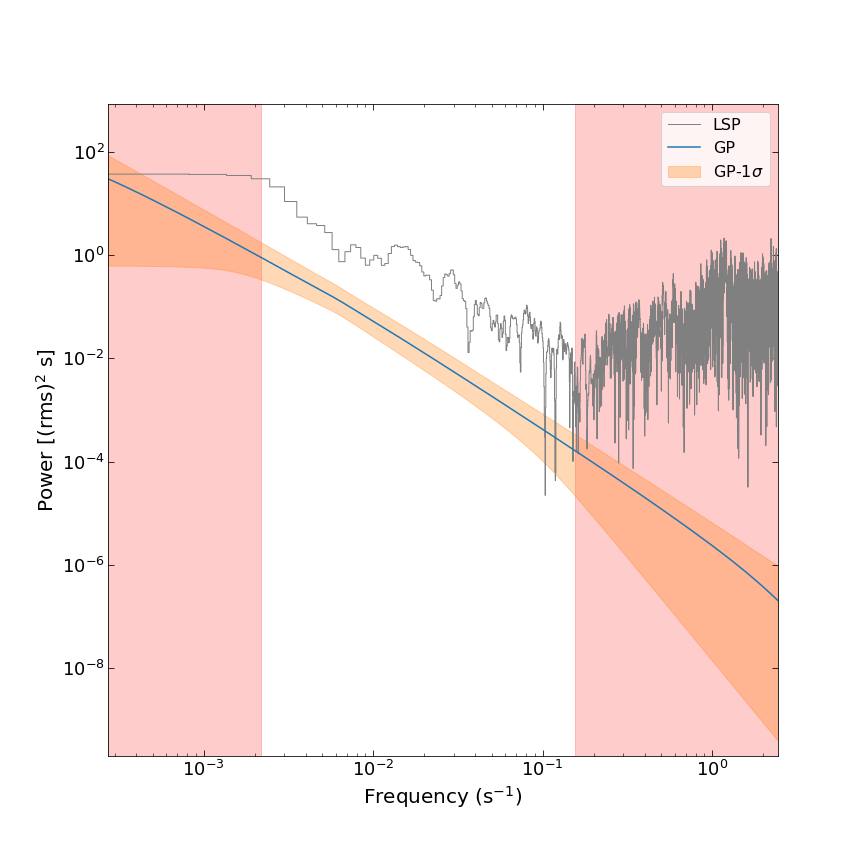}
    \label{fig: detrend PSD}
    {}
    \includegraphics[width=0.45\textwidth]{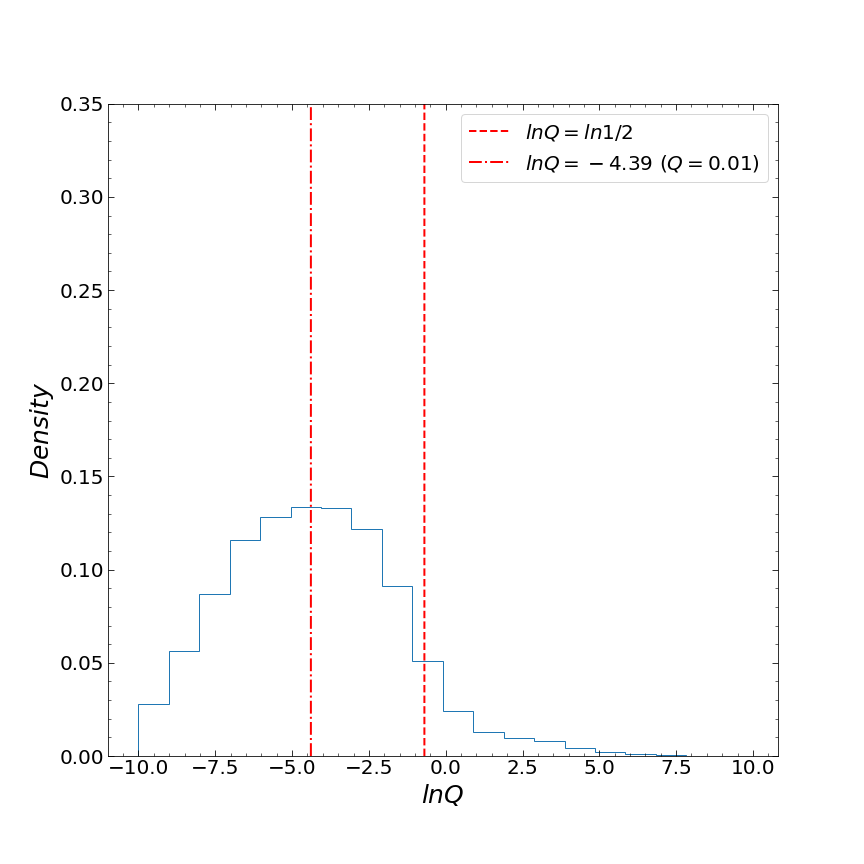}
    \label{fig: detrend Q}
    \caption{The results of the GP(polynomial, SHO) model analysis for GRB 060614. Top-left panel: the X-ray light curve of GRB 060614 in the time interval of [90-500] s. The black points denote the data observed by Swift, while the orange solid line depicts the prediction of the GP(polynomial, SHO) model. The orange-shaded region indicates the 3$\sigma$ uncertainties in the model, and the green line represents the best-fit value of the mean function. Top-right panel: the posterior probability density distribution of each parameter in the GP(polynomial, SHO) model for GRB 060614 The vertical dashed lines indicate the 1$\sigma$ uncertainties. Bottom-left panel: the PSDs of the light curve of GRB 060614. The average PSD of the GP(polynomial, SHO) model and the PSD obtained from the LSP method are represented by the blue line and the gray line, respectively. The orange-shaded area indicate the 1$\sigma$ uncertainties in the GP(polynomial, SHO) model PSD. The red-shaded area represents the unreliable zone, and the white area represents the confidence zone where the upper and lower limits are 1/$P_{min}$ and 1/$P_{max}$, respectively. Bottom-right panel: the distribution of the quality factor $Q$. The dashed line and the dashed-dotted line indicate the critical quality factor ($Q$ = 0.5) and the median of ln $Q$, respectively.}
    \label{fig: 060614 ocl GP(poly,sho)}
\end{figure}
\begin{figure}
    \centering
    \includegraphics[width=0.45\textwidth]{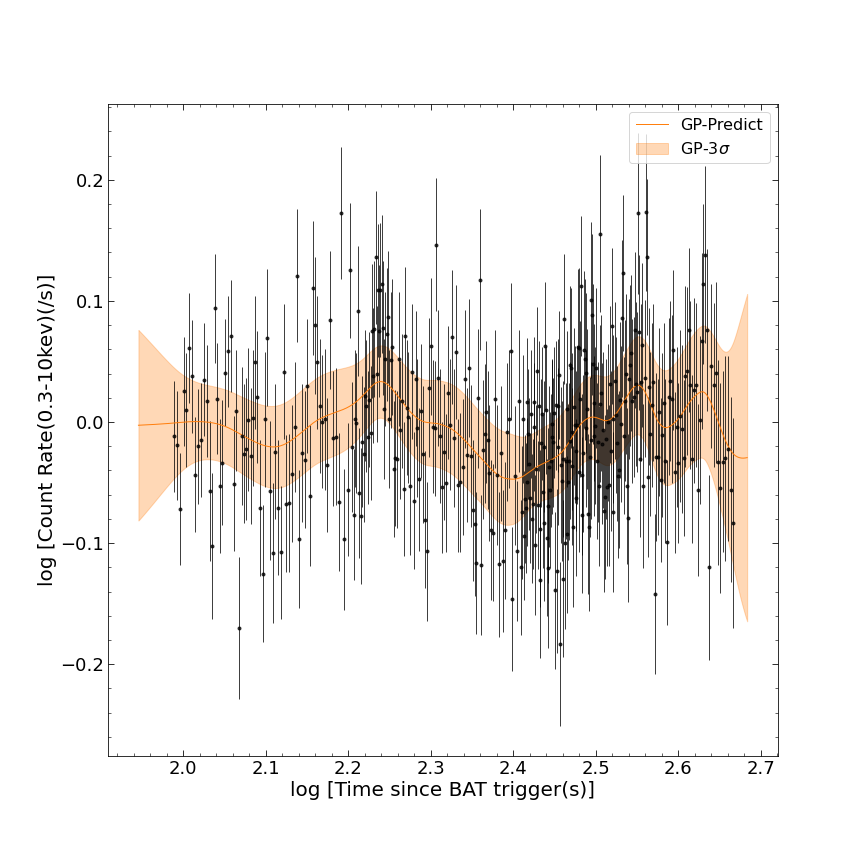}
    \label{fig: detrend MCMC}
    {}
    \includegraphics[width=0.45\textwidth]{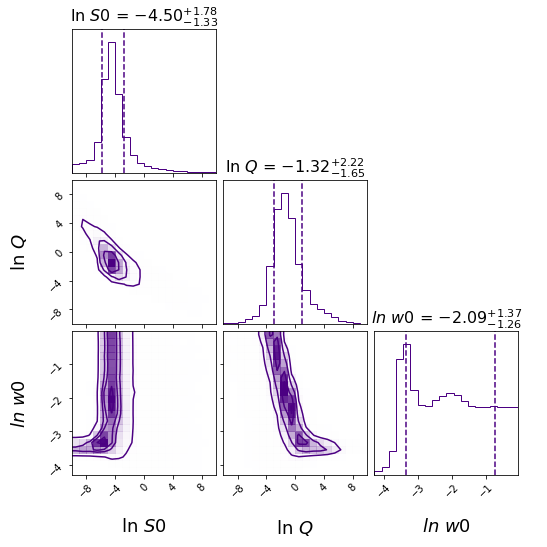}
    \label{fig: detrend MCMC}
    {}
    \includegraphics[width=0.45\textwidth]{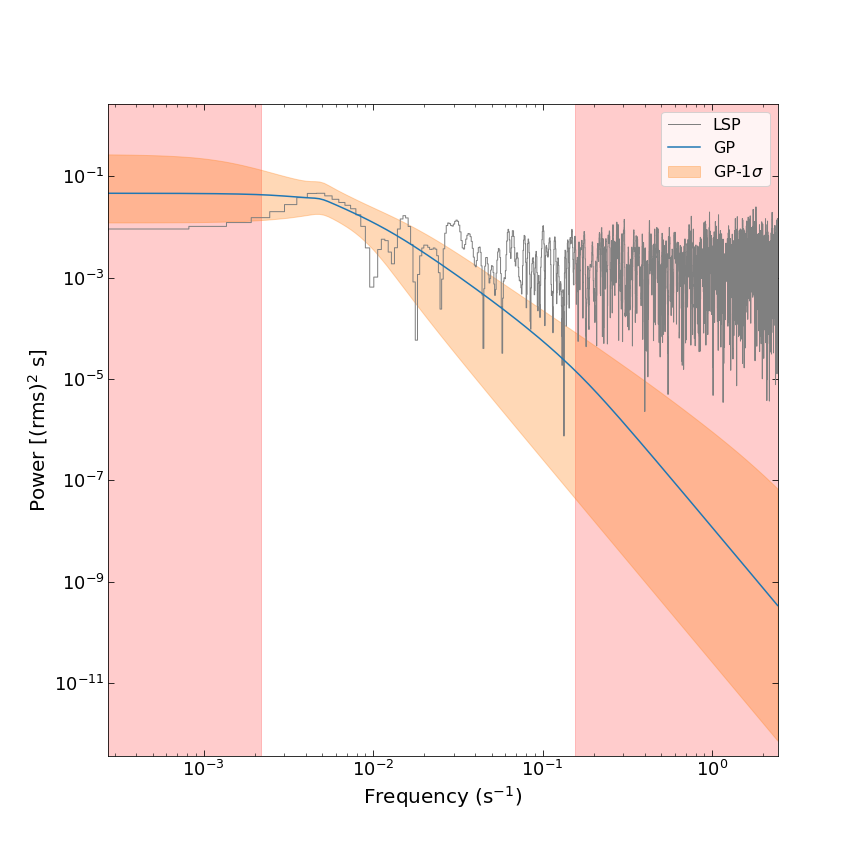}
    \label{fig: detrend PSD}
    {}
    \includegraphics[width=0.45\textwidth]{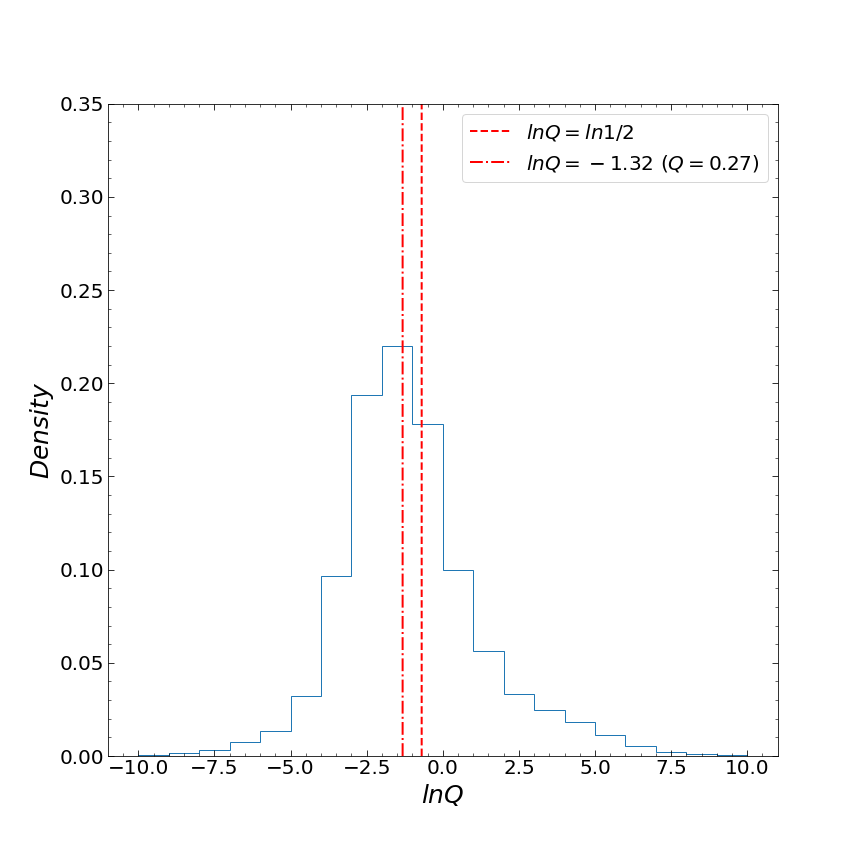}
    \label{fig: detrend Q}
    \caption{The results of the GP(constant, SHO) model analysis for the detrended GRB 060614 light curve. The detrended light curve is obtained by subtracting the best-fit mean function in the GP(polynomial, SHO) model from the original light curve. Top-left panel: the detrended X-ray light curve of GRB 060614 in the time interval of [90-500] s. The black points denote the detrended data, while the orange solid line depicts the prediction of the GP(constant, SHO) model. The orange-shaded region indicates the 3$\sigma$ uncertainties of the model. Top-right panel: the posterior probability density distribution of each parameter in the GP(constant, SHO) model for GRB 060614. The vertical dashed lines indicates the 1$\sigma$ uncertainties. Bottom-left panel: the PSDs of the light curve of GRB 060614. The average PSD of the GP(constant, SHO) model and the PSD obtained from the LSP method are represented by the blue line and the gray line, respectively. The orange-shaded area indicates the 1$\sigma$ uncertainties in the GP(constant, SHO) model PSD. The red-shaded area represents the unreliable zone, and the white area represents the confidence zone where the upper and lower limits are 1/$P_{min}$ and 1/$P_{max}$, respectively. Bottom-right panel: the distribution of the quality factor $Q$. The dashed line and the dashed-dotted line indicate the critical quality factor ($Q$ = 0.5) and the median of ln $Q$, respectively.}
    \label{fig: 060614 dcl GP(poly,sho)}
\end{figure}
\begin{figure}
    \centering
    \includegraphics[width=0.45\textwidth]{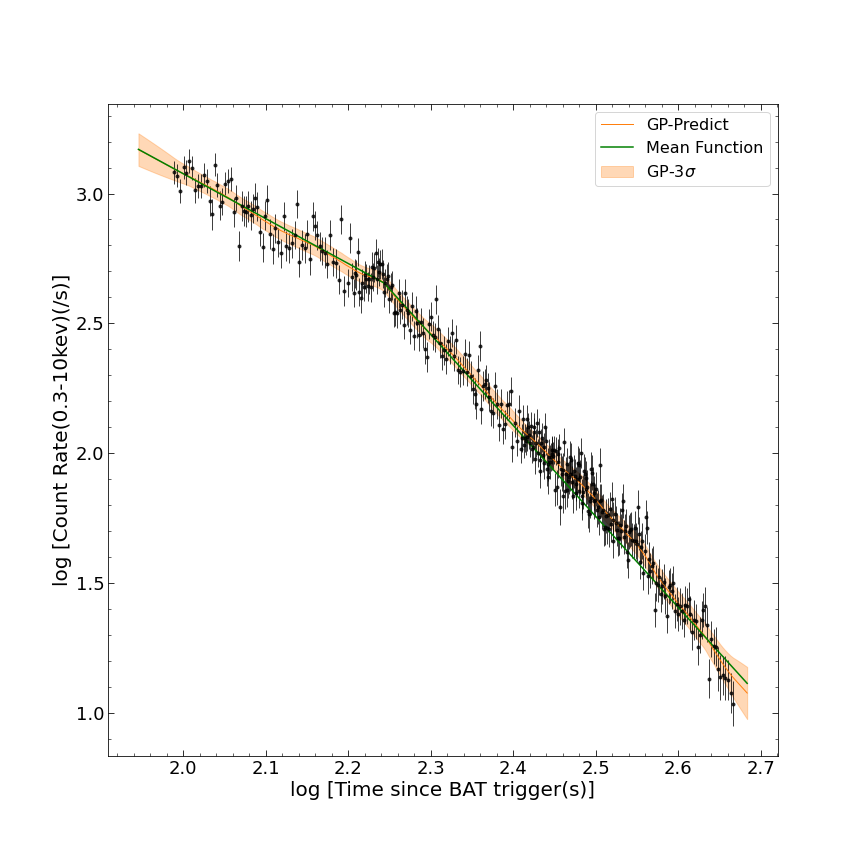}
    {}
    \includegraphics[width=0.45\textwidth]{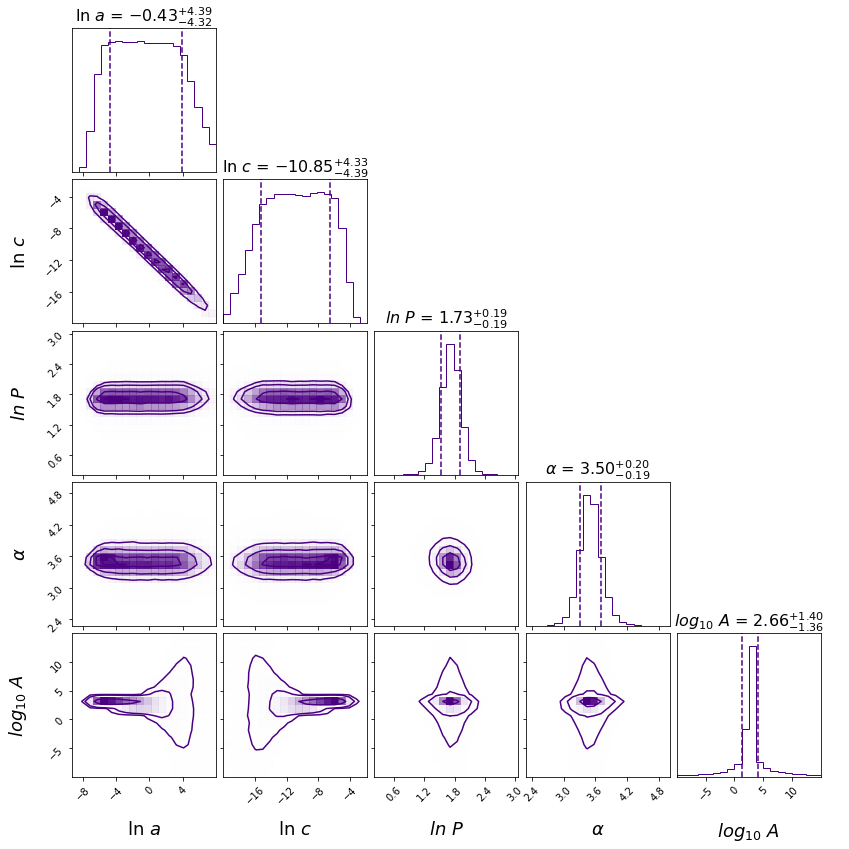}
    \label{fig: detrend MCMC}
    {}
    \includegraphics[width=0.45\textwidth]{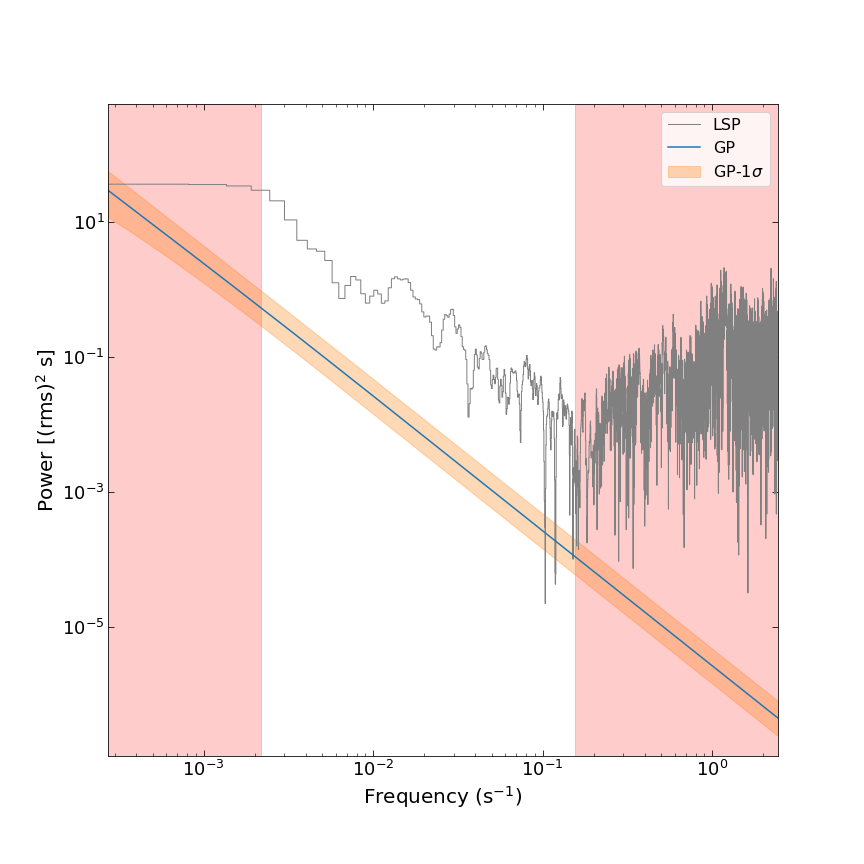}
    \label{fig: detrend PSD}
    {}
    \caption{The results of the GP(power law, DRW) model analysis for GRB 060614. Top-left panel: the X-ray light curve of GRB 060614 in the time interval of [90-500] s. The black points denote the data observed by Swift, while the orange solid line depicts the prediction of the GP(power law, DRW) model. The orange-shaded region indicates the 3$\sigma$ uncertainties in the model, and the green line represents the best-fit value of the mean function. Top-right panel: the posterior probability density distribution of each parameter in the GP(power law, DRW) model for GRB 060614 The vertical dashed lines indicate the 1$\sigma$ uncertainties. Bottom panel: the PSDs of the light curve of GRB 060614. The average PSD of the GP(power law, DRW) model and the PSD obtained from the LSP method are represented by the blue line and the gray line, respectively. The orange-shaded area indicates the 1$\sigma$ uncertainties in the GP(power law, DRW) model PSD. The red-shaded area represents the unreliable zone, and the white area represents the confidence zone where the upper and lower limits are 1/$P_{min}$ and 1/$P_{max}$, respectively. }
    \label{fig: 060614 ocl GP(pl,drw)}
\end{figure}
\begin{figure}
    \centering
    \includegraphics[width=0.45\textwidth]{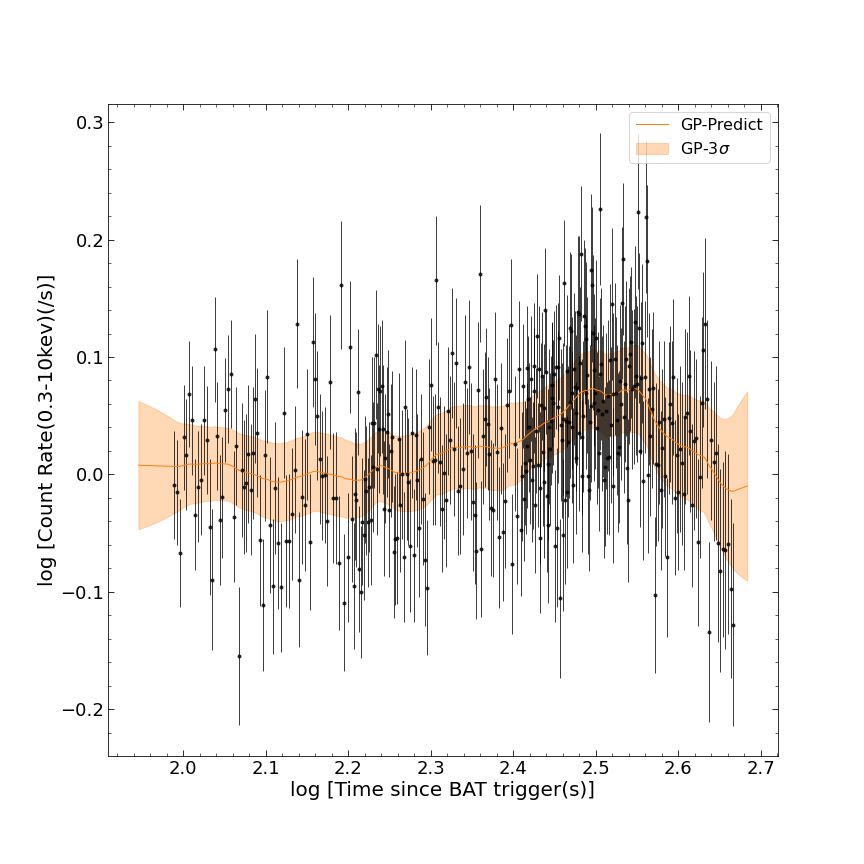}
    {}
    \includegraphics[width=0.45\textwidth]{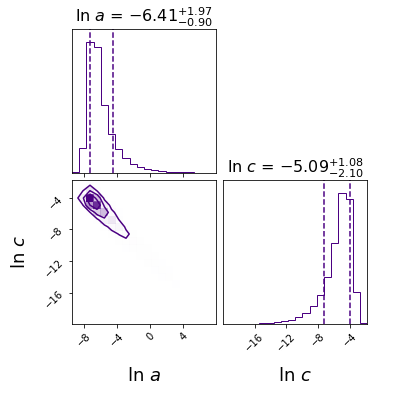}
    \label{fig: detrend MCMC}
    {}
    \includegraphics[width=0.45\textwidth]{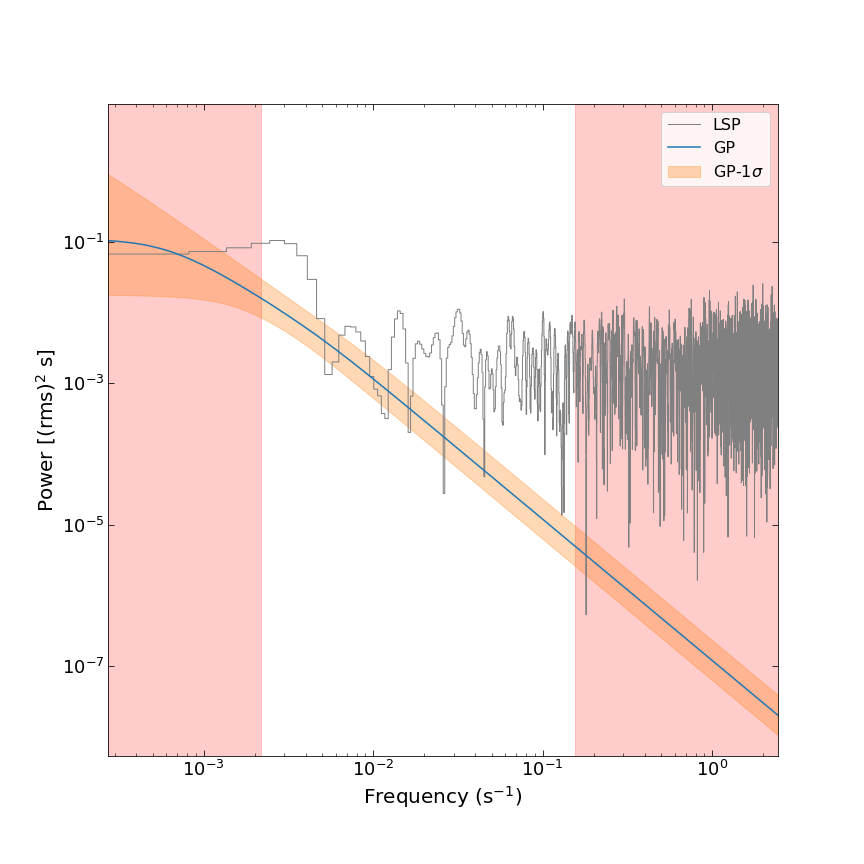}
    \label{fig: detrend PSD}
    {}
    \caption{The results of the GP(constant, DRW) model analysis for the detrended GRB 060614 light curve. The detrended light curve is obtained by subtracting the best-fit mean function in the GP(power law, DRW) model from the original light curve. Top-left panel: the detrended X-ray light curve of GRB 060614 in the time interval of [90-500] s. The black points denote the detrended data, while the orange solid line depicts the prediction of the GP(constant, DRW) model. The orange-shaded region indicates the 3$\sigma$ uncertainties of the model. Top-right panel: the posterior probability density distribution of each parameter in the GP(constant, DRW) model for GRB 060614. The vertical dashed lines indicate the 1$\sigma$ uncertainties. Bottom panel: the PSDs of the light curve of GRB 060614. The average PSD of the GP(constant, DRW) model and the PSD obtained from the LSP method are represented by the blue line and the gray line, respectively. The orange-shaded area indicates the 1$\sigma$ uncertainties in the GP(constant, DRW) model PSD. The red-shaded area represents the unreliable zone, and the white area represents the confidence zone where the upper and lower limits are 1/$P_{min}$ and 1/$P_{max}$, respectively. }
    \label{fig: 060614 dcl GP(pl,drw)}
\end{figure}
\begin{figure}
    \centering
    \includegraphics[width=0.45\textwidth]{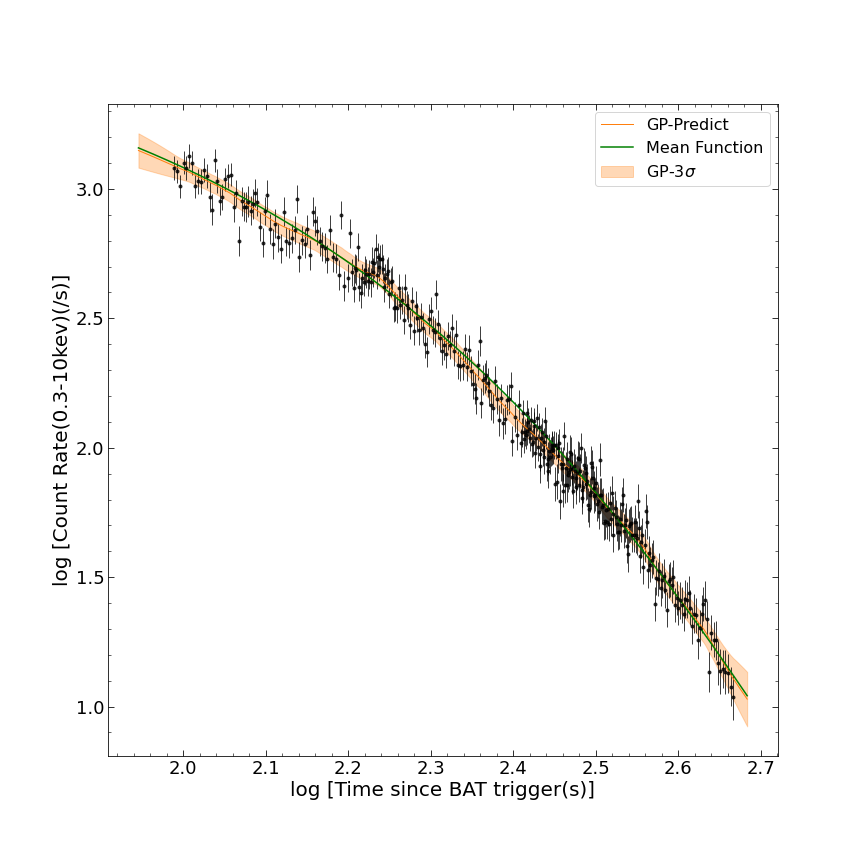}
    {}
    \includegraphics[width=0.45\textwidth]{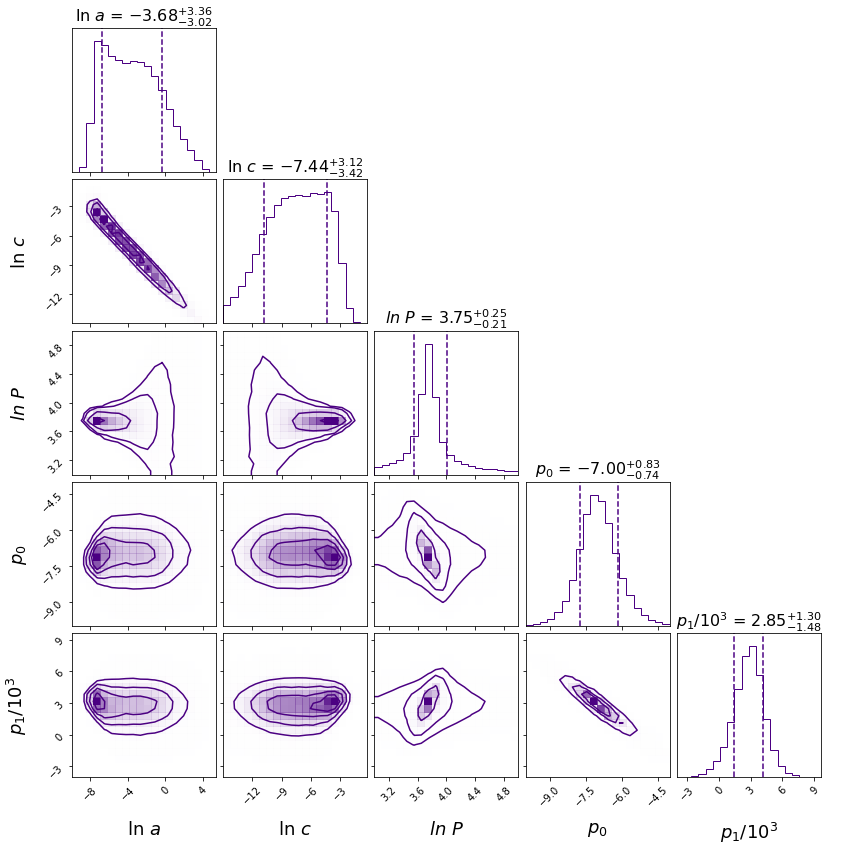}
    \label{fig: detrend MCMC}
    {}
    \includegraphics[width=0.45\textwidth]{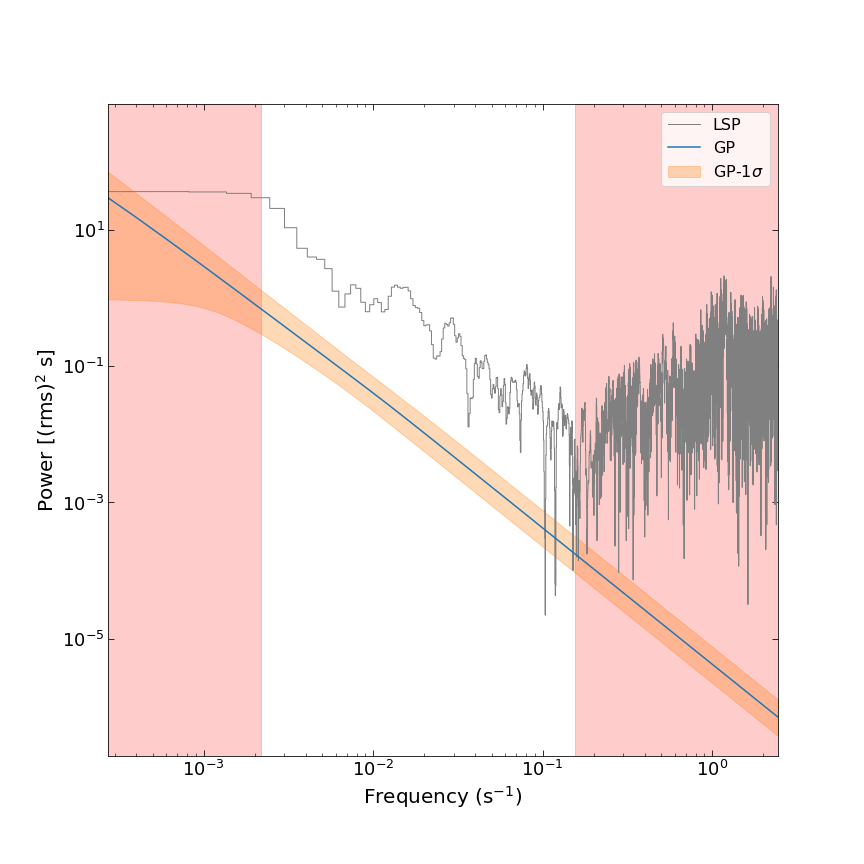}
    \label{fig: detrend PSD}
    {}
    \caption{The results of the GP(polynomial, DRW) model analysis for GRB 060614. Top-left panel: the X-ray light curve of GRB 060614 in the time interval of [90-500] s. The black points denote the data observed by Swift, while the orange solid line depicts the prediction of the GP(polynomial, DRW) model. The orange-shaded region indicates the 3$\sigma$ uncertainties in the model, and the green line represents the best-fit value of the mean function. Top-right panel: the posterior probability density distribution of each parameter in the GP(polynomial, DRW) model for GRB 060614 The vertical dashed lines indicate the 1$\sigma$ uncertainties. Bottom panel: the PSDs of the light curve of GRB 060614. The average PSD of the GP(polynomial, DRW) model and the PSD obtained from the LSP method are represented by the blue line and the gray line, respectively. The orange-shaded area indicates the 1$\sigma$ uncertainties in the GP(polynomial, DRW) model PSD. The red-shaded area represents the unreliable zone, and the white area represents the confidence zone where the upper and lower limits are 1/$P_{min}$ and 1/$P_{max}$, respectively.}
    \label{fig: 060614 ocl GP(poly,drw)}
\end{figure}
\begin{figure}
    \centering
    \includegraphics[width=0.45\textwidth]{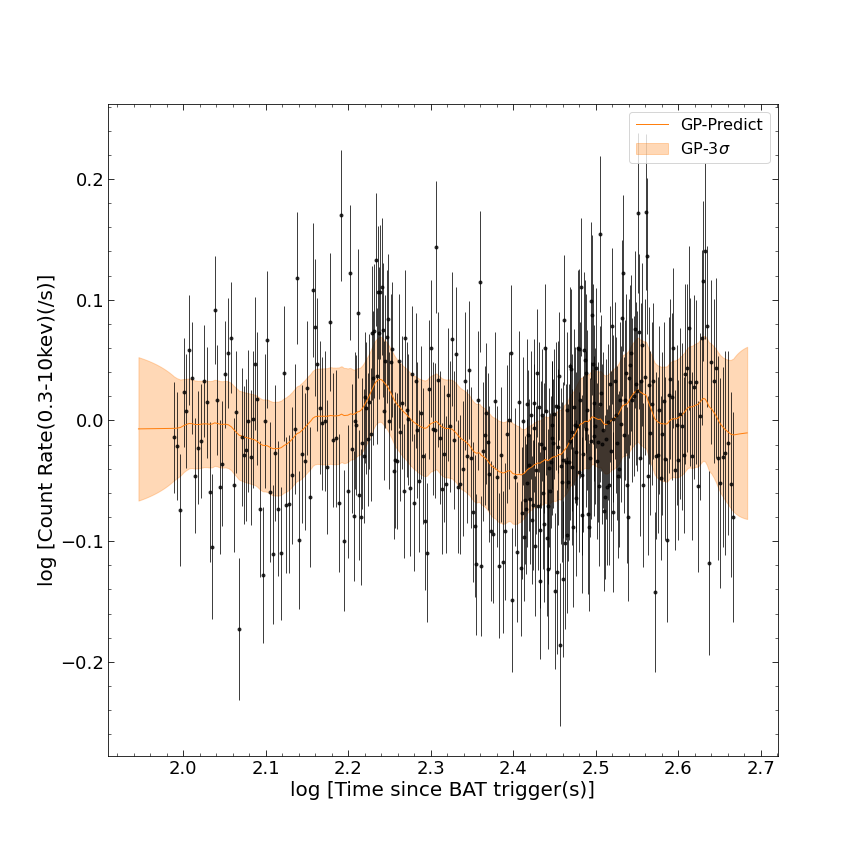}
    {}
    \includegraphics[width=0.45\textwidth]{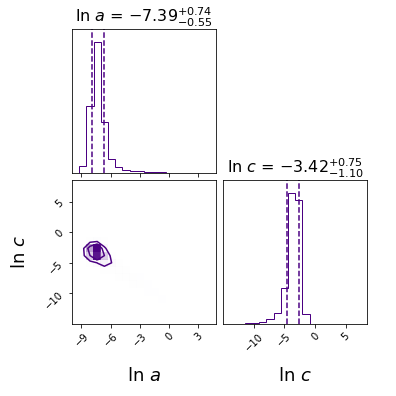}
    \label{fig: detrend MCMC}
    {}
    \includegraphics[width=0.45\textwidth]{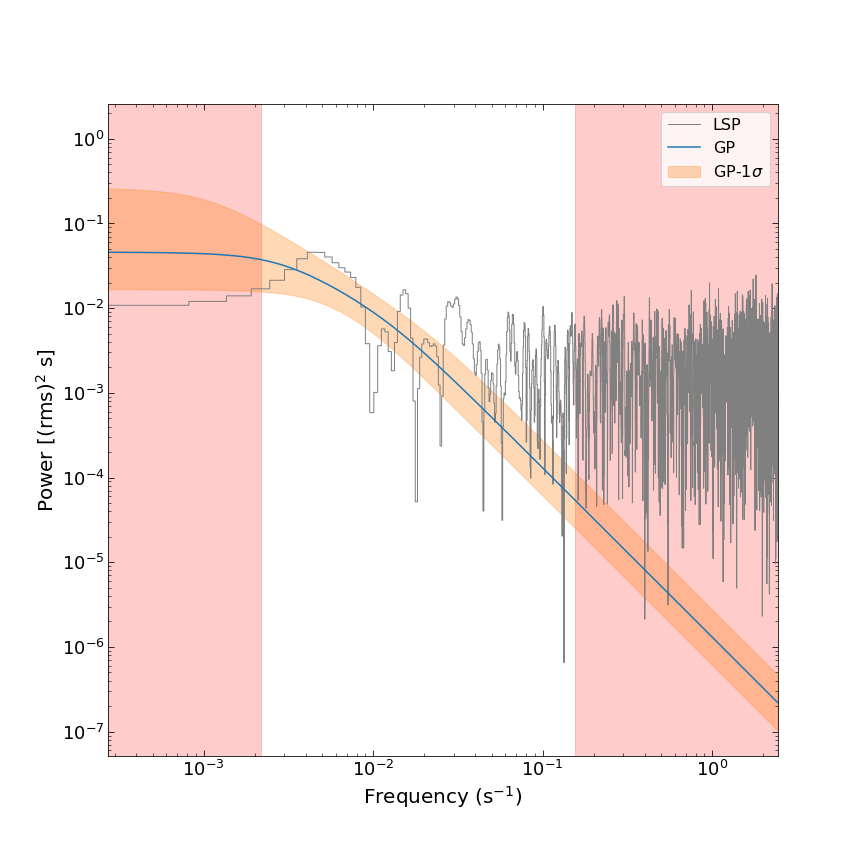}
    \label{fig: detrend PSD}
    {}
    \caption{The results of the GP(constant, DRW) model analysis for the detrended GRB 060614 light curve. The detrended light curve is obtained by subtracting the best-fit mean function in the GP(polynomial, DRW) model from the original light curve. Top-left panel: the detrended X-ray light curve of GRB 060614 in the time interval of [90-500] s. The black points denote the detrended data, while the orange solid line depicts the prediction of the GP(constant, DRW) model. The orange-shaded region indicates the 3$\sigma$ uncertainties in the model. Top-right panel: the posterior probability density distribution of each parameter in the GP(constant, DRW) model for GRB 060614. The vertical dashed lines indecate the 1$\sigma$ uncertainties. Bottom panel: the PSDs of the light curve of GRB 060614. The average PSD of the GP(constant, DRW) model and the PSD obtained from the LSP method are represented by the blue line and the gray line, respectively. The orange-shaded area indicates the 1$\sigma$ uncertainties in the GP(constant, DRW) model PSD. The red-shaded area represents the unreliable zone, and the white area represents the confidence zone where the upper and lower limits are 1/$P_{min}$ and 1/$P_{max}$, respectively.}
    \label{fig: 060614 dcl GP(poly,drw)}
\end{figure}

\begin{figure}
    \centering
    \includegraphics[width=0.45\textwidth]{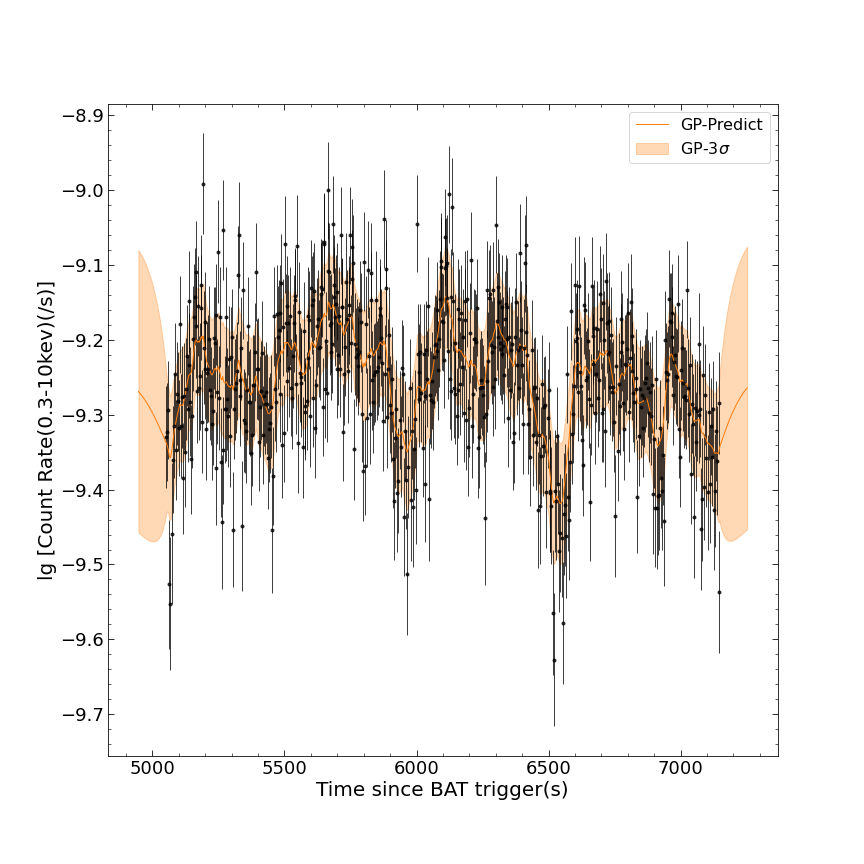}
    \label{fig: detrend MCMC}
    {}
    \includegraphics[width=0.45\textwidth]{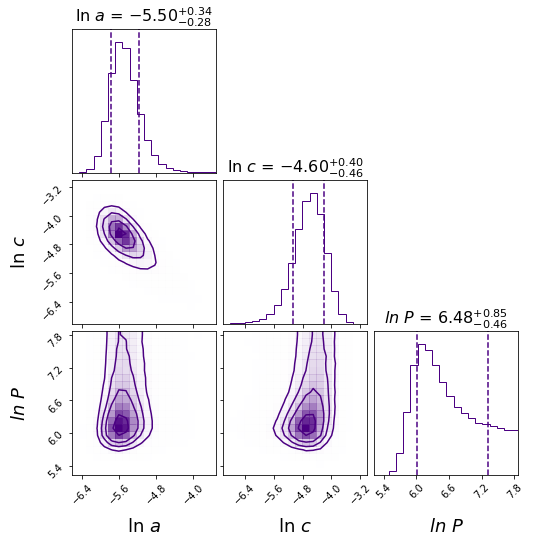}
    \label{fig: detrend MCMC}
    {}
    \includegraphics[width=0.45\textwidth]{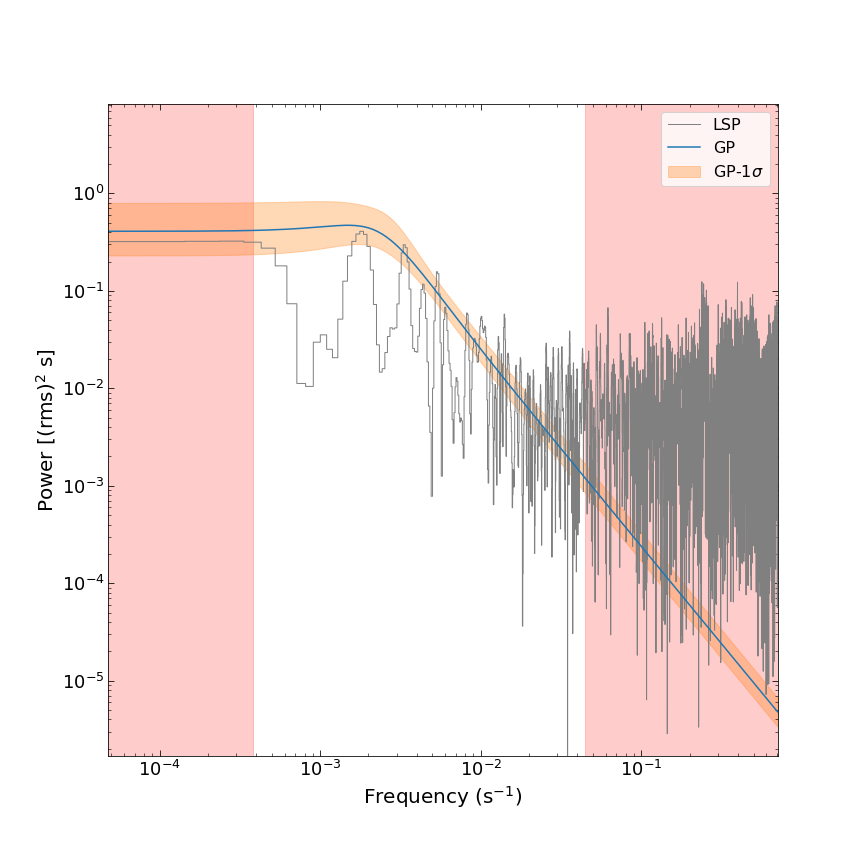}
    \label{fig: detrend PSD}
    {}
    \includegraphics[width=0.45\textwidth]{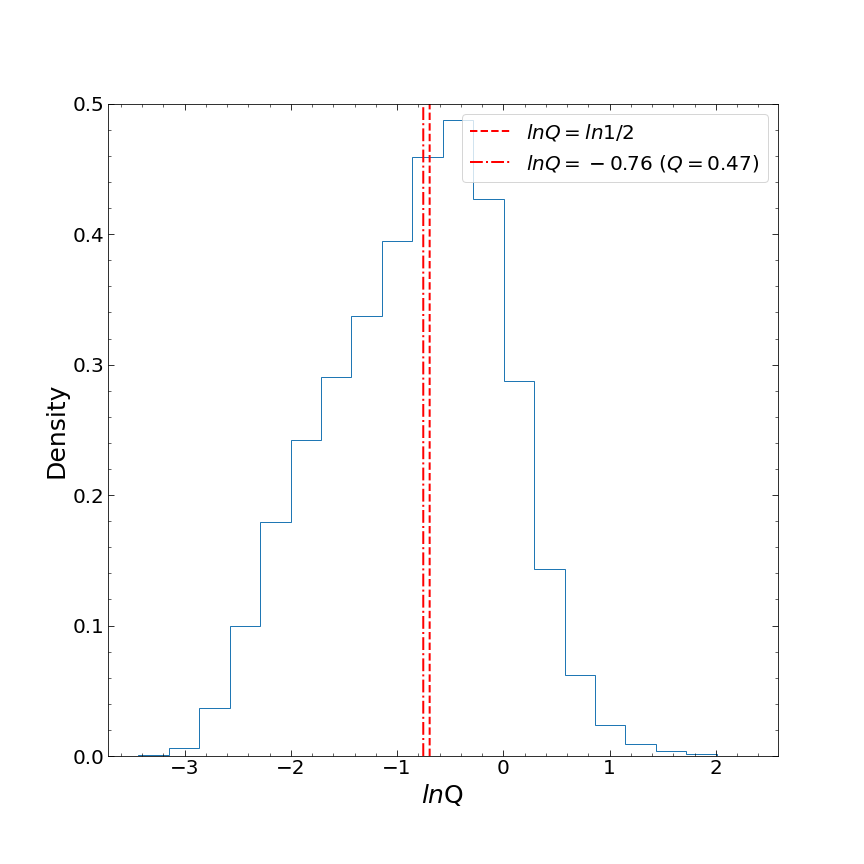}
    \label{fig: detrend Q}
    \caption{The results of the GP(constant, QPO) model analysis for GRB 101225A. Top-left panel: the X-ray light curve of GRB 101225A in the time interval of [5050-7150] s. The black points denote the data observed by Swift, while the orange solid line depicts the prediction of the GP(constant, QPO) model. The orange-shaded region indicates the 3$\sigma$ uncertainties in the model, and the green line represents the best-fit value of the mean function. Top-right panel: the posterior probability density distribution of each parameter in the GP(constant, QPO) model for GRB 101225A. The vertical dashed lines indicate the 1$\sigma$ uncertainties. Bottom-left panel: the PSDs of the light curve of GRB 101225A. The average PSD of the GP(constant, QPO) model and the PSD obtained from the LSP method are represented by the blue line and the gray line, respectively. The orange-shaded area indicates the 1$\sigma$ uncertainties in the GP(constant, QPO) model PSD. The red-shaded area represents the unreliable zone, and the white area represents the confidence zone where the upper and lower limits are 1/$P_{min}$ and 1/$P_{max}$, respectively. Bottom-right panel: the distribution of the quality factor $Q$. The dashed line and the dashed-dotted line indicate the critical quality factor ($Q$ = 0.5) and the median of ln $Q$, respectively.}
    \label{fig: 101225A ocl GP(c,qpo)}
\end{figure}
\begin{figure}
    \centering
    \includegraphics[width=0.45\textwidth]{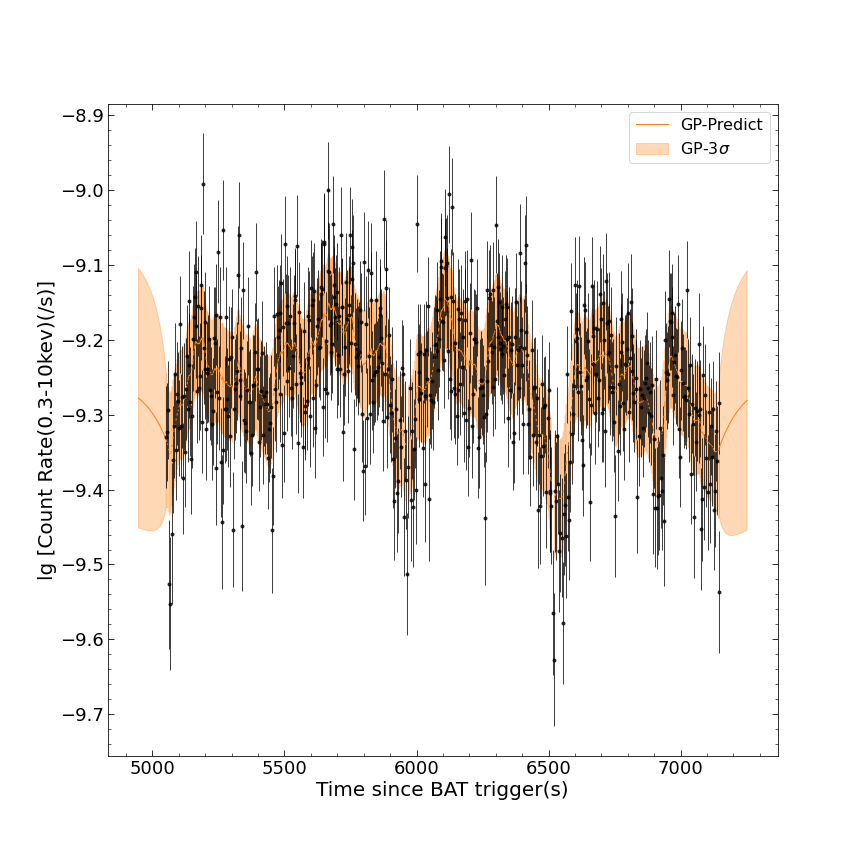}
    \label{fig: detrend MCMC}
    {}
    \includegraphics[width=0.45\textwidth]{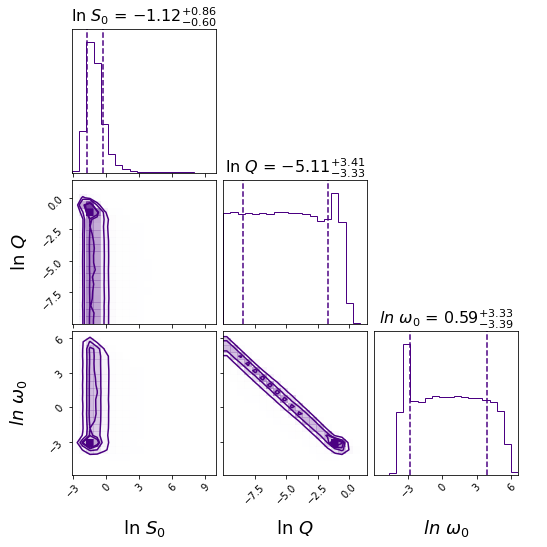}
    \label{fig: detrend MCMC}
    {}
    \includegraphics[width=0.45\textwidth]{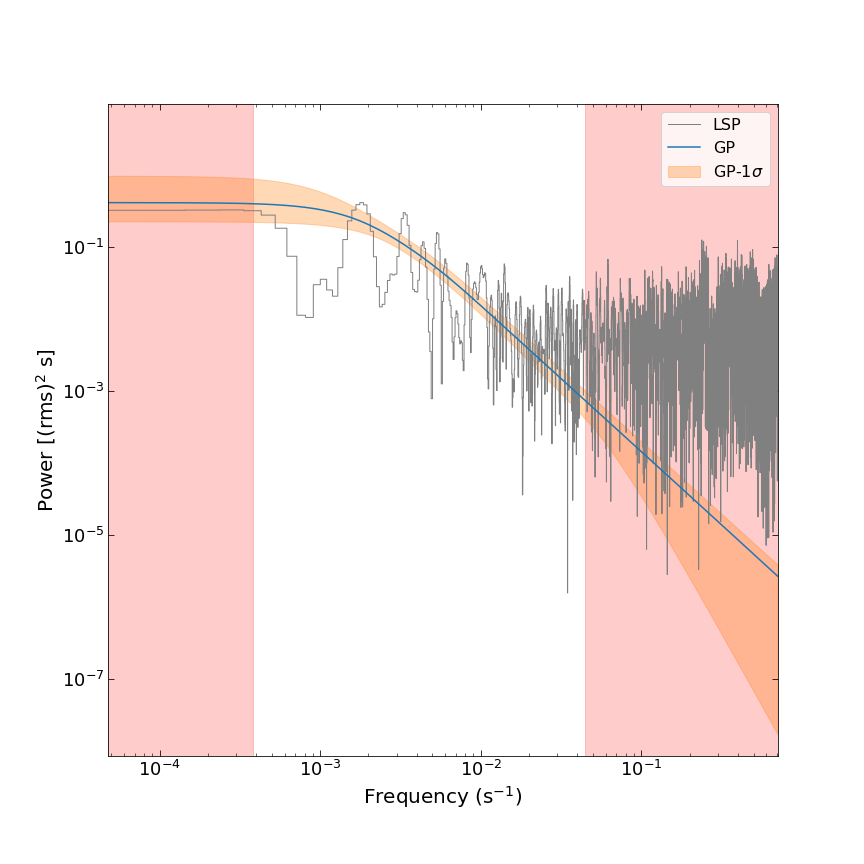}
    \label{fig: detrend PSD}
    {}
    \includegraphics[width=0.45\textwidth]{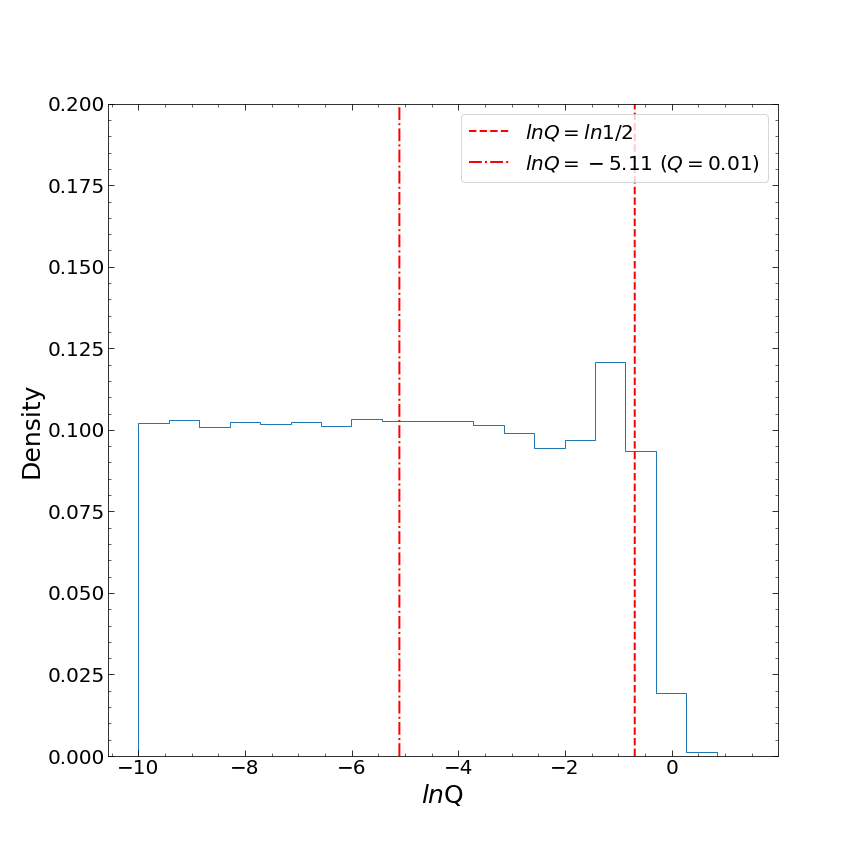}
    \label{fig: detrend Q}
    \caption{The results of the GP(constant, SHO) model analysis for GRB 101225A. Top-left panel: the X-ray light curve of GRB 101225A in the time interval of [5050-7150] s. The black points denote the data observed by Swift, while the orange solid line depicts the prediction of the GP(constant, SHO) model. The orange-shaded region indicates the 3$\sigma$ uncertainties in the model, and the green line represents the best-fit value of the mean function. Top-right panel: the posterior probability density distribution of each parameter in the GP(constant, SHO) model for GRB 101225A. The vertical dashed lines indicate the 1$\sigma$ uncertainties. Bottom-left panel: the PSDs of the light curve of GRB 101225A. The average PSD of the GP(constant, SHO) model and the PSD obtained from the LSP method are represented by the blue line and the gray line, respectively. The orange-shaded area indicates the 1$\sigma$ uncertainties in the GP(constant, SHO) model PSD. The red-shaded area represents the unreliable zone, and the white area represents the confidence zone where the upper and lower limits are 1/$P_{min}$ and 1/$P_{max}$, respectively. Bottom-right panel: the distribution of the quality factor $Q$. The dashed line and the dashed-dotted line indicate the critical quality factor ($Q$ = 0.5) and the median of ln $Q$, respectively.}
    \label{fig: 101225A ocl GP(c,sho)}
\end{figure}
\end{document}